\documentclass[twocolumn]{aastex631}
\usepackage{tabularx} % put in preamble
\usepackage{booktabs} % For formal tables

\usepackage{float}
\makeatletter
\usepackage{hyperref} % Enables hyperlinks
\makeatother
\usepackage{amsmath} % recommended
\usepackage{relsize} % for \mathlarger

\received{June 1, 2019}
\revised{January 10, 2019}
\accepted{\today}
\submitjournal{ApJS}

\shorttitle{JWST observations of 10 and 21~$\mu$m emission in the PHANGS-JWST sample}
\shortauthors{Hassani et al.}
\graphicspath{{./}{figures/}}
\usepackage{xspace}

\begin{document}

\title{The Hidden Life of Stars: Embedded Beginnings to AGB Endings \\ in the PHANGS-JWST Sample. I. Catalog of Mid-IR Sources}

\correspondingauthor{Hamid Hassani}
\email{hhassani@ualberta.ca}

\author[0000-0002-8806-6308]{Hamid Hassani}
\affiliation{Dept. of Physics, University of Alberta, 4-183 CCIS, Edmonton, Alberta, T6G 2E1, Canada}

\author[0000-0002-5204-2259]{Erik Rosolowsky}
\affiliation{Dept. of Physics, University of Alberta, 4-183 CCIS, Edmonton, Alberta, T6G 2E1, Canada}
% \email{rosolowsky@ualberta.ca}
% add your info here, order TBD

\newcommand{\OSU}{\affiliation{Department of Astronomy, The Ohio State University, 140 West 18th Avenue, Columbus, OH 43210, USA}}

\newcommand{\CCAPP}{\affiliation{Center for Cosmology and Astroparticle Physics (CCAPP), 191 West Woodruff Avenue, Columbus, OH 43210, USA}}

\author[0000-0002-2545-1700]{Adam~K.~Leroy}
\OSU
\CCAPP

\author[0000-0002-4378-8534]{Karin Sandstrom}
\affiliation{UCSD}

\author[0000-0003-0946-6176]{Médéric~Boquien}
\affiliation{Université Côte d'Azur, Observatoire de la Côte d'Azur, CNRS, Laboratoire Lagrange, 06000, Nice, France}

\author[0000-0002-8528-7340]{David~A.~Thilker}
\affiliation{Department of Physics and Astronomy, The Johns Hopkins University, Baltimore, MD 21218 USA}

\author[0000-0002-3784-7032]{Bradley~C.~Whitmore}
\affiliation{\STScI}

 \author[0000-0002-5259-2314]{Gagandeep S. Anand}
\affiliation{Space Telescope Science Institute, 3700 San Martin Drive, Baltimore, MD 21218, USA}

\author[0000-0003-0410-4504]{Ashley~T.~Barnes}
\affiliation{European Southern Observatory, Karl-Schwarzschild-Strasse 2, 85748 Garching bei München, Germany}

\author[0000-0001-5301-1326]{Yixian Cao}
\affiliation{Max-Planck-Institut f\"ur Extraterrestrische Physik (MPE), Giessenbachstr. 1, D-85748 Garching, Germany}

\author[0000-0001-8241-7704]{Ryan Chown}
\OSU

\author[0000-0002-8549-4083]{Enrico Congiu}
\affiliation{European Southern Observatory (ESO), Alonso de Córdova 3107, Casilla 19, Santiago 19001, Chile}

\author[0000-0002-5782-9093]{Daniel~A.~Dale}
\affiliation{Department of Physics and Astronomy, University of Wyoming, Laramie, WY 82071, USA}

\author[0000-0002-4755-118X]{
        Oleg V. Egorov}\affiliation{Astronomisches Rechen-Institut, Zentrum f\"{u}r Astronomie der Universit\"{a}t Heidelberg, M\"{o}nchhofstra\ss e 12-14, D-69120 Heidelberg, Germany}

\author[0000-0001-7113-8152]{Ivan Gerasimov}
\affiliation{Université Côte d'Azur, Observatoire de la Côte d'Azur, CNRS, Laboratoire Lagrange, 06000, Nice, France}

\author[0000-0002-3247-5321]{Kathryn~Grasha}
\altaffiliation{ARC DECRA Fellow}
\affiliation{Research School of Astronomy and Astrophysics, Australian National University, Canberra, ACT 2611, Australia}   
\affiliation{ARC Centre of Excellence for All Sky Astrophysics in 3 Dimensions (ASTRO 3D), Australia}

\author[0000-0002-4663-6827]{Rémy~Indebetouw}
\affiliation{National Radio Astronomy Observatory, 520 Edgemont Road, Charlottesville, VA 22903, USA; Department of Astronomy, University of Virginia, Charlottesville, VA 22904, USA}

\author[0000-0003-0946-6176]{Janice C. Lee}
\affiliation{Space Telescope Science Institute, 3700 San Martin Drive, Baltimore, MD 21218, USA}

\newcommand{\STScI}{\affiliation{Space Telescope Science Institute, 3700 San Martin Drive, Baltimore, MD 21218, USA}}

\author[0000-0003-2496-1247]{Fu-Heng Liang}
% \affiliation{European Southern Observatory, Karl-Schwarzschild-Straße 2, 85748 Garching bei München, Germany}
\affiliation{Astronomisches Rechen-Institut, Zentrum f\"{u}r Astronomie der Universit\"{a}t Heidelberg, M\"{o}nchhofstra\ss e 12-14, D-69120 Heidelberg, Germany}

\author[0000-0001-6038-9511]{Daniel Maschmann}
\affiliation{Department of Physics and Astronomy, University of Wyoming, Laramie, WY 82071, USA}

\author[0000-0002-6118-4048]{Sharon E. Meidt}
\affiliation{Sterrenkundig Observatorium, Universiteit Gent, Krijgslaan 281 S9, 9000 Gent, Belgium}

\author[0000-0002-0119-1115]{Elias~K.~Oakes}
\affiliation{Department of Physics, University of Connecticut, 196A Auditorium Road, Storrs, CT 06269, USA}

\author[0000-0002-0873-5744]{Ismael Pessa}
\affiliation{Leibniz-Institut for Astrophysik Potsdam (AIP), An der Sternwarte 16, 14482 Potsdam, Germany}

\author[0000-0003-3061-6546]{J\'er\^ome Pety}
\affiliation{IRAM, 300 rue de la Piscine, 38400 Saint Martin d'H\`eres, France}
\affiliation{LUX, Observatoire de Paris, PSL Research University, CNRS, Sorbonne Universités, 75014 Paris, France}

\author[0000-0002-0472-1011]{Miguel Querejeta}
\affiliation{Observatorio Astron\'{o}mico Nacional (IGN), C/Alfonso XII, 3, E-28014 Madrid, Spain}

\author[0000-0002-9190-9986]{Lise~Ramambason}
\affiliation{Universit\"at Heidelberg, Zentrum f\"ur Astronomie, Institut f\"ur Theoretische Astrophysik, Albert-Ueberle-Str. 2, 69120 Heidelberg, Germany}

\author[0000-0002-0579-6613]{M. Jimena Rodríguez}
\affiliation{Space Telescope Science Institute, 3700 San Martin Drive, Baltimore, MD 21218, USA}
\affiliation{Instituto de Astrofísica de La Plata, CONICET--UNLP, Paseo del Bosque S/N, B1900FWA La Plata, Argentina }

\author[0000-0002-6313-4597]{Sumit K. Sarbadhicary}
\affiliation{Department of Physics and Astronomy, The Johns Hopkins University, Baltimore, MD 21218 USA}

\author[0000-0002-9183-8102]{Jessica Sutter}
\affiliation{Whitman College, 345 Boyer Avenue, Walla Walla, WA 99362, USA}

\author[0000-0001-7130-2880]{Leonardo \'Ubeda}
\affiliation{Space Telescope Science Institute, 3700 San Martin Drive, Baltimore, MD 21218, USA}

\author[0000-0002-0012-2142]{Thomas G. Williams}
\affiliation{Sub-department of Astrophysics, Department of Physics, University of Oxford, Keble Road, Oxford OX1 3RH, UK}

\suppressAffiliations

% \nocollaboration{2}



\begin{abstract}
 We present a multiwavelength catalog of mid-infrared-selected compact sources in 19 nearby galaxies, combining JWST NIRCam/MIRI, HST UV-optical broadband, H$\alpha$ narrowband, and ALMA CO observations. We detect 24,945 compact sources at 21~$\mu$m and 55,581 at 10~$\mu$m. Artificial star tests show 50\% completeness limits of $\sim$5~$\mu$Jy for the 10~$\mu$m catalog, and $\sim$24~$\mu$Jy for the 21~$\mu$m catalog. We find that 21~$\mu$m compact sources contribute $\sim$20\% of the total galaxy emission in that band, but only contribute $5\%$ at 10~$\mu$m. We classify sources using stellar evolution and population synthesis models combined with empirical classifications derived from the literature. Our classifications include H$\alpha$-bright and dust-embedded optically faint clusters, red supergiants (RSGs), oxygen-rich and carbon-rich AGB stars, and a range of rarer stellar types. In sampling a broad range of star-forming environments with a uniform, well-characterized selection, this catalog enables enables analyses of infrared-bright stellar populations. We find that H$\alpha$-faint sources account for only 10\% of dusty (likely young) clusters, implying that the infrared-bright, optically faint phase of cluster evolution is short compared to the H$\alpha$-bright stage. The luminosity functions of 10 and 21~$\mu$m sources follow power-law distributions, with the 21~$\mu$m slope ($-1.7 \pm 0.1$) similar to that of giant molecular cloud mass functions and ultraviolet bright star-forming complexes, while the 10~$\mu$m slope ($-2.0 \pm 0.1$) is closer to that of young stellar clusters.
\end{abstract}
%removed from asbstract:

 % Our goal is to characterize the nature and evolutionary role of 10 and 21~$\mu$m point sources, ranging from embedded star-forming regions to evolved stellar populations such as asymptotic giant branch (AGB) stars.
 % Using a flux-conservative method, we recovered faint sources previously blended with diffuse emission, 

 % Overall, our results highlight the importance of mid-infrared emission in tracing both the earliest dust-enshrouded phases of star formation and the later stages of stellar evolution, providing critical insight into the interconnected lifecycle of stars and dust, from their birth in molecular gas to their eventual enrichment of the interstellar medium by evolved stellar populations.



\section{Introduction} \label{sec:intro}

Dust grains are fundamental components of the interstellar medium (ISM). Along with polycyclic aromatic hydrocarbons (PAHs), they reshape the radiation fields of galaxies \citep{Draine2014}. Most of the mid-infrared (mid-IR) emission from galaxies arises from continuum radiation emitted by stochastically heated small dust grains \citep[][]{Draine2001} and the stretching and bending modes of polycyclic aromatic hydrocarbons (PAHs).  The PAH emission manifests as distinct spectral bands \citep[e.g., 3.3, 7.7, and 11.3~$\mu$m,][and references therein]{Tielens}.

This infrared radiation, which experiences lower extinction than optical light, serves as an effective tracer of the ISM, offering valuable insights into the formation and evolution of star-forming regions as well as the life cycle of dusty stars \citep{Boyer_smc,hassani23}. Mid-IR emission can be a tracer of the (recent) star formation rate (SFR), complementing recombination line measurements (e.g., H$\alpha$), particularly in dust-obscured systems where H$\alpha$ could be faint \citep{Belfiore2023, Calzetti2024}. This emission traces reprocessed short-wavelength radiation in the luminous but dust-obscured early phases of stellar evolution. Star-forming regions in the early stages of evolution can be bright in mid-IR emission, which can persist for a few million years, often in tandem with H$\alpha$ \citep{hassani23,Whitmore2023}. 

% \textbf{\citet{Kim_2023} found that the duration of this embedded phase is approximately 5~Myr, during which the H$\alpha$ emission is heavily obscured for the first 2~Myr.}

The mid-IR emission also traces the evolution of individual stars. Dust in the circumstellar envelopes of Red Supergiants (RSGs) and Asymptotic Giant Branch (AGB) stars contributes to their infrared emission, with typical dust temperatures of order 1000–1500~K \citep{HH}. These ejecta enrich the ISM and contribute significantly to dust production in galaxies \citep{Goldman_m31, Matsuura2009}.
%. In M31, AGB stars alone may account for up to 35 percent of the total dust budget, although this estimate depends on dust grain lifetimes \citep{Goldman_m31}. In the Magellanic Clouds, however, the contribution from AGB stars is likely lower \citep{Matsuura2009}.}
Different AGB types generate dust particles with distinct physical and chemical properties. Oxygen-rich AGB stars form silicates, whereas carbon-rich AGBs produce amorphous carbon and graphite \citep{Karovicova,Gobrecht,HH}. \added{Mid-IR data ($\lambda > 3,\mu$m) remain essential for accurate classification of AGB stars, as many dusty carbon-rich AGB (C-AGB) stars—especially those in the thermally pulsing AGB (TP-AGB) phase—can be faint at optical wavelengths due to heavy circumstellar dust obscuration of the photosphere, while undergoing large thermal pulses and returning a considerable fraction of their material (up to $\sim$80\%) to the interstellar medium via dense stellar winds \citep{Groenewegen2018}.}

Thanks to the \textit{Spitzer} and \textit{Herschel} space telescopes, many studies have targeted nearby galaxies, such as the Magellanic Clouds and M31, exploring both stellar clusters and different types of stellar populations. These studies have yielded insights on the mechanisms of dust production in galaxies \citep{Matsuura2009, Matsuura2013, Jones2017, Goldman_m31} and explored the earliest stages of high mass star-forming regions, making a direct connection to the host clouds \citep{whitney2008, Bonanos, seale2014, Jones2018}. However, the characterization of these stellar sources is challenging in galaxies beyond 3~Mpc, given the limited resolution of \textit{Spitzer} and \textit{Herschel}. New observations with the James Webb Space Telescope (JWST) provide an opportunity to isolate and characterize high mass star-forming regions and bright, dusty stars in in more distant targets. Pushing to these larger distances is essential to place these populations in the context of a more diverse set of host galactic environments.

In this paper, we present catalogs of compact sources emitting at 10 and 21~$\mu$m in JWST imaging of 19 nearby galaxies ($D = 5$-20~Mpc). At these distances, both star-forming regions and evolved stars will appear as point-like or marginally resolved sources that we collectively refer to as ``compact'' \citep{hassani23,Pathak2024}, corresponding to physical scales below 10 to 60 pc at the distances to our targets. These compact source catalogs offer new opportunities to understand both star formation and the later stages of stellar evolution in the context of the larger galactic environment. These observations and catalog are part of the Physics at High Angular Resolution in Nearby GalaxieS (PHANGS) survey, which also provides high resolution ancillary data of molecular gas from the Atacama Large Millimeter/submillimeter Array \citep[PHANGS-ALMA,][]{Leroy2021}, high resolution optical imaging in both broadband filters and in the H$\alpha$ line from the Hubble Space Telescope \citep[PHANGS-HST,][]{phangs-hst, Chandar25}, optical integral field unit spectroscopy \citep[PHANGS-MUSE,][]{phangs-muse} \added{, and resolved ultraviolet imaging from AstroSat UVIT \citep{Hassani_astrosat}.}

This study expands on previous findings of \citet{hassani23}, extending the analysis from \added{4} to 19 targets and adding a 10~$\mu$m catalog to the 21~$\mu$m catalog presented in that paper. In this work, we develop a photometric method and establish a classification framework for these compact sources. A companion paper (Hassani et al., in prep.) will focus on the physical properties of young clusters through SED fitting with physically-motivated models. The present work focuses on how we isolate compact sources from diffuse emission (Section~\ref{sec:diffuse}), how we detect sources and measure photometry in JWST and supporting data (Sections~\ref{sec:source_finder} and \ref{sec:photo}), and how we establish completeness limits for these catalogs (Section~\ref{sec:ccomp}). We then develop a classification framework to use SED shape to identify young clusters, AGB stars, and RSGs in Section~\ref{sec:classify}. We present our results in Section~\ref{sec:res}, including the correlation between F2100W and H$\alpha$ luminosity in Section~\ref{sec:corr}, and the luminosity distribution of young clusters in Section~\ref{sec:physics}.

\begin{figure*}[!t]
    \centering
    \includegraphics[width=0.95\textwidth]{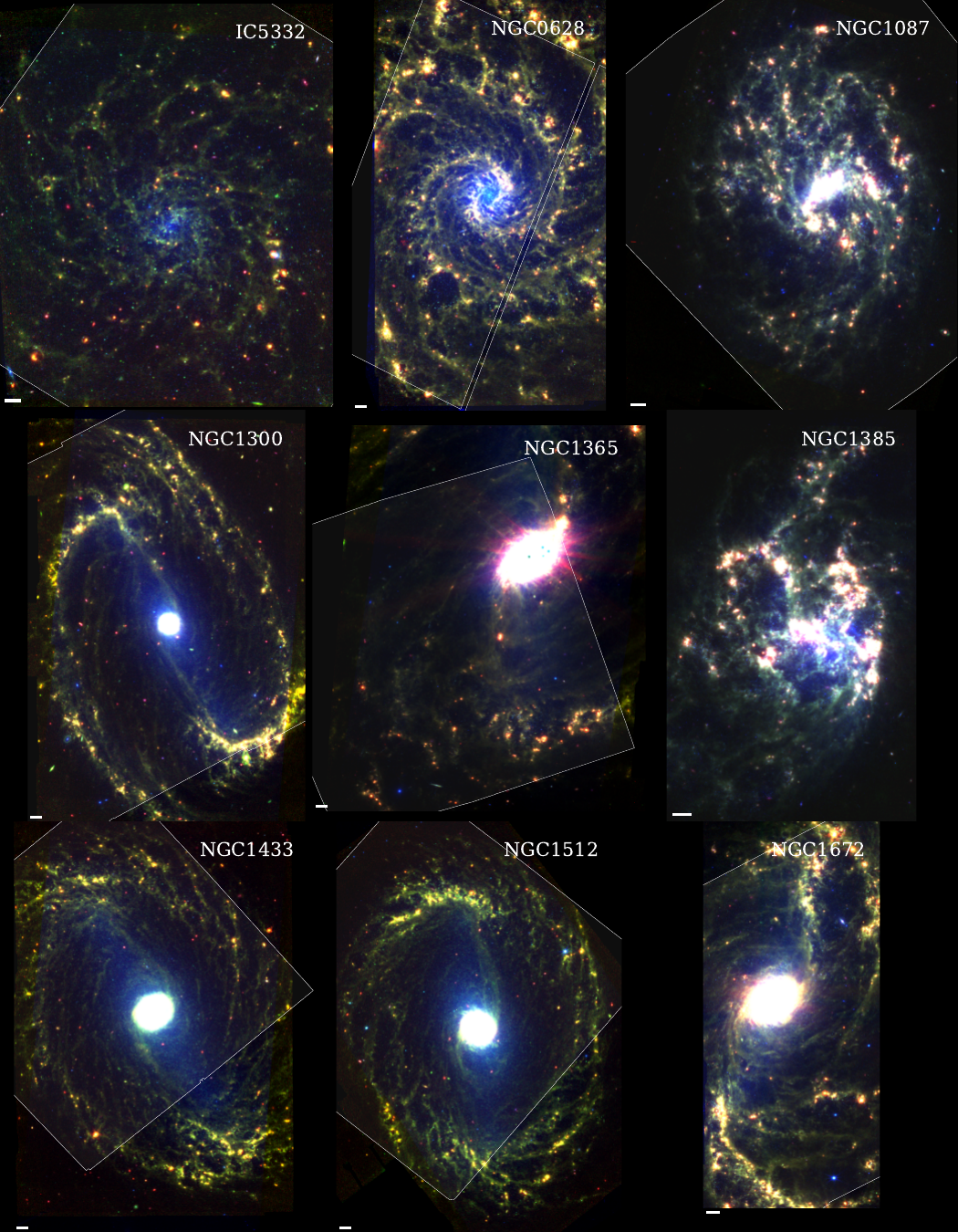}
    \caption{False-color RGB maps of the PHANGS-JWST sample at wavelengths of 21~$\mu$m (red), 10~$\mu$m (green), and 3.35~$\mu$m (blue). The white shaded area represents the HST H$\alpha$ narrowband footprints. The scale bar represents 5 arcseconds.}
    \label{fig:fig1}
\end{figure*}

\begin{figure*}[!t]
    \centering
    \includegraphics[width=0.95\textwidth]{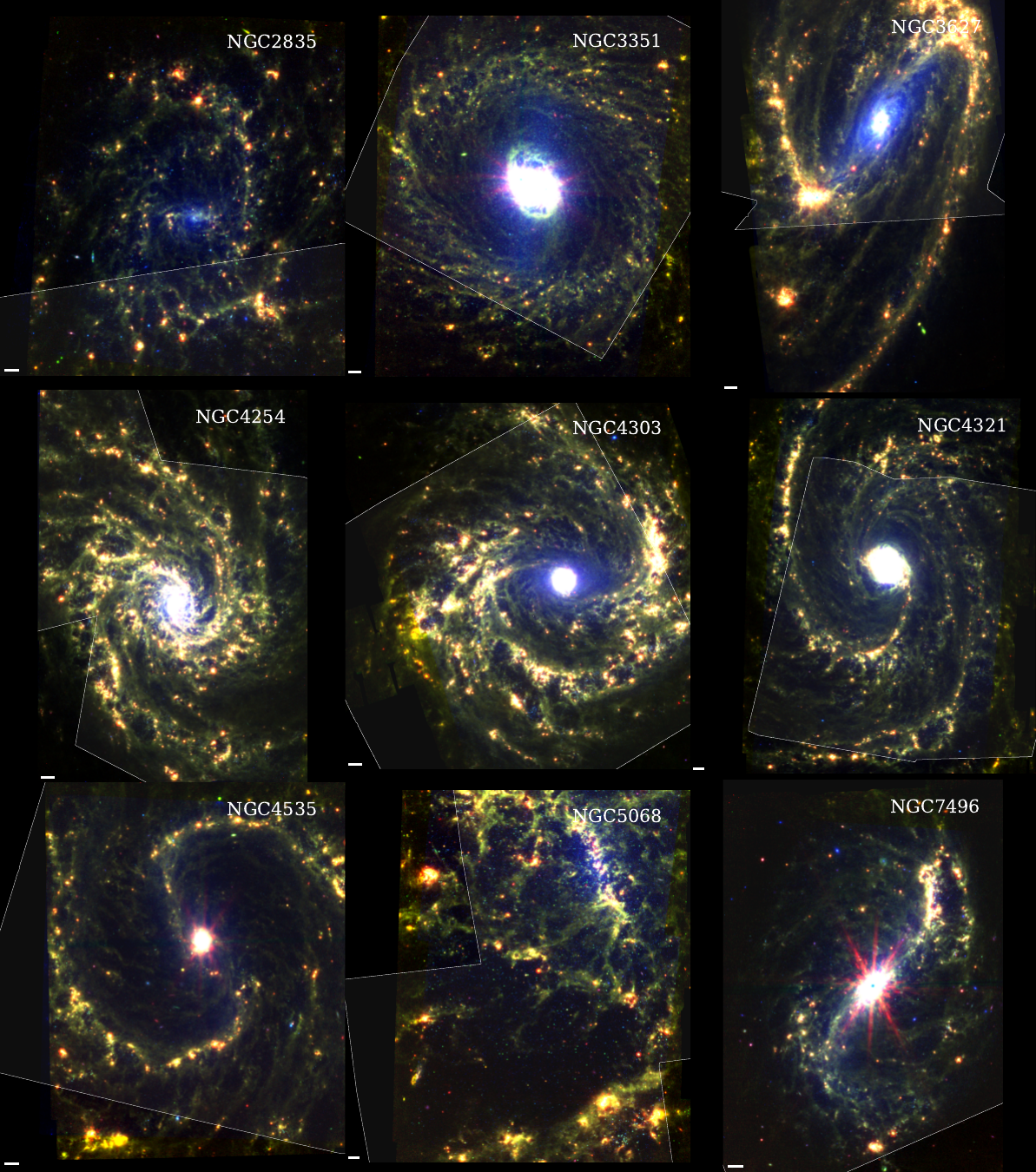}
\caption{Continuation of Figure~\ref{fig:fig1}.}
    \label{fig:fig2}
\end{figure*}

\section{Data}
\label{sec:observations}
\subsection{Sample Selection}
\label{sec:sample_selection}
In this work, we study a set of 19 nearby galaxies at distances between 5 and 20~Mpc, all of which have been observed as part of the PHANGS survey. The full PHANGS sample consists of 90 galaxies, with stellar masses in the range of $10^{9.75} < M_\star / M_\odot < 10^{11.0}$, most of them on the star-forming main sequence, and characterized by a specific star formation rate of $\mathrm{SFR}/M_\star > 10^{-11}~\mathrm{yr}^{-1}$ \citep{Leroy2021}. Additionally, the selection criteria exclude highly inclined systems, restricting the sample to those with an inclination angle of $i < 75^\circ$.  Here, we focus on the subset of 19 galaxies from the original 90, which have been observed with JWST MIRI/NIRCam (Cycle 1), HST broadband \citep{phangs-hst} and H$\alpha$ narrowband imaging \citep{Chandar25}, and MUSE IFU \citep{phangs-muse}. Table \ref{tab:galaxies} lists the galaxies included in this study along with their properties, as reported in \cite{lee2023} and \cite{Leroy2021}.

\subsection{JWST Imaging}

The PHANGS–JWST observations were obtained under JWST program 2107 (PI: J. Lee) \added{and are available from the Mikulski Archive for Space Telescopes (MAST) at the Space Telescope Science Institute via  \dataset[doi: 10.17909/ew88-jt15]{https://doi.org/10.17909/ew88-jt15}}. The imaging program used NIRCam and MIRI in eight filters spanning central wavelengths from 2, 3, 3.3, 3.6, 7.7, 10, 11.3, to 21~$\mu$m \citep{lee2023}. The observations use one to two pointings with NIRCam (Module B only) and two to four pointings with MIRI for each galaxy. The exposure durations per pointing vary by filter, ranging from 6.4 to 20 minutes for NIRCam and 1.5 to 5.4 minutes for MIRI. We use the data reduction pipeline \texttt{pjpipe}, described in \citet{Williams2024}, which reduces $1/f$ noise in NIRCam, improves background flux consistency, and provides better calibration for both relative and absolute astrometry compared to MAST images\footnote{\href{https://archive.stsci.edu/hlsp/phangs/phangs-jwst}{PHANGS-JWST Archive}}. In this work, as we aim to identify bright 10$\mu$m and 21$\mu$m compact sources, we generate two sets of images, each at the resolution of its respective catalog. For a 10~$\mu$m catalog, all maps are convolved to match the resolution of the F1130W filter (0\farcs36) using the \texttt{jwst\_kernels}\footnote{\url{https://github.com/francbelf/jwst_kernels}} code. Similarly, for the 21~$\mu$m catalog, we convolve all maps to the resolution of F2100W (0.67\arcsec). The original map units remain in MJy/sr. Figures \ref{fig:fig1} and \ref{fig:fig2} show false-color RGB images of these galaxies based on NIRCam and MIRI JWST data at their native resolution.

\subsection{HST Imaging}

We use PHANGS-HST broadband observations of our sample \citep{phangs-hst}, \added{which are also available from MAST via \dataset[doi: 10.17909/t9-r08f-dq31]{ https://doi.org/10.17909/t9-r08f-dq31}}. These observations were conducted using the WFC3/UVIS camera in parallel mode with ACS/WFC (Cycle 26, PID 15654), utilizing the filter set F275W (NUV), F336W (U), F438W (B), F555W (V), and F814W (I). The total exposure times for each filter are about: $\sim2200$~s (NUV), $\sim1100$~s (U), $\sim1100$~s (B), $\sim670$~s (V), and $\sim830$~s (I). All images were drizzled and aligned to the native pixel scale of WFC3, which is 0.04\arcsec\ and the astrometry was established using GAIA DR2 sources \citep{GaiaMission}. Further detail regarding the imaging pipeline and photometric calibration are provided in \cite{phangs-hst}. To match the resolution of the 10 and 21~$\mu$m catalogs, we convolved all HST maps accordingly. Because of the large difference in resolution between the HST and the resulting JWST/MIRI resolutions, the shape of the high-resolution PSF has negligible effect on the final images. Thus, for the 10~$\mu$m catalog, all maps were convolved to 0.36\arcsec, using F200W to F1130W kernels. Similarly, for the 21~$\mu$m catalog, we applied F200W to F2100W kernels and converted all units to MJy/sr for our HST maps.

In the case of HST narrowband H$\alpha$ imaging, we use data from the PHANGS-HST H$\alpha$ survey \citep{Chandar25}.  The survey imaged galaxies with the WFC3 camera using either the F657N or F658N filter, depending on redshift. Here, we use the flux-calibrated version of the images, which uses MUSE observations to correct for \ion{N}{2} contamination. Similar to the HST wideband maps, we convolved all narrowband maps to match the 10~$\mu$m and 21~$\mu$m resolutions using kernels from F200W to F1130W and from F200W to F2100W, respectively. The final images are converted to units of MJy/sr.

\begin{table*}
\centering
\begin{tabular}{lccccccc}
\hline
Galaxy & $\alpha^{a}$ & $\delta^{a}$ & $D^{a}$  & SFR$^{a}$  & log $\Sigma_\mathrm{SFR}^{b}$ & ALMA Beam$^{c}$ & MUSE PSF$^{d}$  \\
& (J2000) &  (J2000) & (Mpc) & (M$_\odot$ yr$^{-1}$) &  (M$_\odot$ yr$^{-1}$ kpc$^{-2}$)  & ($\arcsec$) & ($\arcsec$)\\
\hline
IC 5332   & 23h34m27.49s & -36d06m03.9s  & 9.01  & 0.4  & $-2.82$  & 0.74  & 0.87 \\
NGC~0628  & 01h36m41.75s & +15d47m01.2s  & 9.84  & 1.7  & $-2.36$  & 1.12  & 0.92 \\
NGC~1087  & 02h46m25.16s & -00d29m55.1s  & 15.85 & 1.3  & $-2.25$  & 1.60  & 0.92 \\
NGC~1300  & 03h19m41.08s & -19d24m40.9s  & 18.99 & 1.2  & $-2.94$  & 1.23  & 0.89 \\
NGC~1365  & 03h33m36.37s & -36d08m25.4s  & 19.57 & 17.0 & $-2.49$  & 1.38  & 1.15 \\
NGC~1385  & 03h37m28.85s & -24d30m01.1s  & 17.22 & 2.1  & $-2.10$  & 1.27  & 0.77 \\
NGC~1433  & 03h42m01.55s & -47d13m19.5s  & 18.63 & 1.1  & $-3.04$  & 1.10  & 0.91 \\
NGC~1512  & 04h03m54.28s & -43d20m55.9s  & 18.83 & 1.3  & $-3.05$  & 1.03  & 1.25 \\
NGC~1566  & 04h20m00.42s & -54d56m16.1s  & 17.69 & 4.6  & $-2.15$  & 1.25  & 0.80 \\
NGC~1672  & 04h45m42.50s & -59d14m49.9s  & 19.40 & 7.6  & $-2.30$  & 1.93  & 0.96 \\
NGC~2835  & 09h17m52.91s & -22d21m16.8s  & 12.22 & 1.3  & $-2.36$  & 0.84  & 1.15 \\
NGC~3351  & 10h43m57.70s & +11d42m13.7s  & 9.96  & 1.3  & $-2.81$  & 1.46  & 1.05 \\
NGC~3627  & 11h20m14.96s & +12d59m29.5s  & 11.32 & 3.9  & $-2.00$  & 1.63  & 1.05 \\
NGC~4254  & 12h18m49.60s & +14d24m59.4s  & 13.10 & 3.1  & $-2.11$  & 1.78  & 0.89 \\
NGC~4303  & 12h21m54.90s & +04d28m25.1s  & 16.99 & 5.4  & $-2.02$  & 1.81  & 0.78 \\
NGC~4321  & 12h22m54.83s & +15d49m18.5s  & 15.21 & 3.5  & $-2.42$  & 1.67  & 1.16 \\
NGC~4535  & 12h34m20.31s & +08d11m51.9s  & 15.77 & 2.2  & $-2.47$  & 1.56  & 0.56 \\
NGC~5068  & 13h18m54.81s & -21d02m20.8s  & 5.20  & 0.3  & $-2.20$  & 1.04  & 1.04 \\
NGC~7496  & 23h09m47.29s & -43d25m40.6s  & 18.72 & 2.2  & $-2.49$  & 1.68  & 0.89 \\
\hline
\end{tabular}
\caption{Main physical properties of the PHANGS sample.\\ $^{a,b}$ Values taken from \cite{lee2023}. The SFR represents the total star formation rate of each galaxy, derived from GALEX far-UV and WISE W4 maps \citep{z0mgs}.\\ 
$^{b}$ The data are sourced from \cite{Santoro,z0mgs}, based on measurements from GALEX and WISE observations of entire galaxy. \\
$^{c}$ Adopted from \cite{Leroy2021}, where all targets have been observed using the 12~m+7~m+Total Power (TP) array configuration. \\
$^{d}$ Based on the FWHM of the Gaussian PSF of the homogenised mosaic from \cite{phangs-muse}.}
\label{tab:galaxies}
\end{table*}

\subsection{Supporting Data Sets}

\paragraph{ALMA CO(2-1)} -- We used CO\,(2-1) emission line maps from the PHANGS–ALMA project \citep{Leroy2021}. These emission lines were observed at a rest frequency of $\nu=230.538$~GHz \citep[Band 6,][]{Leroy2021} in 2.5~km~s$^{-1}$ channels. The typical 1$\sigma$ noise level in the spectral cubes for all targets is  $\sim 6.2$~mJy~beam$^{-1}$ at the native resolution, corresponding to a brightness temperature of  $\sim 0.17$~K. The field of view (FoV) of ALMA largely coincides with that of JWST NIRCam/MIRI, as both datasets primarily probe the inner disk regions of the targets. The ALMA beam size varies from \added{about} 0.74\arcsec\ to 1.81\arcsec\ across different targets.  The survey data thus have a wide range of physical resolutions: \added{NGC 5068 has a resolution of approximately 25~pc}, while NGC~1672 is approximately 180~pc. We summarized the ALMA beam size for our targets in Table \ref{tab:galaxies}.

\paragraph{VLT-MUSE} -- We also incorporated data from the PHANGS-MUSE survey (program IDs: 1100.B-0651, PI: E. Schinnerer; 095.C-0473, PI: G. Blanc; and 094.C-0623, PI: K. Kreckel). This dataset covers the same set of 19 nearby, star-forming spiral galaxies and as described in \citet{phangs-muse}. Each galaxy is observed with 3 to 15 different pointings to ensure sufficient coverage of its disk. The observations were conducted in the wide-field mode (WFM) of MUSE. The data cubes cover wavelengths from 4800 to 9300~\AA\ with a spectral resolution of $\sim 2.75$~\AA\ (FWHM) but this varies with wavelength with a Resolving Power of $\sim2000$ around H$\alpha$ emission line.  For this work, we use emission line maps from \citet{phangs-muse}, which were generated using the MUSE Data Analysis Pipeline (DAP). This pipeline simultaneously fits emission lines and models the underlying stellar continuum. In particular, we rely on H$\alpha$ and H$\beta$ emission line maps that are corrected for the Milky Way Foreground extinction as well as the $E(B-V)$ reddening maps from \cite{Belfiore2023}, which also allow us to correct our HST H$\alpha$ narrowband images for dust attenuation. We also use the stellar continuum of the MUSE data combined with the H$\alpha$ to measure the H$\alpha$ equivalent width (EW), since this line-to-continuum measure is an empirical proxy for stellar population age \citep{Levesque_2, Leitherer99}.  We note the FWHM of the Gaussian PSF of the mosaics is $\gtrsim 0.9\arcsec$~for most of our targets, which is larger than our working resolution at 10 or 21 $\mu$m (see Table \ref{tab:galaxies}).

\paragraph{Morphological Environments} -- We further use the morphological masks of sub-galactic environments from \cite{Querejeta2021}. These maps are derived from Spitzer 3.6~$\mu$m images and classify different environments within our sample, including the center, bar, spiral arms, interarm regions, and the disk. We note that the central region primarily corresponds to stellar structures within $R < 10\arcsec$, while the bar ends are defined as the areas where the bar overlaps with the spiral arms, if such overlap occurs. More details on the classification criteria and the size of each environment can be found in \cite{Querejeta2021}.

\section{Catalog Generation}

\begin{figure*}[!t]
    \centering
        \includegraphics[width=0.9\linewidth]{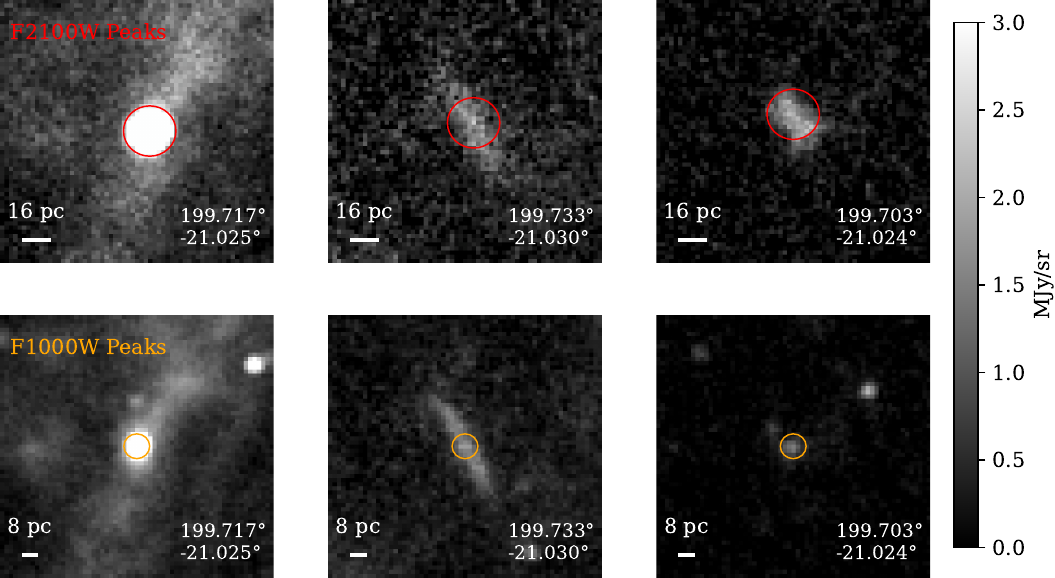}
    \caption{Three peaks with various shapes at F2100W (top) and F1000W (bottom) in NGC~5068 with a 7\arcsec\ cutout size. The sources, depicted as circular (left), elongated (center), and irregular (right), are highlighted with red and orange circles corresponding to the PSF sizes of F2100W and F1000W filters. The position in the respective catalogs is noted in the bottom right corner, and the physical resolution is in the left.}
    \label{fig:sources_type}
\end{figure*}

In this study, our goal is to generate a catalog of compact mid-infrared continuum (i.e, F1000W and F2100W) emission sources, referred to as ``peaks.'' We first visually inspect the images of the different galaxies to examine different types of compact sources and inform the design of our catalog algorithm. Figure \ref{fig:sources_type} illustrates various peak shapes in our nearest target, NGC~5068, in both the F2100W and F1000W bands.

We find that most 21~$\mu$m sources are unresolved or marginally resolved. Most peaks are round, but some of them appear elongated or irregular.  The elongated peaks occur when a peak is found on a filamentary dust structure or on the edge of a bubble or shell-like feature. Edge-on background galaxies can also appear as elongated sources.  The irregularly shaped peaks appear to mostly be stellar clusters, where intense \ion{H}{2} regions interact with multiple radiation sources, and higher-resolution NIRCam bands reveal densely crowded stellar cluster regions.  These irregular sources are also frequently a blend of neighboring sources, which could be resolved in the higher-resolution NIRCam bands. We also note that the diffraction pattern around bright sources can cause sources to appear irregular or even result in the detection of false sources.

Based on visual inspection of the F1000W and F2100W images, we find that many more isolated stars are visible in the F1000W image. Thanks to the smaller PSF size, we also see that many compact clusters appear as marginally resolved at F1000W. We also find that diffuse dust emission is prominent in both the F1000W and F2100W bands, which makes separating compact peaks from the background particularly challenging.

Informed by this inspection, we develop a source identification method that finds a range of source morphologies while avoiding the detection of artifacts. The key components are to filter out diffuse emission (Section \ref{sec:diffuse}) and then identify sources using a non-parametric approach using dendrograms (Section \ref{sec:source_finder}). We provide more information about catalog properties in the Appendix \ref{app:catalog}.

\subsection{Diffuse Emission Subtraction}
\label{sec:diffuse}

Diffuse emission has an important contribution to the F1000W and F2100W maps. Previous studies by \citet{Crocker} have demonstrated that by masking the brightest \ion{H}{2} regions in 8~$\mu$m maps, more than 30\% of the emission is unrelated to recent star formation. Removing this diffuse component is essential for improving the detection of faint compact sources. We accomplish this by selectively filtering out large-scale ``diffuse'' emission and performing source detection on the filtered images. This separation of scales can be accomplished using Fourier filtering or wavelets \citep{Starck}, which effectively distinguish structures of varying sizes across the images. However, these methods can produce artifacts in decomposed maps, especially in areas with large contrasts in emission, resulting in negative, non-physical emission \citep{Coifman1995}. In this study, we use Constrained Diffusion \citep[CD,][]{Li2022}, a related approach that ensures the sum of all decomposed component fluxes matches the original input map and are non-negative. In CD, the size of the different scales refers to the dispersion of the Gaussian function used for smoothing.

\begin{figure*}[!t]
    \centering
    \includegraphics[width=0.95\textwidth]{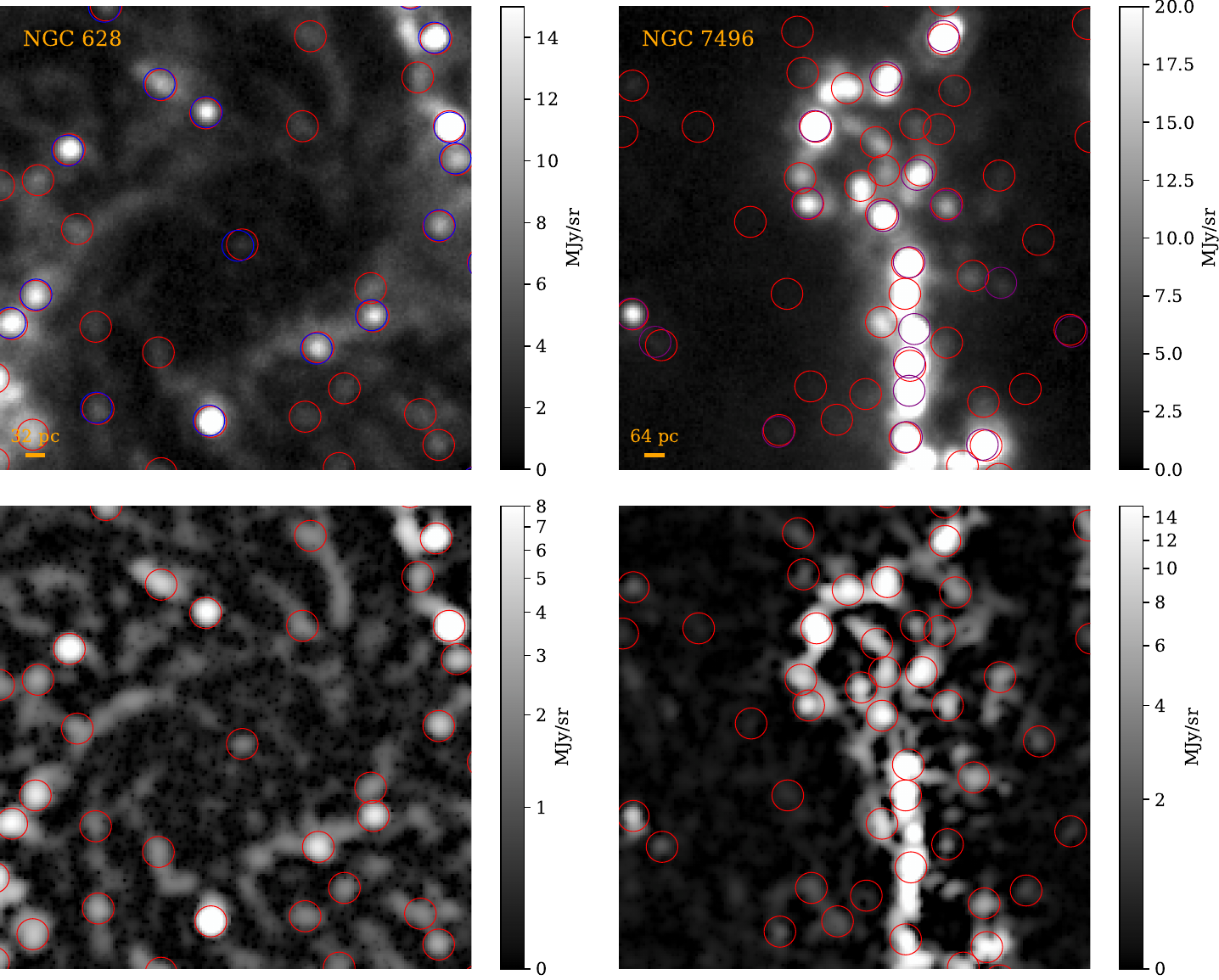}
    \caption{The 21~$\mu$m sources detected in the central region of NGC~628 (left) and spiral arm of NGC~7496 (right). The top panel displays the original F2100W maps, while the bottom panel features the second scale of the multiscale diffusion output of the same F2100W map in asinh scaling. The red circles indicate sources identified in this study, while the blue circles mark peaks detected by \citet{hassani23} without diffuse emission subtraction.}
    \label{fig:source}
\end{figure*}

The CD method decomposes the emission into multiple scales (typically 8 or 9 in our images, starting from 0), where the smallest scales capture compact sources, and larger scales capture more diffuse, extended structures. In this study, we use the second scale of our F2100W images for source detection in the 21~$\mu$m maps, as smaller scales are dominated by pixel-to-pixel noise. At this scale, the detected sources primarily have radii of about 6 pixels. In the case of the 10~$\mu$m maps, we used the sum of the first and second scales, which improved the detection of faint sources. However, we note that including the noisy smallest scale (zeroth scale) resulted in poorer recovery of faint sources. Furthermore, We note that \citet{Williams2024} convolved the F2100W maps to a resolution of 0.85$\arcsec$, resulting in a typical noise level of 0.08~MJy/sr (an improvement of 50\% in NGC0628) and enhanced the signal-to-noise ratio. We compare the second component of our F2100W maps with the sum of the 0, 1, and 2 components from the F2100W maps convolved to 0.85$\arcsec$ in \cite{Pathak2024}. Our analysis shows that both sets of maps recover bright sources similarly. However, the convolved maps result in the detection of about 20\% more faint sources below about 10~$\mu$Jy flux limit in NGC~628. Despite this advantage, the convolution to 0.85$\arcsec$ introduces a 20~pc increase in physical resolution for the most distant targets. Given this limitation, we opted to use the filtered, native-resolution maps.

Compared to the approach of \citet{hassani23}, the CD method effectively suppresses diffuse emission, detecting thousands of new sources when implemented. Using CD to remove the diffuse emission makes it possible to detect two main types of sources that are inaccessible without some background treatment: faint, isolated objects that were previously blended with the background, and sources embedded within filaments. However, one limitation is that saturated sources may be fragmented into several parts, where the wings and core of the PSF blend together, potentially confusing the source detection algorithm. Additionally, this method can exacerbate diffraction spike artifacts in the central region of galaxies and result in numerous false detections, notably impacting two of our targets: NGC~1365 and NGC~7496. Hence we used \textsc{stpsf} \citep{Perrin} to generate a PSF matching the pixel scale of the MIRI images, positioned it at the center of the galaxy oriented to match the saturated sources, and subsequently applied a mask to the images. The radius of the masked region is approximately 18\arcsec. Figure~\ref{fig:source} shows the second component of the CD with detected sources from this study overlaid alongside sources from the previous 21~$\mu$m study by \citet[][]{hassani23}, where no diffuse emission was removed. The CD approach detects many more fainter sources while still identifying the brighter sources. We also find some fluffy sources in our sample, many of which are located in dust filaments. Since many young clusters exhibit such mid-infrared emission, we do not exclude any of these from our catalog.

\subsection{Source Identification}
\label{sec:source_finder}

Various source-finding algorithms have been developed to identify and characterize sources with diverse morphologies (e.g., circular, elongated, irregular) and to extract them within regions dominated by diffuse emission \citep[and references therein]{menshchikov}. Among these, \textsc{clumpfind} \citep{Williams94} and \textsc{fellwalker} \citep{BERRY201522} are widely used; however, both have limitations. In the case that medium consists of structured diffuse emission blended with compact sources, CLUMPFIND will often identify larger structures that reflect the diffuse emission rather than cleanly isolating the compact sources \citep{Williams}. Other algorithms, such as \textsc{SExtractor} \citep{Bertin}, are better suited for identifying elliptical sources but struggle with irregular structures \citep{Williams}. Infrared observations of galaxies reveal extended, continuous structures that are hierarchically nested, unlike broad-band optical observations, where sources appear more as isolated compact sources \citep{Pathak2024,Leroy_2023_co_ir}.

Given the limitations of the source-finding algorithms discussed above, we adopt a dendrogram-based approach for source identification, following \citet{Williams} and \citet{hassani23}. Dendrograms were initially applied to millimeter observations to analyze the hierarchical structure of molecular clouds \citep{Rosolowsky2008}. This complexity, where high-density cores are embedded within lower-density envelopes, introduces challenges for detecting individual sources. Mid-infrared observations of galaxies exhibit similar hierarchical structures to millimeter observations, capturing both distinct compact sources and extended emission regions. Physically, mid-infrared structures also resemble molecular gas emission \citep{Leroy_2023_co_ir}. We use the \textsc{astrodendro}\footnote{\href{https://dendrograms.readthedocs.io/en/stable}{https://dendrograms.readthedocs.io/en/stable}} package to identify and measure the structure of mid-infrared emission peaks.

 % Furthermore, dendrograms use contours to trace the natural structure in the emission and do not identify sources of specific shapes like the PSF or Gaussians. 

% , adapting the concept introduced by \cite{HS92} and further developed by \cite{Rosolowsky2008}, with an emphasis on the isosurfaces for three-dimensional data. Dendrograms serve as graphical representations of primitive structures within images of any dimension, where local maxima define the “leaves” or top levels of the dendrogram. Each local maximum has a unique surrounding region devoid of larger values, forming distinct isosurfaces. This non-parametric approach allows us to effectively describe and filter the emission contours, helping to identify key high-density features. 

We compute the dendrograms of the CD-filtered image, which has two important parameters: the minimum intensity threshold and the independence criteria for identifying distinct structures. The minimum intensity threshold is set by taking the lowest valid value in the data and adding a scaled ``step'' above it. This ``step'' represents the typical variation or noise level in the CD-filtered images and is calculated as the difference between the median value and the 16th percentile, which would correspond to a $\approx1\sigma$ distribution for a Gaussian distribution. The ``step'' values range between $\sim$ 0.05 to $\sim$ 0.13~MJy/sr in the F2100W band and 0.09 to 0.27~MJy/sr in the F1000W band for different galaxies. By scaling this step by a factor of 5, the function ensures that only features significantly brighter than the background noise are included in the analysis. We note that since the ``step''  values vary for each galaxy, the corresponding noise levels and detection limits also differ across our sample, resulting in a non-homogeneous sensitivity. For example, in our F2100W images, the step value is 0.09 MJy/sr for NGC 5068 and 0.21 MJy/sr for NGC 1365—the nearest and farthest targets in our sample—corresponding to luminosity of about $0.5 \times 10^{16}$ and $15 \times 10^{16}$ W/Hz, respectively.

The independence criteria, on the other hand, determine whether a region in the data should be classified as its own structure or as part of a larger surrounding feature. We adopt three criteria to define independent structures: (1) each leaf must contain at least as many pixels as the area subtended by the PSF, which corresponds to $\sim$ 41 pixels for F2100W and $\sim$ 10 pixels for F1000W, calculated from the PSF FWHM \citep[see Table 4 of][]{lee2023}, (2) the difference between the minimum value in a leaf and the peak intensity is at least three times as large as the ``step'' size defined previously, and (3) the local maximum of the leaf is separated by all other local maxima by at least one half-width of the PSF for the band we are using: 0.33\arcsec\ ($\sim$3 pixels) for F1000W and 0.17\arcsec\ ($\sim$1.5 pixels) for F2100W. This final criterion prevents bright diffraction structures from causing a single source to be detected as multiple sources. More details about the handling of these sources are provided in \citet{hassani23}.

We ran this algorithm on both the F1000W and F2100W bands across our targets and found 24945 sources in the F2100W maps and 55581 in the F1000W maps. 

\begin{table*}[ht]
\centering
\begin{tabular}{@{}lrrrrrrrrrr@{}}
\toprule
Galaxy & F1000W & HST H$\alpha$ & $f_{\text{LS}}$ & $f_{\text{Catalog}}$ & $f_{\text{Missed}}$ & F2100W & HST H$\alpha$ & $f_{\text{LS}}$ & $f_{\text{Catalog}}$ & 
$f_{\text{Missed}}$  \\
\midrule
Resolution & 0.32\arcsec & & & &  & 0.67\arcsec\\ 
\midrule
 &~$\mu$Jy & (10$^{36}$ erg/s) & (\%) & (\%) & (\%) &~$\mu$Jy  & (10$^{36}$ erg/s) & (\%) & (\%) & (\%) \\ \midrule
IC5332 & 4 & 1 & 18 & 15 & 67 & 18 & 3 & 14 & 20 & 67 \\
NGC0628 & 6 & 2 & 47 & 8 & 45 & 24 & 2 & 46 & 24 & 30 \\
NGC1087 & 6 & 7 & 65 & 5 & 30 & 23 & 12 & 64 & 22 & 14 \\
NGC1300 & 3 & 2 & 45 & 7 & 47 & 21 & 4 & 36 & 19 & 44 \\
NGC1365 & 4 & 9 & 52 & 6 & 42 & 21 & 19 & 67 & 10 & 23 \\
NGC1385 & 7 & 6 & 71 & 5 & 24 & 43$^{*}$ & 13 & 70 & 19 & 11 \\
NGC1433 & 3 & 5 & 54 & 8 & 38 & 21 & 10 & 36 & 16 & 48 \\
NGC1512 & 3 & 6 & 53 & 7 & 40 & 19 & 12 & 39 & 19 & 42 \\
NGC1566 & 4 & 9 & 65 & 6 & 29 & 22 & 14 & 60 & 25 & 15 \\
NGC1672 & 5 & 7 & 73 & 6 & 22 & 21 & 11 & 69 & 25 & 5 \\
NGC2835 & 5 & 4 & 42 & 9 & 49 & 25 & 7 & 37 & 26 & 36 \\
NGC3351 & 4 & 2 & 56 & 9 & 35 & 25 & 4 & 48 & 26 & 26 \\
NGC3627 & 8 & 6 & 74 & 6 & 20 & 36 & 8 & 66 & 24 & 10 \\
NGC4254 & 8 & 9 & 69 & 5 & 26 & 29 & 13 & 69 & 20 & 11 \\
NGC4303 & 7 & 9 & 69 & 6 & 25 & 28 & 19 & 67 & 21 & 11 \\
NGC4321 & 5 & 6 & 64 & 5 & 31 & 24 & 9 & 61 & 20 & 19 \\
NGC4535 & 5 & 7 & 53 & 6 & 41 & 21 & 13 & 49 & 29 & 22 \\
NGC5068 & 4 & 1 & 45 & 11 & 44 & 27 & 1 & 45 & 25 & 30 \\
NGC7496 & 4 & 7 & 46 & 6 & 48 & 28 & 13 & 60 & 12 & 28 \\
\bottomrule
\end{tabular}
\caption{The completeness limits flux of the mid-IR continuum compact sources with the corresponding HST H$\alpha$ level at different resolutions. The columns $f_\mathrm{LS}$, $f_\mathrm{Catalog}$, $f_\mathrm{Missed}$ indicate, respectively, the fractions of the flux from that galaxy found in large-scale structures, the compact source catalog, and the remaining flux in the image. \\
* We chose a median value of 24 for this galaxy. }
\label{tab:comp}
\end{table*}

\subsection{Source Photometry}
\label{sec:photo}

 \begin{figure}[!t]
    \centering
        \includegraphics[width=1\linewidth]{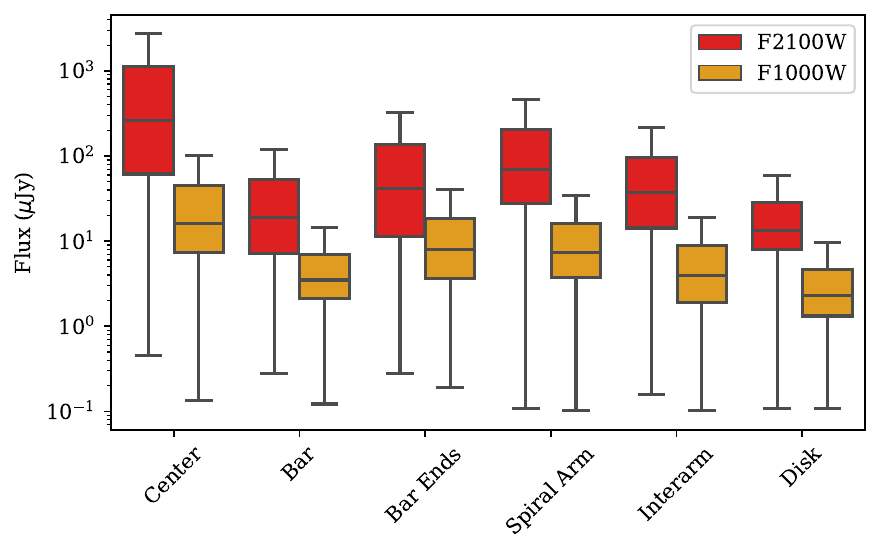}
\caption{The flux distribution of 21 and 10~$\mu$m peaks across various galactic environments is visualized, where the inter-quartile range and median are highlighted for each environment. Sources with fluxes below 0.1~$\mu$Jy are not included here.}
    \label{fig:envs}
\end{figure}

% To minimize influence of the background emission, we pinpoint the location of the sources by subtracting diffuse emission, as detailed in Section \ref{sec:source_finder}, using the maps provided in Section \ref{sec:diffuse}. 

% We first convert HST map units to surface brightness units in MJy/sr. Using circularized convolution kernels similar to those in \cite{hassani23}, we smoothed all maps (including narrowband HST H$\alpha$) to a common resolution suitable for each catalog. For the F1000W catalog, we convolved all HST and JWST maps to match the 0.36\arcsec\ resolution of the F1130W band, which closely approximates the 0.32\arcsec\ resolution of F1000W. This target resolution then includes all different, available wavebands except the worse-resolution F2100W images. For the F2100W catalog, we convolved all maps to the 0.67\arcsec\ resolution of F2100W and use all the images. In this convolution process, we used the band-specific kernels generated from the \textsc{WebbPSF} \citep{Perrin} to convert from the native resolution of each map to the target resolution \citep{Williams}. For HST maps, we used the F200W band kernels to match the target resolution (either F1130W or F2100W) since the effect of their original PSF in convolving to these longer wavelength bands is small. 

After identifying the locations of compact sources (Section \ref{sec:source_finder}), we perform photometry on those locations using the original, unfiltered data across all bands in our analysis. Using the original data avoids concerns about how the CD filtering affects source fluxes. After convolving each image to the target resolution (that of F1130W or F2100W) for each catalog, we first measured the surface brightness of each source across all JWST and HST bands at the local maximum positions of the 10 or 21~$\mu$m sources. After subtracting the median background, we convert the background-subtracted surface brightness to flux units ($\mu$Jy) by multiplying it by the solid angle subtended by the target PSF (see Table 4 of \citealt{lee2023}), which was calculated using the \textsc{stpsf} package \citep{Perrin}. The background was estimated using an annulus with radii set to $2\times$ and $3\times$ the PSF size: 1.34\arcsec\ and 2.01\arcsec\ for the F2100W catalog, and 0.72\arcsec\ and 1.08\arcsec\ for the F1000W catalog. We assess the local surface brightness uncertainty using the median-absolute-deviation in the background annulus.

We also included the surface brightness of CO(2-1) and H$\alpha$ EW from \cite{Leroy2021} and MUSE data from \cite{phangs-muse}, respectively. These measurements are sampled from the native resolution maps of those tracers without background subtraction. We also note that local background subtraction was not applied to our HST H$\alpha$ fluxes.

In Figure \ref{fig:envs}, we present the flux distributions of both catalogs across various galactic environments. The median flux in the 21~$\mu$m catalog is 21.1~$\mu$Jy with an uncertainty of 4.8~$\mu$Jy, while in the 10~$\mu$m catalog, the median flux and uncertainty are 3.8~$\mu$Jy and 0.85~$\mu$Jy, respectively. Notably, the brightest regions, aside from the galactic centers, are found near the bar ends and spiral arms, a trend more pronounced in the F2100W band. 

\subsection{Completeness Limits}
\label{sec:ccomp}
To establish a detection limit for our 10 and 21~$\mu$m catalog, we use artificial source tests (AST), as previously conducted for infrared observatories of nearby galaxies \citep[e.g.,][]{Bolatto}. We also performed this test on the HST H$\alpha$ maps, convolved to match the 10$\mu$m and 21$\mu$m catalog resolutions. Understanding a detection limit for the H$\alpha$ bands is crucial, as we will use these bands to differentiate between \added{optically} embedded and exposed star formation regions (see section \ref{sec:classify}). For the HST H$\alpha$ maps, we used the convolved versions obtained in Section \ref{sec:photo}, regridding them to the MIRI pixel size (0.11$\arcsec$).

 \begin{figure}[!t]
    \centering
    \includegraphics[width=0.9\linewidth]{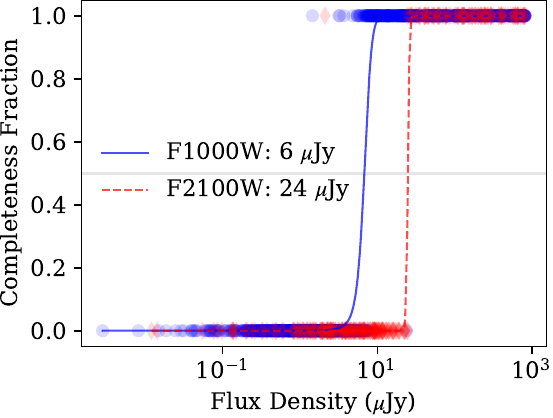}
    \caption{The 50\% completeness limit estimates for NGC~628 in F1000W and F2100W bands. The individual symbols indicate the flux density of an artificial star and the completeness is a boolean flag corresponding to a detection (1) or not (0). The curves are logistic fits to the data and the completeness limits are set by where a given completeness fraction intersects these fits.}
    \label{fig:sources}
\end{figure}

First, using the PSF provided by the \textsc{WebbPSF} package, we regridded the circularized PSF from 0.2\arcsec\ to align with the MIRI pixel scale of 0.11\arcsec. We then adjusted the brightness of the PSF by assigning random values from a log-uniform distribution ranging between 0.1 and 30,000. We generated 2000 simulated or ``fake'' point sources and placed them at random positions across each galaxy in the original maps, maintaining approximately a 2.5-PSF separation from ``real'' sources (0.8\arcsec\ for the F1000W catalog and 1.5\arcsec\ for the F2100W catalog), as identified in Section \ref{sec:source_finder}. 

Following the methods described in Sections \ref{sec:diffuse} to \ref{sec:photo}, we applied our source-finding algorithm using the CD method to the diffuse-subtracted ``fake''  maps, which include both fake and real sources. We performed photometry on the unfiltered maps, not the diffuse-subtracted version. These ASTs are designed to measure the sensitivity of the source detection algorithm in the presence of diffuse emission rather than the effects of source blending. This motivates our choice to avoid injecting sources near real sources in the map. Blending can still occur between fake sources, especially when two injected sources are close together, so we remove fake sources whose separation is less than twice the PSF size.

These procedures were carried out for all targets in the JWST F1000W and F2100W bands, as well as the HST H$\alpha$ maps at two distinct resolutions. The estimated completeness limit should be an approximate function of the injected position (or galactic environment) and the background level. We expect higher recovery rates for fake sources at a given flux in regions with low background emission, where the background level is minimal (e.g., inter-arm regions or disk). In contrast, in crowded areas with higher background levels, the completeness flux limit should increase. However, we note that in such crowded fields, fewer fake sources were injected due to the separation criteria defined between fake–fake and fake–real sources.

We assess the recovery rate of injected fake sources in relation to their flux by fitting the probability of detection $P$ with a logistic function
\begin{equation}
    P(f_\nu) = \frac{1}{1+\exp[-c_1(f_\nu-c_2)]}
\end{equation}
where $c_1$ is a shape parameter and $c_2$ is the 50\% completeness limit. We summarize our detection limits for the JWST mid-infrared bands and HST H$\alpha$ at various resolutions in Table \ref{tab:comp}.

% This increase aligns with the relationship between sensitivity and resolution, where sensitivity scales with the square of the resolution ratio: $(0.67/0.32)^2 \approx 4.5$ combined with slightly higher noise levels in MIRI at longer wavelengths.
The median detection limit at the F1000W resolution is 5~$\mu$Jy, while at F2100W, it is about five times higher at 24~$\mu$Jy. In luminosity units, the median detection limit in the 21~$\mu$m catalog is \(7 \times 10^{17}\) W Hz\(^{-1}\), spanning roughly an order of magnitude—from \(9 \times 10^{16}\) to \(1 \times 10^{18}\) W Hz\(^{-1}\)—depending on the distance of the galaxies. This median value is lower by a factor of 7 in our 10~$\mu$m catalog and spans the same order of magnitude, depending on the galaxy distance. This variation with distance is most obvious in the nearest target, NGC 5068, where we can detect sources that are approximately 10 times fainter than those in the more distant galaxies (D $\sim$ 15 Mpc) in our sample.

The uncertainty in the completeness limit due to fitting error is less than 0.2~$\mu$Jy for the F1000W catalog and 0.7~$\mu$Jy for the F2100W catalog, as calculated from the covariance matrix of the curve fitting. However, in the case of NGC1385, the completeness limit at F2100W exhibits unusually high uncertainty, so we adopt the median value of 24$\mu$Jy for this target. These 50\% completeness limits could be about at least 30 percent higher if we do not exclude cases where fake sources overlap with each other. In F1000W, our measured detection limits are $\sim 10\times$ the point-source sensitivities reported by \cite{lee2023} for the F1000W band and $15\sigma$ for the F2100W band established for point-source aperture photometry using small (50\% enclosed power) apertures. Our empirical determination of these large detection thresholds relative to the nominal noise likely arises from the contributions of background emission.  Since our thresholds are established relative to the fluctuations in the filtered maps, these test how significant a source needs to be for detection relative to the diffuse ISM in the bands.

We observe only 10 to 30 percent variation in the detection limit across different galactic environments from the integrated values reported in Table \ref{tab:comp} for the F1000W catalog. \added{In the spiral arms of NGC~4321, we find an F1000W completeness limit of $\sim$8~$\mu$Jy, which is approximately 30\% higher than the completeness limit measured for the galaxy as a whole.} We attribute this increased threshold to the crowded regions and elevated background levels. For the F2100W catalog, we were unable to inject an adequate number of fake sources into the galaxy centers due to the separation criteria. However, in the cases of NGC~628 and NGC~1512, we observe a completeness limit exceeding 40~$\mu$Jy with unreliable fits, which suggests issues from crowding and high backgrounds in these regions.  \added{We also find that the completeness limit increases in the spiral arms of NGC~3627 by approximately 25\% relative to the galaxy-wide value, indicating the effects of increased crowding and higher background emission. We also note that the completeness limit decreases in the bar regions with lower diffuse emission; for example, it is reduced by approximately 15\% relative to the galaxy-wide value in NGC 1300. We present the completeness limits for all galaxies, further separated by galactic environment, for the F1000W and F2100W bands in Appendix \ref{app:comp}.  Our cataloging algorithm requires sources to be spatially separated from each other, but allows for irregularly shaped sources.  Hence, the primary effect of crowding on our catalog will be source blending.  In most environments, the sources are separated by larger than a PSF scale, but crowded regions, notably galaxy centers (Figure \ref{fig:envs}) and bar ends are likely to have some source blending, particularly at 21 $\mu$m where the size of the PSF is worse.}

% The modest exceptions occur in the spiral arms of NGC~1365 and NGC~1672 at the F1000W band where we find 50\% completeness limits of 29~$\mu$Jy and 14~$\mu$Jy, respectively, with uncertainties under 2~$\mu$Jy.

We specifically check whether the completeness limit varies with local background level by dividing the data into four bins of background level (units of MJy/sr): $(0, 10^{-0.5})$, $(10^{-0.5}, 1)$, $(1, 10^{0.5})$, and $(10^{0.5}, \infty)$, then calculating the completeness limit for fake sources within these background ranges.  We see that the completeness limit (Table \ref{tab:comp}) is relatively stable with local background, which we attribute to the uniform performance of the CD image filtering combined with the aperture photometry. There remain a few small variations of note. For the F1000W catalog, we find that the completeness limit does not change for background levels below $10^{0.5}$~MJy/sr. However, at background levels above $10^{0.5}$~MJy/sr, the completeness limit reaches $18 \, \mu$Jy for NGC~1385, $14 \, \mu$Jy for NGC~3627, and $11 \, \mu$Jy for NGC~4303. In other cases, we did not obtain a reliable fit, or changes remained within a $10\text{--}20$ percent range. For the F2100W catalog, we occasionally find that lower background levels correspond to lower completeness limits (e.g., NGC~3627 has a $23 \, \mu$Jy completeness limit in the $10^{-0.5}$ and 1 MJy/sr background range compared to 36~$\mu$Jy for the galaxy as a whole).  

% We do not see this sensitivity to local background in the F1000W analysis. 
% suggests that restricting completeness calculations to fake sources located in low-background fields could reduce the completeness limit though this effect is not observed in the F1000W catalog. 
 % In the case of BKGs, we future require a range between 0 to 1.2 for F200W/F300M.\\

We also use CD-filtered maps to estimate the large-scale (LS) emission, providing a rough estimate of the diffuse emission fraction ($f_{\text{LS}}$) in JWST observations of our targets. By summing the CD scales from 3 to 9 (the maximum), which capture more large-scale structure in the CD decomposition, and dividing by the total galaxy flux, we estimate how much of the integrated flux comes from diffuse emission. This filtering captures large scale structure in the map on scales larger than $\sim 2\times$ the PSF at F2100W ($\sim 4\times$ the PSF at F1000W). Additionally, we report the fraction of total flux from cataloged peaks divided by the integrated flux of each galaxy, denoted as $ f_{\text{Catalog}} $. We anticipate that the sum of $ f_{\text{LS}} $ and $ f_{\text{Catalog}} $ should approximate the total recoverable flux, with the remainder representing missed undetected, compact structures that can include faint compact sources and small filaments. The median $ f_{\text{LS}} $ is about 50\% in the F2100W band and slightly higher, around 56\%, in the F1000W band. Our results are in good agreement with \cite{Belfiore2023}, who measured the diffuse F2100W emission outside \ion{H}{2} regions to be, on average, 60\%. \citet{Leroy2023,Pathak2024} also reported nearly the same percentage of diffuse mid-IR flux at 7–21 $\mu$m.

% deb found that f_catalog is about 40 percent in HII regions, it is double of my numbers. 

The median $ f_{\text{Catalog}} $ is around 20\% in the F2100W band, which is about three times higher than in the F1000W band. The flux at these wavelengths primarily originates from bright star-forming regions, which tend to be more luminous at longer wavelengths. In these regions, dust heated by intense radiation from nearby massive stars reaches higher temperatures, leading to stronger mid-IR emission as the Planck distribution shifts more IR emission toward shorter wavelengths. Finally, the percentage of missed peaks varies from 5\% to 67\%, with a median of 22\% in the F2100W band, showing a narrower variation in the F1000W band, where about 38\% of sources are missed.

% We expect a higher fraction of diffuse emission in the F1000W mid-infrared continuum band due to the presence of more PAH dust. In most cases, this increase is a few percent, but in galaxies such as NGC~1300, NGC~1433, and NGC~1512, it is around 10\%. These galaxies have slightly higher metallicity than the sample median (see Table 2 of \cite{lee2023}), which results in a greater fraction of PAH emission and probably a larger contribution to the F1000W band.

\begin{table*}[!t]
\centering
\setlength{\tabcolsep}{2pt}
\begin{tabular}{lcccccccccc}
\hline
\textbf{Flux Ratio}   & \textbf{Emb} &   \textbf{Expo}       & \textbf{RSG$^{1}$}          & \textbf{O-AGB$^{1}$}         & \textbf{C-AGB$^{1}$}     & \textbf{B[e]$^{2}$}     & \textbf{WR$^{2}$}   & \textbf{CPN$^{2}$} &   \textbf{BKG} &    \textbf{FG}$^{3}$  \\
\hline
$r_{2-3.6}=$$f_{\text{F200W}}$/$f_{\text{F360M}}$        & - &   -  & $> 1.5$               & $< 2.5$               & $< 1.5$  &  $*$  &   -   & $*$ & $<1.2$  & $>1$            \\
\textbf{$r_{3-3.6}=$$f_{\text{F300M}}$/$f_{\text{F360M}}$}& - &   -  & $<1.4$                & $0.4-1.3$               & $< 1$  & -  &   -   & - & - & -            \\
$r_{2-7}=$$f_{\text{F200W}}$/$f_{\text{F770W}}$      & -&-         & $<11 $                & $<6$                & $< 3$  & $< 1$ &  $< 3$    & $<0.1$ & $-$ & $>1$            \\
$r_{3}=$$f_{\text{F335M}}$/$f_{\text{F300M}}$       &-&-        & $> 0.8$              & $0.9-2$              & $> 0.9$  & -  & -   & $-$&    $ <1.2$ & -             \\
$r_{3.6}=$$f_{\text{F335M}}$/$f_{\text{F360M}}$    & $>1.7–0.5r_{3}$&   $>1.7–0.5r_{3}$    & $< 1.2$   & $<1.1$            & $< 1.1$              & - & $- $ &  $<1$ &   $ <1.2$  & -          \\

$r_{3-7}=$$f_{\text{F300M}}$/$f_{\text{F770M}}$    & -&  -    & - & $> 0.65\,\left(r_{3.6-11}\right)^{1.2}$            &  $< 0.65\,\left(r_{3.6-11}\right)^{1.2}$ & - & - &  -&  -   & -         \\

$r_{7}=$$f_{\text{F770W}}$/$f_{\text{F1000W}}$  & -  &  -          & $<1.7$                  & $< 1.3$                & $< 1.3$ & $< 1.2$ & $<1.5$   & $>0.5$   & - &$>1$        \\
$r_{11}=$$f_{\text{F1130W}}$/$f_{\text{F1000W}}$      & $>2.5-$$r_{7}$   &  $>2.5-$$r_{7}$      & $>0.5$                  & $< 1.2$                & $0.5 - 2.5$  & $< 1.5$ & $>1$   & $>1$  &  -  & $<1$        \\
$r_{7-11}=$$f_{\text{F770W}}$/$f_{\text{F1130W}}$      & -  & -    &-    & -                & -  & $<1.5$ & $<0.6$    & $>0.2$ &  -  & -   \\
$r_{21-10}=$$f_{\text{F2100W}}$/$f_{\text{F1000W}}$   & - & - & - & - &- & $<3$ & $>2$ & $>1$  &  - &-  \\
\hline
\textbf{H$\alpha$ Detection}    &  No &  Yes$^{4}$  & No         & No         & No     &   Yes$^{5}$   &   -  & - &   -&    -  \\
\hline
\end{tabular}
\caption{Band ratio classifications used in this catalog. We classify sources as: sources associated with ISM emission that are optically embedded (Emb) or exposed (Expo) based on the detection of H$\alpha$ emission, red supergiants (RSG), oxygen and carbon-rich asymptotic giant branch stars (O-AGB, C-AGB respectively), B[e] stars, Wolf-Rayet Star candidates (WR), sources that can be compact planetary nebula or carbon-rich post-AGB stars (CPN), background galaxies (BKG), or foreground stars (FG).\\
*We adopt F200W/(F335M+F360M)$<$0.5 for CPNe and $<1$ for B[e] stars.\\ $^{1}$Adopted based on stellar track predictions combined with SAGE-LMC observations.\\
$^{2}$Only based on the observations of the SAGE-LMC sources from \cite{Jones2017}.  \\
$^{3}$Identified by cross-matching to \textit{Gaia} DR3.
\\$^{4}$Based on Table \ref{tab:comp} for each galaxy.
\\$^{5}$ $5\times$ brighter than the detection limited provided in Table \ref{tab:comp}.\\  
% $^{6}$ No condition for H$\alpha$ detection.
}
% $^{D}$\textbf{Emb} stands for \textit{Embedded} and refers to ISM sources that are bright in the mid-IR but lack H$\alpha$ emission, while \textbf{Expo} stands for \textit{Exposed} and indicates sources also detected in H$\alpha$.}  }
\label{tab:flux_ratios_sources}
\end{table*}

\section{Source Classification}
\label{sec:classify}

We use the Spectral Energy Distribution (SED) of the sources to classify them into different categories of objects. The population of sources that we recover in the mid-infrared includes dusty young stellar clusters, evolved dusty stars, foreground stars, and background galaxies, which were visually classified in \citet{hassani23}. Practically, we complete this classification using a set of flux ratios that reflect the shape of the predicted SED, and then apply cuts derived either from models or previous observations. We define flux ratios as $ r_\mathrm{X} $, where $ X $ denotes the relevant band(s), expanding on the notation from \citet{hassani23}: 
\begin{eqnarray}
    r_{2{-}3.6} &=& f_\mathrm{F200W}/f_\mathrm{F360M} \nonumber \\
    r_{2{-}7} &=& f_\mathrm{F200W}/f_\mathrm{F770W} \nonumber \\
    r_{3-3.6} &=&   f_\mathrm{F300M}/f_\mathrm{F360M} \nonumber \\
    r_{3} &=&     f_\mathrm{F335M}/f_\mathrm{F300M} \nonumber \\
    r_{3.6} &=&   f_\mathrm{F335M}/f_\mathrm{F360M} \nonumber \\
    r_{3-7} &=&   f_\mathrm{F300M}/f_\mathrm{F770W} \nonumber \\
    r_{3.6-11} &=&   f_\mathrm{F360M}/f_\mathrm{F1130W} \nonumber \\
    r_{7} &=&     f_\mathrm{F770W}/f_\mathrm{F1000W} \nonumber \\
    r_{11} &=&    f_\mathrm{F1130W}/f_\mathrm{F1000W} \nonumber \\
    r_{7-11} &=&   f_\mathrm{F770W}/f_\mathrm{F1130W} \nonumber \\
    r_{21{-}10} &=& f_\mathrm{F2100W}/f_\mathrm{F1000W} \nonumber
    \end{eqnarray}
Table \ref{tab:flux_ratios_sources} presents the source classifications used in the catalog. \added{We note that some of the classifications overlap, and thus source classifications are not mutually exclusive. At 10~$\mu$m, 1,089 out of 55,581 sources (2.0\%) exhibit multiple possible classifications, while 321 out of 24,945 sources (1.3\%) do so at 21~$\mu$m.}

We use the \texttt{CIGALE} code (section \ref{sec:s_cigale}) to model dusty, young stellar clusters and establish band ratio thresholds to find ``ISM sources'' that we divide into optically exposed or embedded based on the presence of H$\alpha$ emission (Section \ref{sec:ismsources}).  We use the \texttt{PARSEC} stellar evolutionary tracks (section \ref{sec:parsec}) to classify RSGs, O-AGBs, and C-AGBs, which we validate with observations from the SAGE-LMC survey \citep[][Section \ref{sec:stars}]{Jones2017,Jones2018}. We also define empirical band ratio thresholds from the SAGE-LMC catalog to classify B[e] stars, Wolf-Rayet (WR) stars, and carbon-rich planetary nebulae (CPN). \added{We note that the B[e], WR, and CPN classifications in this study are likely biased toward the lower-metallicity regime characteristic of the LMC, as our reference samples are drawn from \citep{Jones2017}, which provides the largest set of spectroscopically confirmed sources. We therefore rely exclusively on color-based criteria for classification, rather than absolute magnitudes, in order to avoid excluding sources with either lower or higher dust content.} We \added{also} cross match sources with \textit{Gaia} catalogs to find foreground stars (FGs; Section \ref{sec:fg}).  Finally, we identify background galaxies (BKGs) visually to establish their typical flux ratios, and then we use those ratios and the source morphology in the F200W imaging (Section \ref{sec:bkg}).

\begin{figure*}[!t]
    \centering
\includegraphics[width=1\textwidth]{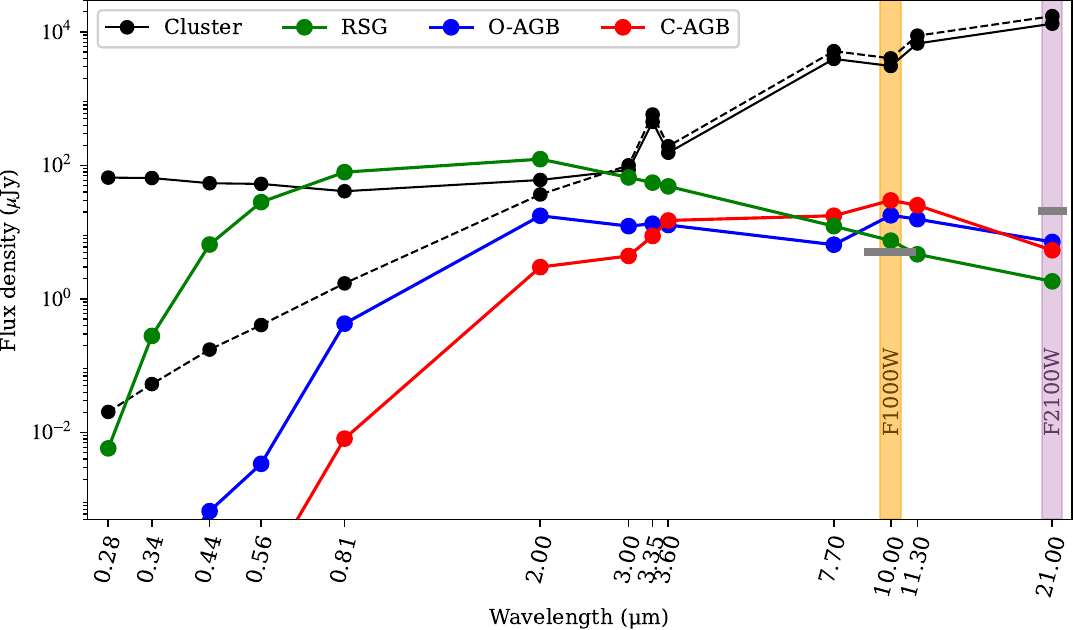}
   \caption{The modeled SEDs of different sources at a distance of 5.2~Mpc, corresponding to our closest target, NGC~5068. A young stellar cluster with an age of 1~Myr is shown in black, with $A_V = 1$~mag as a solid line and $A_V = 10$~mag as a dashed line. \added{We set $q_{\mathrm{PAH}} = 2.5$ for the two stellar cluster models presented here.} Predictions from \texttt{PARSEC} stellar evolution models are included for a RSG (green), carbon-rich AGB (red), and oxygen-rich AGB (blue). \added{We used a single model for each stellar type at inital $Z = 0.0152$, adopting a dust composition of 85\% AMC and 15\% SiC for the C-AGB (see \ref{sec:cagb_models} for details), and 60\% silicate and 40\% AlOx for the O-AGB (see \ref{sec:rsg-oagbs} for details).} The detection limits for compact sources at this distance are marked by thick gray bars in the F1000W and F2100W bands.}
    \label{fig:models}
\end{figure*}

\begin{figure*}[!t]
    \centering
\includegraphics[width=1\textwidth]{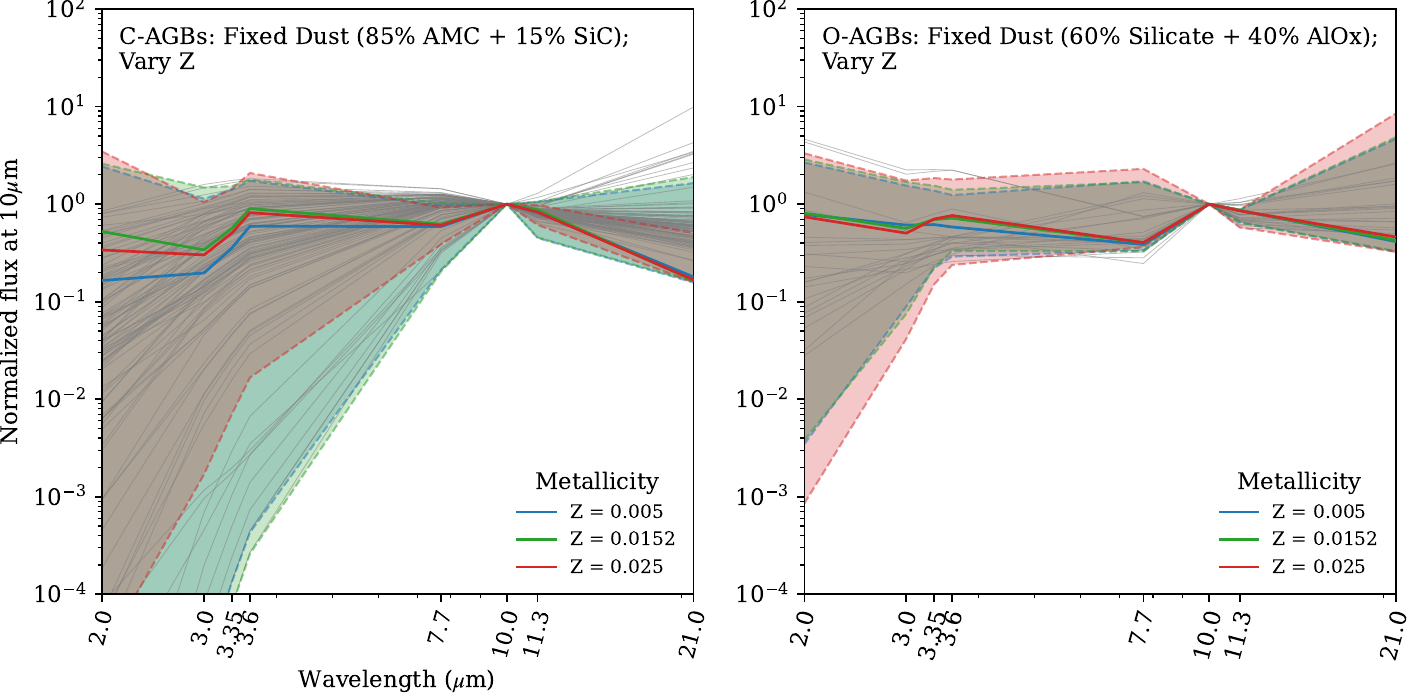}
   \caption{\added{Predicted SEDs for C-AGB stars (left) and O-AGB stars (right) normalized to 10~$\mu$m. The panels show models spanning different metallicities  at fixed dust chemistry. In each panel, the solid curve represents the median SED, and the dashed curves indicate the 0.01 and 99.99 percentiles of the models. Only models detectable in the F1000W band at a distance of 5.2~Mpc are shown. Carbon stars and M-type stars from the LMC, taken from \citet{Groenewegen2018}, that would be detectable at a distance of 5.2~Mpc are overplotted in gray for comparison.}}
    \label{fig:models_agbs}
\end{figure*}

\subsection{Source Modeling}
\label{sec:models}

To inform our source classification framework, we use models to interpret our observed SEDs. In this section we present the different models required to intepret our SEDs and establish the band ratio cuts (Table \ref{tab:flux_ratios_sources}) used in source classification.

\subsubsection{Young Stellar Clusters}
\label{sec:s_cigale}
We use the \texttt{CIGALE} code to generate models for young stellar clusters \citep{cigale}. We simulate a young stellar cluster with a simple stellar population using an exponential star formation history (\textit{sfh2exp}) with a short e-folding time ($\tau_{\text{main}} < 0.001$~Myr) and no contribution from a late starburst phase. We generate models for a range of stellar ages. We adopt the BC03 stellar population synthesis models from \citet{Bruzual03} with a metallicity of $Z = 0.02$ and a Chabrier initial mass function \citep{Chabrier}. 
The models include nebular emission, assuming an electron density of 100~cm$^{-3}$. Additionally, we apply dust attenuation using the \texttt{dustatt\_modified\_starburst} model with two different visual extinction values, $A_V = 1$~mag and $A_V = 10$~mag, adopting the Milky Way extinction curve and $R_{V} = 3.1$. We also incorporate the dust emission model from \citet{DraineLi2007,Draine2014}, considering a PAH mass fraction, $q_{\mathrm{PAH}}$, ranging from 0.95 to 3.8 \%, and a minimum radiation field intensity of $U_{\mathrm{min}} = 10$. We further consider $\alpha=2$, which is the power-law index for the distribution of radiation field intensity that is illuminating the dust. These parameter choices are typical for young stellar clusters \citep[e.g.,][]{lee2023} and have been applied to fit young clusters in \citet{hassani23}, as well as in recent works by \citet{henny2025} and Hannon et al.~(subm).

\subsubsection{Evolved Stars}
\label{sec:parsec}
We classify stellar sources by combining \texttt{PARSEC} predictions with literature studies, identifying RSGs, O-AGBs, and C-AGBs using flux ratio criteria. We base our analysis on \texttt{PARSEC} v1.2s tracks, which model the evolution of stars with metallicities ranging from $0.0001 \leq Z \leq 0.02$ for masses $0.1 \leq M/M_\odot < 350$, $0.03 \leq Z \leq 0.04$ for $0.1 \leq M/M_\odot < 150$, and $Z = 0.06$ for $0.1 \leq M/M_\odot < 20$ \citep{chen2014,chen2015,Tang}. These tracks cover stellar evolution from the pre-main sequence (PMS) phase up to either the first thermal pulse (TP) or carbon ignition. For stars in the thermally pulsating AGB (TP-AGB) phase, we incorporate the TP-AGB evolution using the COLIBRI code \citep{Pastorelli2019,Pastorelli2020,Marigo,Rosenfield}.  \added{We consider a wide range of initial metallicities in our models, spanning sub-solar (LMC-like; $Z=0.005$), solar ($Z=0.0152$), and super-solar ($Z=0.025$) values, in order to capture the diversity of evolved-star populations across different galactic environments. 

In addition to metallicity, the dust output of AGB stars is influenced by their circumstellar dust chemistry. For C-rich AGB stars, we consider two dust compositions following the models of \citet{Groenewegen2006}: a purely amorphous carbon (AMC; 100\%) model, and a mixed model consisting of 85\% AMC and 15\% silicon carbide (SiC). For O-rich AGB stars, we adopt three dust-chemistry prescriptions: a mixed composition of 60\% silicate and 40\% aluminium oxide (AlOx; Al$_2$O$_3$), a purely AlOx (100\%) model, and a purely silicate (100\%) model, all assuming a uniform dust grain size of 0.1 $\mu$m, again following \citet{Groenewegen2006}. These dust-chemistry prescriptions have previously been shown to successfully reproduce the SEDs of Galactic O-AGB stars (see references in \citealt{Groenewegen2006}) as well as carbon stars in the Local Group \citep{Groenewegen2018}. We note that more sophisticated dust compositions, such as mixtures of olivine and metallic iron, have been shown to reproduce the SEDs of O-rich AGB stars in the Local Group \citep{Groenewegen2018}. However, these dust prescriptions are not available through the CMD 3.7 interface\footnote{\url{http://stev.oapd.inaf.it/cgi-bin/cmd_3.7}}. We therefore restrict our O-rich AGB models to the AlOx and silicate-based dust compositions described above. 

The atmospheres of the AGB stars are modeled using \textsc{COMARCS}, incorporating an extensive set of molecular species for both O-rich AGB and C-rich AGB stars \citep{Aringer,Aringer_2016}. The models include key molecular absorbers that are accessible with our JWST bandset, such as the HCN and C$_2$H$_2$ absorption features, which can be traced with the F300M filter (see Fig.~3 in \citealt{Aringer}). We further note that the PARSEC evolutionary tracks for RSGs, as provided by the CMD interface, do not include circumstellar dust. Consequently, these models describe dust-free stellar photospheres and are unable to reproduce the dust emission features commonly observed in dusty RSGs, particularly at $\sim10$ and $11.3\,\mu$m \citep{Jones2017}.
}
%In this phase, we account for circumstellar dust, assuming a composition of 60\% silicate and 40\% aluminum oxide for M stars, and 85\% amorphous carbon and 15\% silicon carbide for C stars, based on \cite{Groenewegen2006}. REMOVED FROM TEXT.

For simplicity, we assume no line-of-sight interstellar extinction and foreground extinction correction, which should only be a small effect on evolved stars in the near and mid-infrared. We also adopt the Kroupa IMF which is corrected for unresolved binaries \citep{Kroupa2001,Kroupa2002}, which was used to compute the integrated magnitudes. Photometric fluxes for HST and JWST bands are retrieved in the Vega system using the CMD 3.7 web interface. We convert these to AB magnitudes and flux densities using the Vega magnitude offsets defined in the JWST pipeline \citep{jwst-pipeline}.

% \textbf{To categorize evolved stars such as RSGs and AGBs, we use the \texttt{PARSEC} stellar evolution tracks, selecting the brightest stellar sources at 10 $\mu$m  \citep{chen2014, chen2015}.  We focus on this part of the SED since more individual stars are visually identifiable in the F1000W maps than in the F2100W maps. We discuss stellar source classification further in Section~\ref{sec:class}.}

\subsubsection{Example Model SEDs}

Figure \ref{fig:models} shows the model SEDs for young clusters and stellar sources, along with detection limits from Table~\ref{tab:comp}. We scale the model SEDs to a distance of 5.2~Mpc, corresponding to our nearest target, NGC 5068.  While many types of evolved stars are expected to be detectable in this nearby galaxy, we are unlikely to detect RSGs and O-rich AGBs in our more distant targets ($D\gtrsim 10~\mathrm{Mpc})$ (see Section \ref{sec:rsg-oagbs} for discussion). Figure \ref{fig:models} shows that the dusty young stellar clusters exhibit strongly increasing flux densities at longer wavelengths. While dusty stars also show increasing flux densities with longer wavelength in the near infrared, these SEDs turn over between 3 and 10~$\mu$m making it possible to distinguish these sources from young clusters.  The young clusters also show significant emission features from PAHs, which manifests in bright 3.35, 7.7 and 11.3~$\mu$m emission. However, the regions around young clusters may experience PAH ionization and destruction, which could reduce their PAH emission  \citep{Egorov,Sutter2024}. PAH destruction processes are not included in our current modeling of young clusters. \added{Furthermore, we show AGB models in Figure~\ref{fig:models_agbs} for C-AGBs on the left and O-AGBs on the right, which are normalized to the F1000W band. The SEDs of C-AGBs indicate that the effect of metallicity variations is more pronounced at wavelengths $<10 \mu$m and becomes less significant at longer wavelengths. In contrast, the median SEDs of O-AGBs are largely insensitive to metallicity. On the other hand, we note that changes in dust chemistry do not affect the median SEDs, but instead primarily influence the spread of the models.}

\begin{figure*}[!t]
    \centering
    \includegraphics[width=0.70\textwidth]{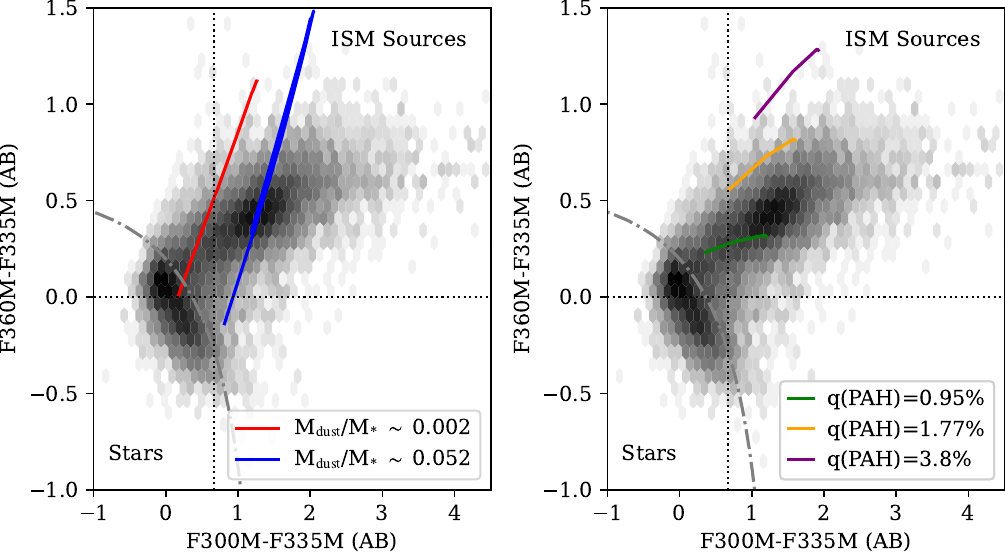}
    \includegraphics[width=0.70\textwidth]{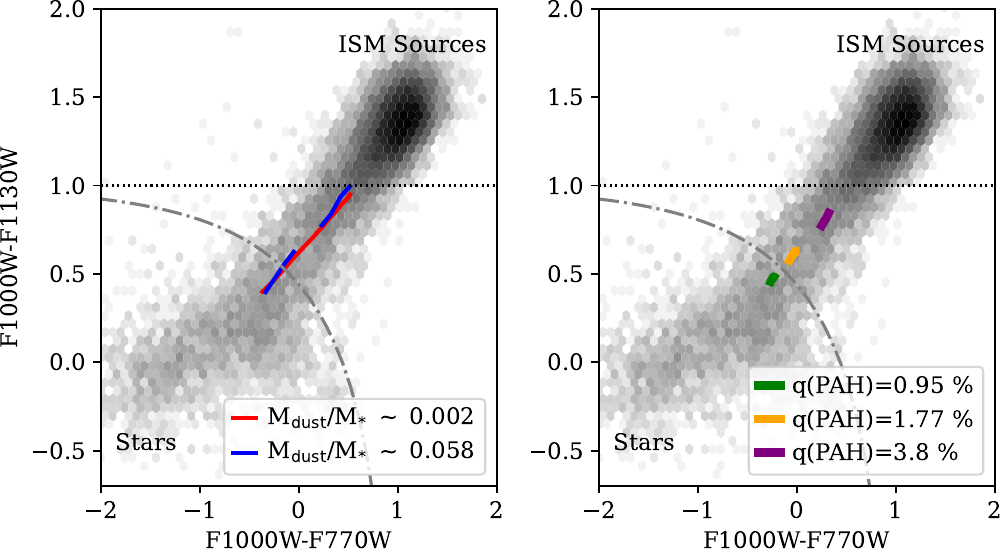}
    \label{fig:data}
\caption{\added{Color-color diagrams of F360M-F335M vs. F300M-F335M (top) and F1000W-F1130W vs. F1000W-F770W (bottom) in AB mag. Observed 10~$\mu$m peaks across 19 galaxies are displayed as gray hexagons, with darker regions indicating higher population densities. Using the CIGALE models, we overlay key physical parameters that explain variations in the flux ratios, M$_\text{dust}$/M$_{*}$ and $q_\text{PAH}$. The dash-dotted gray line represents the source classification from \cite{hassani23}. The dotted line indicates the criteria for identifying 3.3~$\mu$m (i.e. strong) PAH emitters from \cite{Rodriguez2024}, defined as F300M-F335M $>$ 0.67, F360M-F335M $>$ 0, and F1000W-F1130W $>$ 1.}}
\end{figure*}

\subsection{Sources with ISM Emission}
\label{sec:ismsources}

We define ISM sources by the presence of PAH emission detected in both the NIRCam and MIRI bands adopting the band ratios given in \ref{tab:flux_ratios_sources} \citep[see also Section 4 of ][]{hassani23}. In Figure \ref{fig:data}, we present IR color-color diagrams for these sources showing these band ratios. We further use the \texttt{CIGALE} models presented in Section~\ref{sec:s_cigale} to investigate which physical parameters drive the variations observed in the color-color diagram of the identified peaks.

The \texttt{CIGALE} models suggest that the color of the NIRCam F335M PAH band relative to its nearby bands (F300M or F360M) depends on the dust-to-stellar mass ratio and the fraction of dust found in PAHs ($q_\mathrm{PAH}$; Figure \ref{fig:data}). Increasing either $M_\mathrm{dust}/M_\star$ or $q_\mathrm{PAH}$ shifts the sources up and to the right in these color-color diagrams. Increasing the foreground dust screen from \( E(B-V) = 0.3 \) to \( E(B-V) = 3 \)~mag does not significantly shift the models; the effect on the infrared colors is relatively weak ($<5\%$). 
 
Furthermore, the variation in  the $q_{\mathrm{PAH}}$ values results in a vertical shift in the F335M PAH feature relative to the continuum bands. Comparing CIGALE models to observations indicates that most of the observed data could be described by $q_{\mathrm{PAH}}$ values between 1\% and 2\%.  \citet{henny2025} found the same $q_{\mathrm{PAH}}$ by fitting HST and NIRCam observations of optically selected clusters, although these bands only provided constraints on the abundance of smaller PAHs. We note that the F335M PAH feature relative to its continuum does not completely trace $q_{\mathrm{PAH}}$ and requires longer-wavelength data \citep{Sutter2024}.

\cite{Sutter2024} indicate that about 80\% of the emission in the MIRI bands F770W and F1130W is dominated by PAHs. We find that the MIRI PAH band ratios $r_{7}$ and $r_{11}$ are primarily governed by dust mass, or the dust-to-stellar mass ratio, with little or no effect from stellar age, and are further influenced by the PAH mass fraction ($q_{\mathrm{PAH}}$) (see Figure~\ref{fig:data}, second row). 

We further classify these ISM sources into optically embedded and exposed young clusters based on the detection of the associated H$\alpha$ emission. Using the detection limits outlined in Table \ref{tab:comp}, we define embedded regions as those that are bright in the mid-infrared bands (i.e., F1000W for the 10~$\mu$m catalog and F2100W for the 21~$\mu$m catalog) but remain undetected in H$\alpha$. In contrast, exposed regions are detected in both mid-infrared and H$\alpha$ emission. There is a factor of $\sim 4$ variation in distance to the galaxies, so setting flux detection thresholds for the IR source (to be included in the catalog) and H$\alpha$ sources (to be classified as optically bright or faint) will lead to different luminosity thresholds in different galaxies.  However, as shown in Section \ref{sec:corr}, sources in different galaxies have a similar range in the ratio of H$\alpha$/F2100W. In the following, we provide further details about these sources.

 \begin{figure*}[!t]
    \centering
        \includegraphics[width=1\linewidth]{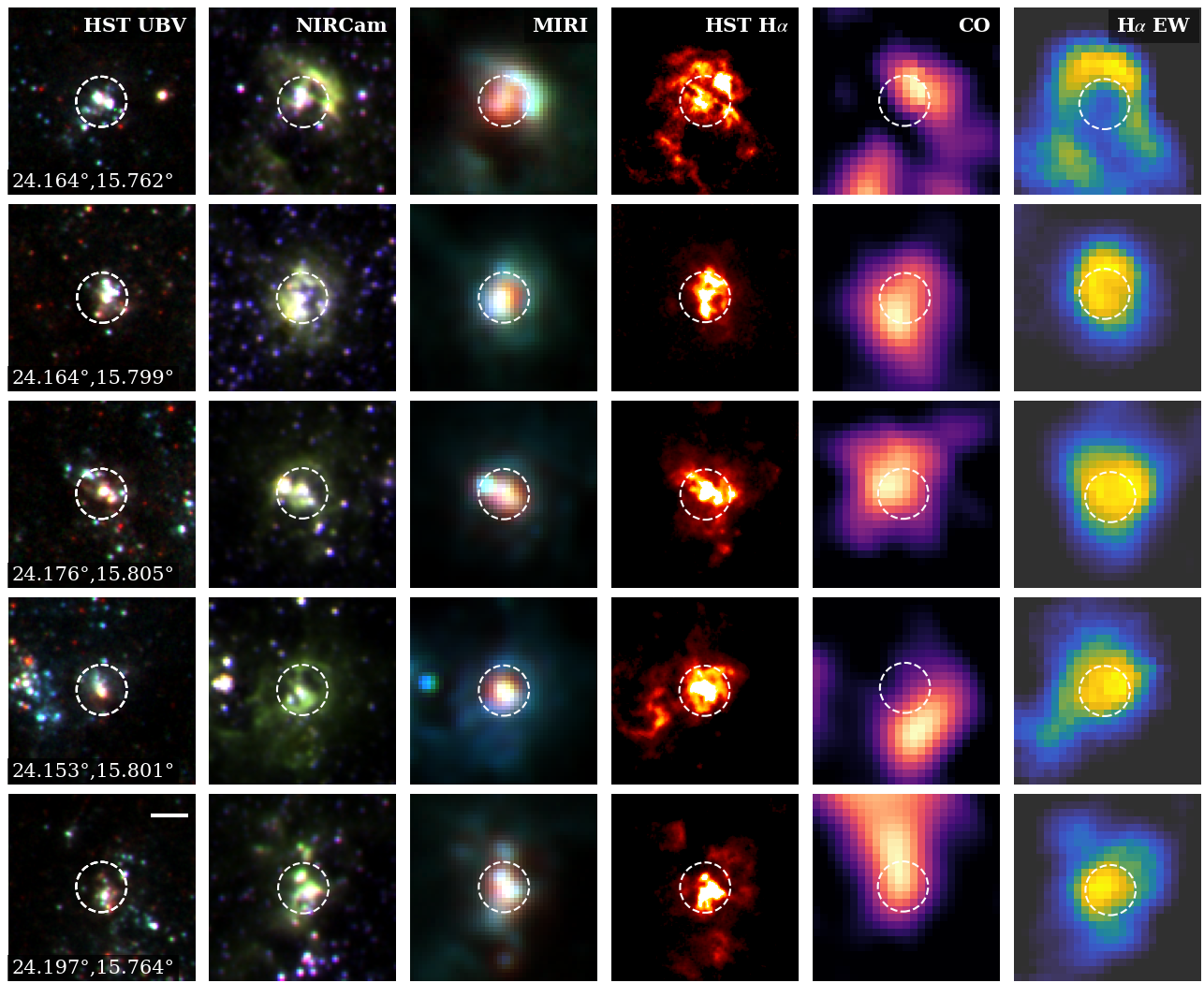}
\caption{Five example ``exposed ISM sources'' in our catalog. These 21~$\mu$m-bright sources are young clusters associated with bright H$\alpha$, high H$\alpha$ EW ($\gtrsim 500$~\AA), and neighboring GMCs in NGC 0628. The three-color HST RGB maps are composed of F438W, F555W, and F814W. NIRCam filters are F300M, F335M, and F360M, and MIRI observations show F770W, F1000W, and F2100W as RGB images. The solid white line in the bottom-left panel indicates a scale of  1\arcsec. The dashed white circle marks the center of the F2100W peak with a radius of 0.67\arcsec.} 
    \label{fig:1myr_objects}
\end{figure*}
\subsubsection{Exposed Sources}
We illustrate several exposed ISM sources in NGC 0628 in Figure~\ref{fig:1myr_objects}, which are young stellar clusters.  These clusters are closely associated with nearby molecular gas traced by the CO and exhibit H$\alpha$ equivalent width (EW) values derived from the MUSE data exceeding 120~\AA, consistent with ages younger than 5~Myr \citep{hassani23,Whitmore2025}. These sources are typical of objects of ``exposed'' objects in our catalog. 

In total, we identified 6,355 exposed regions in the 21~$\mu$m catalog and 10,345 in the 10~$\mu$m catalog, corresponding to approximately 25\% and 20\% of the regions in the 21~$\mu$m and 10~$\mu$m catalogs, respectively. If we limit our catalog to ISM sources, we find that about 50\% are young exposed clusters, $<10\%$ are embedded clusters, and the rest are not detected (i.e., not brighter than the detection limits in the F2100W band). For example, in the 21~$\mu$m catalog of NGC~0628, about 80\% of ISM sources are young exposed clusters, only 5\% are embedded clusters, and the remaining ISM sources fall below the detection limit at F2100W.

We find that 60\% and 74\% of all exposed clusters detected at 10 and 21~$\mu$m, respectively, show H$\alpha$ emission with SNR$>5$.  We regard these sources as our best estimate of the fraction of young clusters with bright optical emission. In the 21~$\mu$m catalog, only 10\% of exposed clusters, and in the 10~$\mu$m catalog, 18\%, exhibit an H$\alpha$ SNR below 3. These sources often display diffuse or spatially extended H$\alpha$ emission, likely originating from nearby regions that contaminate the photometric aperture. This finding is enabled by comparing the native-resolution HST-H$\alpha$ map with the convolved version used for each catalog, an analysis that was not feasible in the previous study by \cite{hassani23} due to the coarser resolution of the MUSE H$\alpha$ data.

By cross-matching our exposed sources with the ``nebular'' catalog from \cite{neb_catalog} using a maximum separation of 1\arcsec\ (comparable to the typical MUSE PSF), we find that about 77\% of the exposed clusters are associated with MUSE nebular regions across our sample of 19 galaxies. \cite{hassani23} found this association to be approximately 90\% in four galaxies: IC 5332, NGC 0628, NGC 1365, and NGC 7496 for their ISM sources.  We attribute the lower fraction in this work to the use of CD in our source-finding algorithm, which finds fainter sources and more embedded clusters. We also benefit from using the spatially resolved HST H$\alpha$ observations, convolved to match the JWST F2100W resolution. The latter is helping to find H$\alpha$ emission localized to the location of the mid-IR source, while the MUSE observations with PSF $\gtrsim 0.9 \arcsec$ might also include other nearby regions.

We find that the average H$\alpha$ equivalent width (EW) measured by MUSE in exposed clusters is 120~\AA, with many clusters exceeding 200~\AA. These values show some variation across different galaxies. In dustier systems, the H$\alpha$ EW tends to be lower due to differential attenuation, which affects the nebular emission and the stellar continuum in distinct ways \citep{Calzetti_rs}. This effect is particularly evident in galaxies such as NGC1365 and NGC3627, where the mean H$\alpha$ EW drops to below 80~\AA. In contrast, less dusty galaxies, such as NGC~5068, have higher values, with H$\alpha$ EW around 160~\AA.  We show H$\alpha$ EW since it is a useful proxy for the ages of stellar clusters \citep{Levesque_2, Leitherer99}.  In forthcoming work (Hassani et al., in prep.), we will perform full SED fitting using combined HST and JWST data.
% However, in this study, we do not rely on EW-based models for cluster age-dating \citep[e.g.,][]{Levesque_2}, such as those provided by Starburst99 \citep{Leitherer99}. Instead, a more detailed analysis will be presented in a forthcoming paper, where we will perform SED fitting using combined HST and JWST data.}

 % The Starburst99 model predicts that for an instantaneous burst star formation history, a Kroupa IMF, and solar metallicity ($Z = 0.02$), H$\alpha$ EW $\gtrsim 400$ \AA, corresponds to an age of less than 5 Myr \citep{Levesque}. Assuming our median EW value of 120~\AA\ and a typical reddening of $E(B-V) = 0.5$~mag (i.e., $A_V = 1.5$ mag), we apply the extinction correction using $A_{H\alpha} = \kappa_{\text{H}\alpha} \times E(B-V)$, following \citep{Kreckel_2013}, with $\kappa_{\text{H}\alpha} = 2.38$. This results in an attenuation of $A_{H\alpha} \approx 1.2$~mag and an attenuation-corrected H$\alpha$ equivalent width of EW$(H\alpha)_{\text{corr}} \approx 400$~\AA, indicating that many of the exposed clusters are very young.

 \begin{figure*}
    \centering
        \includegraphics[width=1\linewidth]{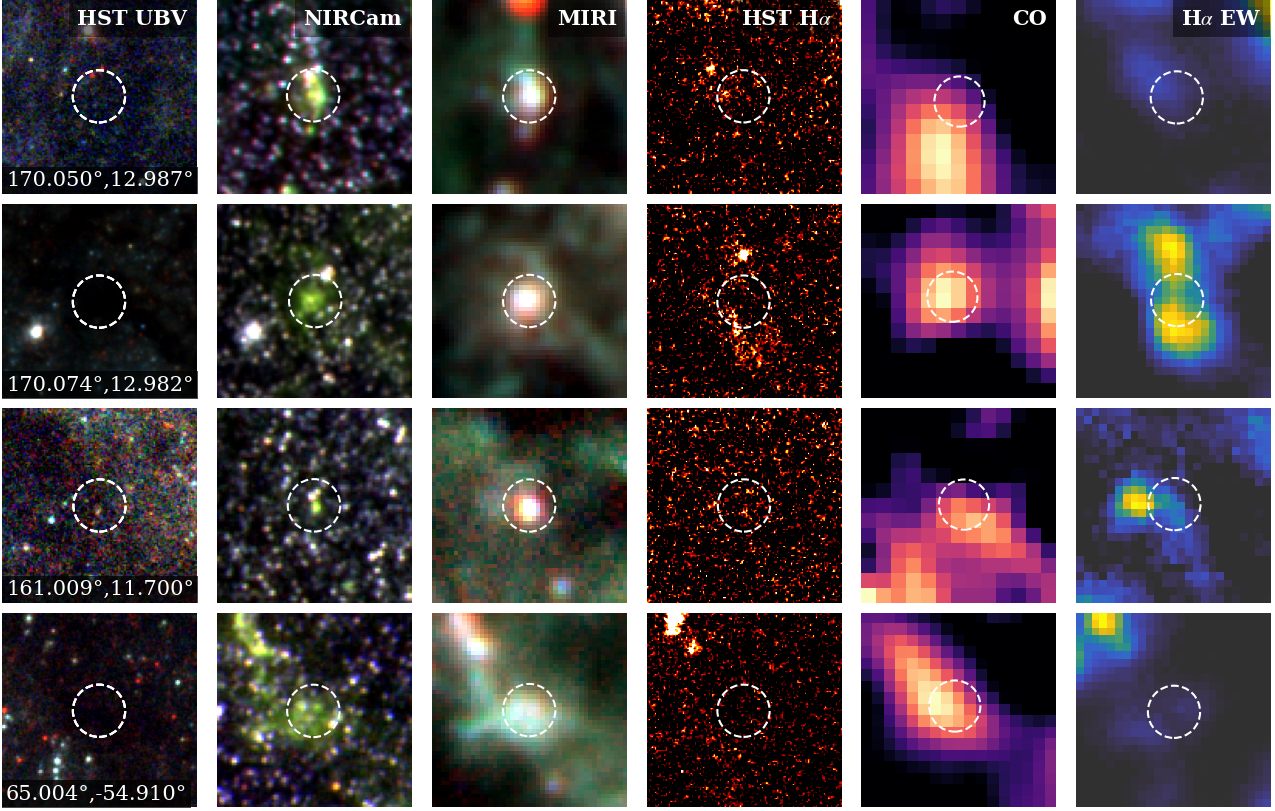}
\caption{Similar to Figure \ref{fig:1myr_objects}, this figure shows ``embedded ISM sources'': the first two panels are in NGC3627, the third in NGC3351, and the fourth in NGC1566. These are stellar clusters without HST H$\alpha$ detections, with low MUSE H$\alpha$ EW ($<$30\AA), and appearing as compact, mid-IR bright clusters ($F_{F2100W} > 50$~$\mu$Jy).}
    \label{fig:emb_sources}
\end{figure*}

\subsubsection{Embedded Sources}
%cross match need to be add.
\label{sec:emb_sources}
Figure \ref{fig:emb_sources} shows three embedded sources: two in NGC3627, one in NGC3351, and one in NGC1566. These sources are very bright in the F2100W band ($\gtrsim 50\mu$Jy), exhibit a strong 3.3~$\mu$m PAH feature, show little to no H$\alpha$ emission in HST, have weak H$\alpha$ EW from MUSE, and remain undetected in the optical. Many embedded regions show stellar clusters within the $0.67\arcsec$ aperture in NIRCam bands, whereas exposed regions sometimes show PAH as shell-like, feedback-powered structures extending beyond the aperture.

We found 939 and 1411 embedded clusters in our 10 and 21~$\mu$m catalogs, which account for $<5\%$ of the whole catalogs. If we restrict our analysis to ISM sources, 5\% and 10\% of them are identified as embedded clusters in the 10~$\mu$m and 21~$\mu$m catalogs, consistent with the findings for four galaxies in \citet{hassani23}. Our results indicate that most embedded clusters are not particularly bright in the F2100W mid-infrared band, with the majority exhibiting luminosity around or below $L_{\nu} < 10^{18}$ W~Hz$^{-1}$ (see Figure~\ref{fig:sf_relation}) while the median luminosity for exposed ISM sources is 3 $\times$ 10$^{18}$  W~Hz$^{-1}$.
% In fact, the median luminosity of embedded sources are $L_{\nu} = \times 10^{18}$ W~Hz$^{-1}$ in the 21 $\mu$m catalog and $L_{\nu} = 2 \times 10^{17}$ W~Hz$^{-1}$ in the 10 $\mu$m catalog.}

Our classification of exposed and embedded clusters is primarily based on $L_{\text{H}\alpha}$. However, it is possible for a 21~$\mu$m embedded cluster to have H$\alpha$ emission originating from a nearby source, which could artificially boost the measured H$\alpha$ flux even if there is no H$\alpha$ directly associated with the cluster. Based on this consideration, we aim to identify additional IR-bright embedded clusters that were missed in our original classification. Hence, we conducted a visual inspection of the SEDs for many sources and identified several IR-bright objects classified as exposed regions in our catalog that show little or no H$\alpha$ emission within the MIRI aperture. Although these sources have H$\alpha$ luminosity above those completeness limits in Table~\ref{tab:comp}, their H$\alpha$ emission is weak because it is the extended emission from a source outside the 21~$\mu$m aperture. We find that the majority of these sources have HST H$\alpha$ SNR$<3$.

\begin{figure*}
    \centering
    \includegraphics[width=0.82\textwidth]{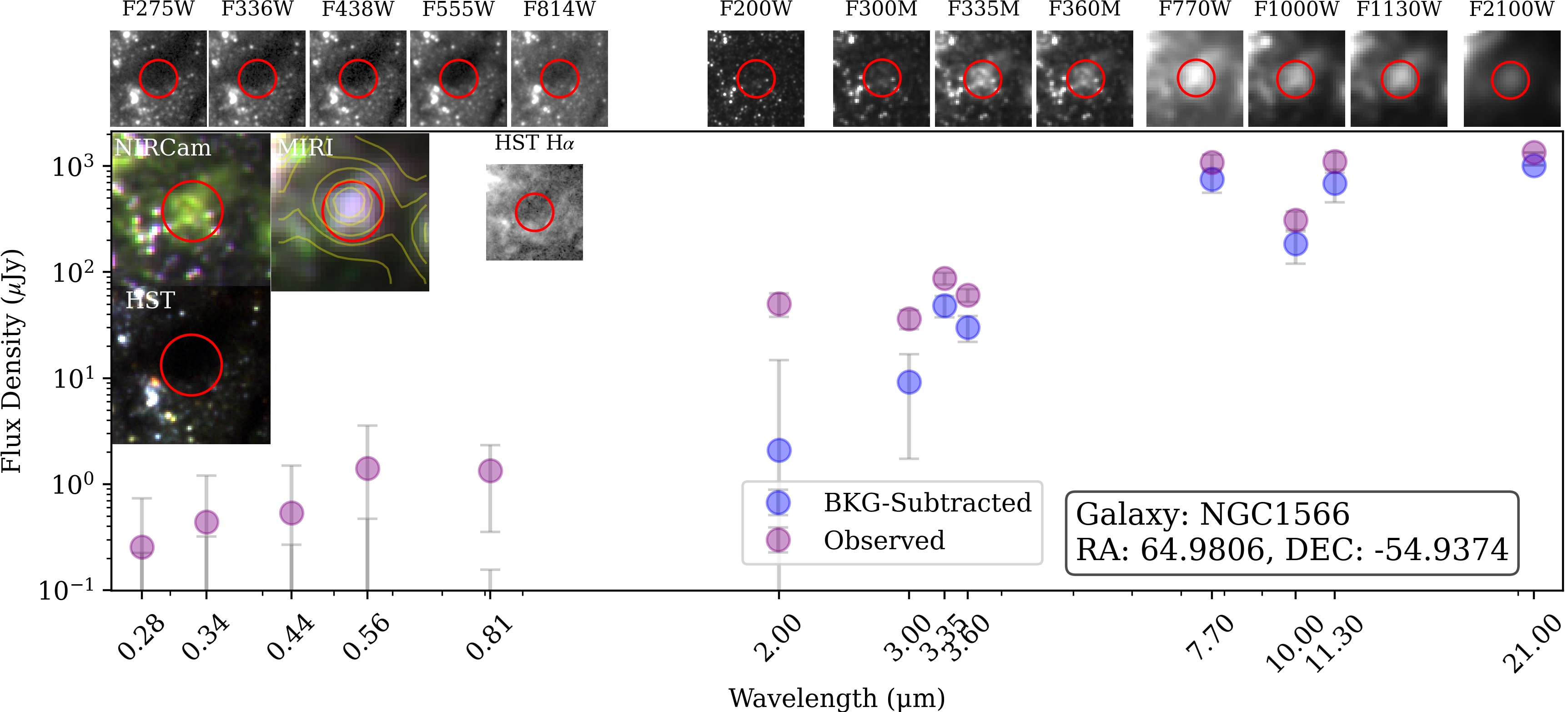}
    \caption{A deeply embedded cluster that is classified as an exposed cluster in our 21~$\mu$m catalog, one of $<1\%$ of such cases. At the top, we present 1.8\arcsec cutouts for each wavelength, using a linear scale for JWST images and a square root scale for HST broad band and H$\alpha$ images to better highlight faint emission. Additionally, we show false-color RGB composites of the source using NIRCam (F300M/F335M/F360M), MIRI (F770W/F1000W/F1130W), and HST (F336W/F438W/F555W).  The red circle highlights the F2100W extraction region. The source has high H$\alpha$ luminosity of $1.2 \times 10^{38}$ erg s$^{-1}$ in the MIRI aperture, despite the absence of a clear H$\alpha$ source (HST H$\alpha$ SNR = 2.5). Its mid-IR luminosity of $L_{\mathrm{F2100W}} \sim 3.8 \times 10^{19}$ W Hz$^{-1}$ and $r_3 \sim 5.2$ indicate that the cluster is in an early evolutionary stage. We note that this source is also detected in the 10~$\mu$m catalog, exhibiting nearly identical properties. Yellow contours overlaid on the RGB MIRI image represent ALMA CO (2–1) emission, with intensity levels ranging from 6 to 30 K km/s.}
    \label{fig:deep_emb_cluster}
\end{figure*}

Consequently, we define an additional class of deeply embedded clusters (``deep\_emb'') that exhibit H$\alpha$ luminosity above the detection limits (similar to exposed clusters) and mid-IR luminosity ($L_{\nu}$) at least three times above the detection thresholds, yet have HST H$\alpha$ SNR$<3$. We find that the number of such clusters is even lower than that of the typical embedded clusters identified above, with only 86 sources in the 21~$\mu$m catalog having a median luminosity of $L_{F2100W} = 10^{19}$ W Hz$^{-1}$, and 228 sources in the 10~$\mu$m catalog with a median luminosity of $L_{F1000W} = 9 \times 10^{17}$ W Hz$^{-1}$. It is important to note that we cannot rule out the possibility that evolved stars (e.g., RSGs) within the MIRI aperture contribute to boosting the mid-IR flux in these clusters, and that they may not differ from the rest of the embedded clusters in our catalog. We present an example SED of such a source from the 21~$\mu$m catalog in NGC 1566 (Figure~\ref{fig:deep_emb_cluster}).

% and another from the 10~$\mu$m catalog in NGC 1672 (Figure~\ref{fig:deep_emb_cluster_1672}), respectively. 
% We further highlight this source and a similar source from the 10 $\mu$m catalog in Figure \ref{fig:sf_relation}, where we examine the relationship between the luminosity of attenuation-corrected H$\alpha$ and F2100W (see Section \ref{subsec:source_pro}). 
% }

% We found that the median F2100W/F1000W ratio is ~3.4 in exposed regions, which it is 40\% lower in embedded regions, likely due to  several reasons.

% % dust temperatures, as these bands traces warm dust, which is more strongly heated in exposed regions where su
% % rrounding dust has been cleared. 

% \textbf{We also note that by comparing the median fluxes of embedded sources detected in the 10 and 21~$\mu$m catalogs, we find that the F1000W flux is about twice as faint in the 10~$\mu$m catalog. We emphasize that the median ratio of F335M relative to its continuum, F300M or F360M, is remarkably similar for both exposed and embedded regions in 21$\mu$m catalog, with $ r_{3} \approx 3 $ and $ r_{3.6} \approx 1.5 $. A similar trend is observed in both catalogs for $ r_{7} $ and $ r_{11.3} $, where both phases have ratios of 2.7 and 3.1, respectively.} This indicates that PAH emission is present in both embedded and exposed phases, and it is much stronger than for any dusty stars (See Section \ref{sec:class}). We also found that the median H$\alpha$ EW in the MUSE data is 40~\AA\ for embedded clusters, with many below 20~\AA, as there is almost no H$\alpha$ emission in these clusters.}

\subsection{Stellar Sources}
\label{sec:stars}

% Using the detection limits provided in Table~\ref{tab:comp}, 
In this section, we present the case for the photometric ratio cuts given in Table \ref{tab:flux_ratios_sources} that motivate our classification of mid-IR bright stellar sources. We rely on both the \texttt{PARSEC} models (section \ref{sec:parsec}) and the spectroscopic classifications of LMC sources from \citet{Jones2017,Jones2018}, which are primarily used for the CPNs, WR, and B[e] stars. 

% \paragraph{RSGs and O-AGBs}
% RSGs and O-AGBs are challenging to distinguish in the optical and near-infrared wavelengths, as both have the same V-I color range of 0 to 1, but mostly RSGs are brighter at V band \citep{Bolatto,Jones2017}. Many RSGs in the Galaxy and Magellanic Clouds show the F1130W PAH feature, with some also exhibiting the 7.7$\mu$m PAH feature \citep{mw_rsg,rsg_mcs}. Additionally, the NIR-to-MIR slope of RSGs is mainly attributed to continuous opacity, likely from amorphous carbon, with enhanced 10$\mu$m emission due to amorphous silicates with olivine stoichiometry \citep{mw_rsg}. O-AGB stars have dust composed of Olivine, amorphous alumina, and metallic iron \citep{Jones2014_oagb}, and also show the 10~$\mu$m silicate feature \citep{Jones2017}.

\subsubsection{RSGs and O-AGBs} 
\label{sec:rsg-oagbs}
RSGs and O-AGBs are challenging to distinguish in the optical and near-infrared wavelengths, as both have the same $V-I$ color range of 0 to 1~mag, but mostly RSGs are brighter at $V$ band \citep{Bolatto,Jones2017}. We note that some RSGs are faint in optical bands, making an infrared classification framework essential to find these sources. In this section, we use stellar models to (i) assess whether a star would be detectable at the distances of our targets and (ii) define flux ratio cuts for classification.

\paragraph{RSGs} We follow the prescription provided by \cite{Johnson_23}, using the \texttt{PARSEC} tracks to follow the evolution of stars from ages of 1 to 40~Myr with a timestep of 0.1~Myr. We limit the effective temperature to a range of $\log(T_{\text{eff}}/\text{K}) = 3.53$–$3.63$ \citep{Levesque} and select bright RSGs with luminosities in the range $4.5 < \log(L/L_\odot) < 5.6$. Additionally, to avoid contamination from AGB and super-AGB stars, we include only stars with initial masses $\geq 10~M_\odot$. Under these criteria, the selected RSGs have masses ranging from $10~M_\odot$ to $28~M_\odot$, ages between 5.6 and 28~Myr, and mass-loss rates between 1 and $15 \times 10^{-6}~M_\odot~\mathrm{yr}^{-1}$.  Figure~\ref{fig:models} shows one of the brightest 10~$\mu$m models \added{with an initial metallicity of $Z=0.0152$ } for a \added{blue} RSG with an age of 5.6~Myr, mass of $28~M_{\odot}$, and $\log(L/L_{\odot})=5.58$. \added{As noted in Section~\ref{sec:parsec}, the RSG models adopted in this work do not include circumstellar dust. Consequently, these models primarily trace the detection limits for relatively dust-free (bluer) RSGs and are not representative of heavily dust-enshrouded objects. Using observed RSGs in the LMC \citep{Groenewegen2018} as empirical benchmarks, we find that a substantial fraction of RSGs would remain detectable in F1000W across our sample. Specifically, we would detect $\sim$90\% (50/57) of LMC-like RSGs at our nearest target, while this fraction decreases to $\sim$30\% (18/57) at 15~Mpc and to $\sim$15\% (8/57) at 19~Mpc. In contrast, detectability in the F2100W band is more limited. At our nearest distance (5.2~Mpc), only $\sim$20\% (13/57) of LMC-like RSGs would be detected. At 10~Mpc, only about two LMC RSGs are expected to be detectable in F2100W, and at larger distances the detection of RSGs is unlikely.}

% The 10~$\mu$m RSGs are expected to have $M \gtrsim 20~M_\odot$, and predominantly those with mass-loss rates exceeding $5 \times 10^{-6}~M_\odot~\text{yr}^{-1}$. Our detection limit in NGC~5068 means we are likely capable of identifying RSGs with $\log(L/L_{\odot}) > 5$, corresponding to an initial mass of $15~M_{\odot}$. These models demonstrates that RSGs are detectable in the F1000W band but is unlikely to be detected in the F2100W band. The brightest, most massive, and youngest RSGs do not exceed a flux of 10~$\mu$Jy at F2100W at the distance of our closest target, NGC~5068, which is lower than the detection limit for the F2100W band. Hence, we do not expect many RSGs to be detected in our 21 $\mu$m catalog or in our more distant targets in either catalog. For comparison, we note here that the studies in the LMC can reach as deep as $\log(L/L_{\odot}) = 4$, which corresponds to an initial mass of $9~M_{\odot}$ \citep{Neugent}. # remved

\begin{figure*}[!t]
    \centering
    \includegraphics[width=0.95 \textwidth]{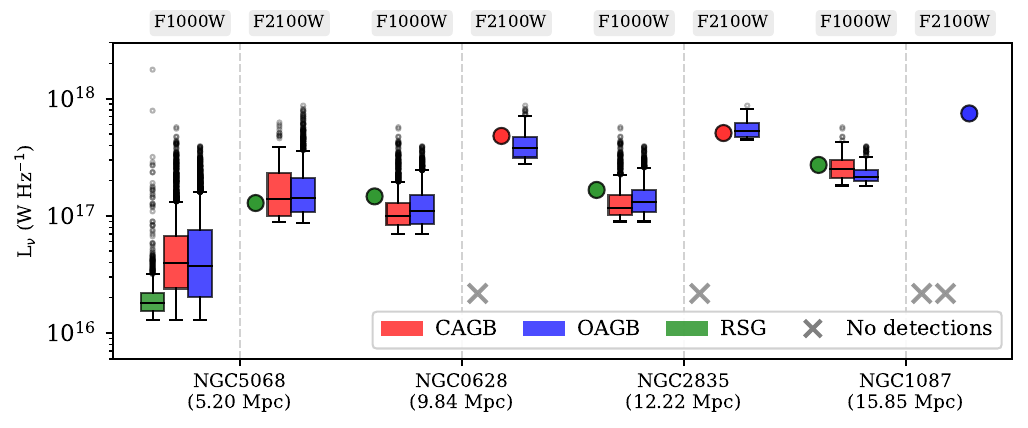}
    \caption{
    Distribution of predicted luminosities from PARSEC stellar tracks at F1000W (left) and F2100W (right) filters for detectable sources. \added{The PARSEC tracks include a wide range of metallicities and different dust chemistries for AGB stars (see Section~\ref{sec:parsec}). The RSG models are complemented by LMC sources from \citet{Groenewegen2018} to include more dusty RSGs as well.} Box plots show the 25th and 75th percentiles of the specific luminosities that are above our completeness limits.  The whiskers extend to 1.5 times the interquartile range, and points beyond indicate extreme values outside the typical range. \added{Circular markers show the median luminosity when there are $<50$ sources and the $\times$ symbol indicates $<5$ of that type are detected.}  We show four targets spanning a range of distances to show how completeness changes catalog make-up depending on the distance to the target. }
    \label{fig:models_all}
\end{figure*}

\added{Figure~\ref{fig:models_all} summarizes the luminosities of modeled, relatively blue RSGs that are brighter than the detection limits, combined with LMC sources from \citet{Groenewegen2018}, for four of our targets.}

% As expected, RSGs are predominantly detectable in our nearest galaxy, NGC 5068, but remain undetected in the more distant galaxies.# removed

The \texttt{PARSEC} stellar evolutionary tracks indicate that these stars show strong emission in the F200W band, with $r_{2-3.6} \gtrsim 2$ and $r_{2-7} \gtrsim 8$ for RSGs. \added{For comparison, the median value for SAGE-LMC RSGs is $r_{2-7} = 4$, and LMC sources from \cite{Groenewegen2018} predominantly show $r_{2-3.6} \gtrsim 1.5$. We therefore adopt selection criteria of $r_{2-7} > 3$ and $r_{2-3.6} > 1.5$ to identify RSGs. We also find that $r_{3-3.6}$ for RSGs lies in the range $\sim 1.2$--$1.4$ from PARSEC tracks and can decrease to $\sim 0.6$, based on LMC sources from \citep{Groenewegen2018}.} Our evolutionary tracks for RSGs predict $ r_{3} > 0.8 $ and $ r_{3.6} < 1.2 $. 

\added{Our models predict $1 < r_{7} < 1.6$ and $r_{11} \sim 0.6$ for RSGs (see Figure~\ref{fig:models}). In contrast, \citet{Jones2018} report that $r_{7}$ spans a broader range of $0.4 \lesssim r_{7} \lesssim 1.6$, while $r_{11}$ ranges from $\sim 0.6$ to 2, in agreement with LMC sources from \citet{Groenewegen2018}. This discrepancy likely arises because our RSG models assume dust-free photospheric emission, with no circumstellar dust grains (e.g., silicates) included. We note that our AGB modeling explicitly accounts for different dust chemistries, which affect the infrared fluxes. The absence of circumstellar dust in the PARSEC RSG tracks is likely explanation for the offset between our model predictions and the observed RSG colors reported by \citet{Jones2018} and \citet{Groenewegen2018}. Therefore, we adopt \added{the empirically-motivated} selection criteria of $r_{7} < 1.7$ and $r_{11} > 0.5$ for identifying RSGs.}  Due to the lack of $V$ and $I$ band detections for many stars, we do not apply optical color cuts in our study. 

%This depth may also include foreground stars such as nearby foreground red dwarfs and AGB stars \citep{Neugent}. We note that RSGs are unlikely to be detected in the more distant targets.
% At a distance of 9.8 Mpc, the brightest RSGs reach a flux of about 4~$\mu$Jy. This flux level makes it unlikely to detect RSGs in galaxies such as NGC~0628 or those located farther away.

\paragraph{O-AGBs} For AGB stars, we evolve the \texttt{PARSEC} models from an age of 100~Myr to 10~Gyr \added{with a time step of 30 Myr}, classifying the stars as either O-AGB or C-AGB based on their carbon-to-oxygen ratio (C/O) and corresponding stage of stellar evolution. 

For O-AGB stars, we require a C/O$<1$ in the early AGB (EAGB) and TP-AGB phases, with an effective temperature range of $\log(T_{\text{eff}}) = 3.4$–$3.6$. This results in O-AGBs with masses ranging from $0.6$ to $5.3~M_{\odot}$, ages between $100~\mathrm{Myr}$ and $10~\mathrm{Gyr}$, and $2.5 < \log(L/L_{\odot}) < 4.7$, with mass loss up to $75 \times 10^{-5}~M_{\odot}$ per year.

\begin{figure}[!t]
    \centering
    \includegraphics[width=0.99\linewidth]{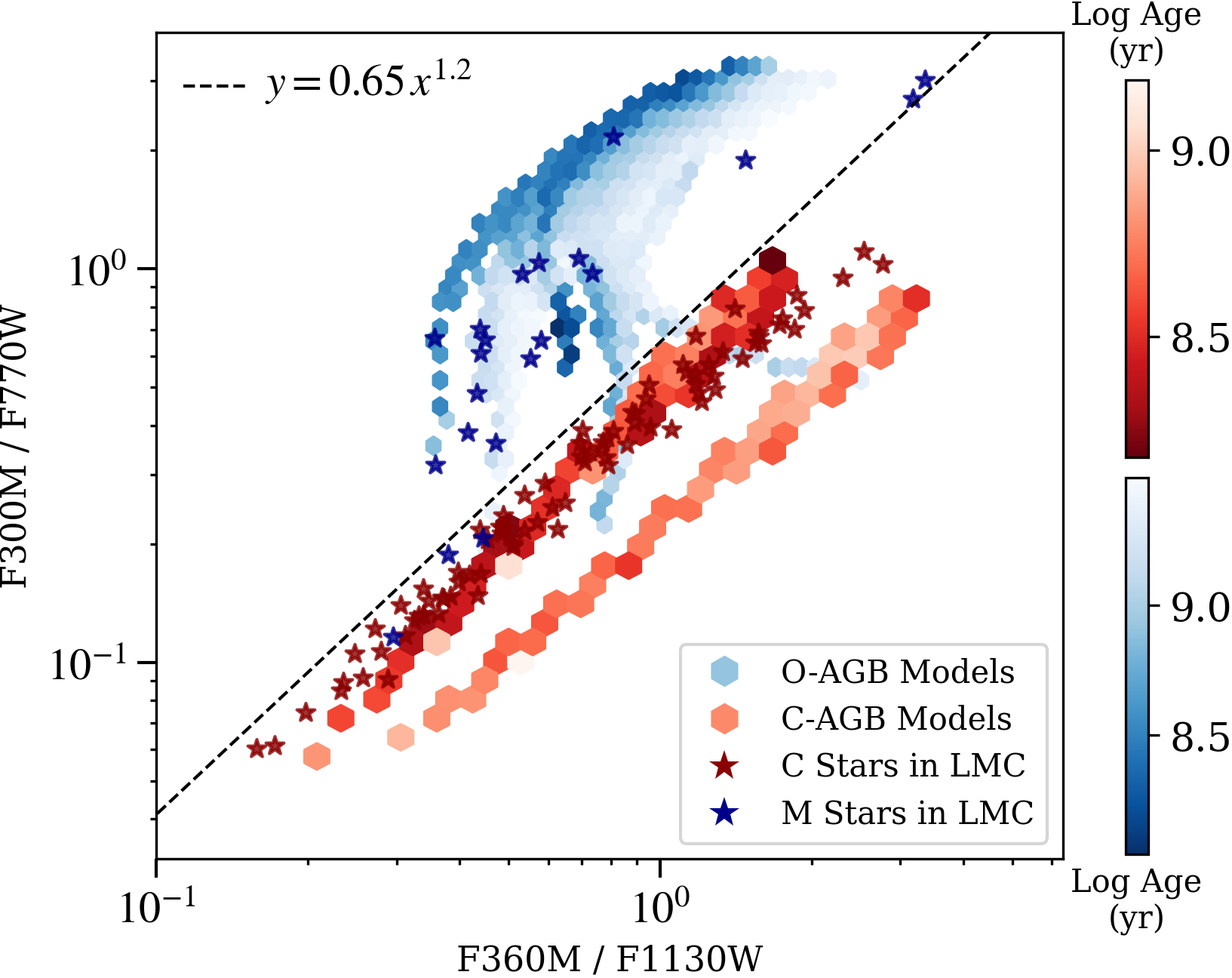}
\caption{\added{Ratio of F300M/F770W versus F360M/F1130W for C-AGB and O-AGB models from PARSEC tracks, shown as hexagonal bins. The models span three metallicities (Z = 0.005, 0.0152, and 0.025) and multiple dust chemistries (see Section \ref{sec:parsec}). Only models that would be detectable in F1000W at a distance of 5.2 Mpc are displayed (see Table \ref{tab:comp}). In each hexagonal bin, the color represents the mean logarithmic stellar age, computed using the average of the contributing models; only bins containing at least five models are displayed. Stars are taken from \cite{Groenewegen2018}, restricted to Large Magellanic Cloud sources that would be detectable at the distance of our nearest target (5.2 Mpc). The dashed line indicates the empirical division between C-AGB and O-AGB stars in this diagram.}}  
\label{fig:models_flux_ratio}
\end{figure}

We should be able to identify O-AGB in many of the nearest targets though the portion of the population we will select will change. In NGC~5068, we are able to detect O-AGBs with $3.7\lesssim \log(L/L_{\odot})<4.8$. Most detectable O-AGBs are $\lesssim 2-3$ Gyr old, with the most probable ages being  $<200$ Myr, where C/O $< 0.5$, showing a maximum flux of \added{120~$\mu$Jy} at F1000W. At the F2100W band, their flux can exceed \added{250}~$\mu$Jy. This enables the detection of relatively young O-AGBs with ages $<1~\mathrm{Gyr}$ in NGC~628 at both the F1000W and F2100W bands. At a distance of $12~\mathrm{Mpc}$ (e.g., NGC~2835 and NGC~3627), it is still possible to detect the youngest O-AGBs with ages below $300~\mathrm{Myr}$ and masses greater than $4~M_{\odot}$ with both bands. At greater distances, only the most massive O-AGBs will be marginally detectable in the F1000W band and possibly in the F2100W band. We show an O-AGB with age of about 130~Myr and mass of 5~$M_{\odot}$ in Figure \ref{fig:models}. Figure \ref{fig:models_all} summarizes the range of detectable luminosities for targets at varying distances. Similar to the RSGs, only models brighter than the detection limits are included, showing that O-AGBs are likely to be detectable in our more distant targets. In more distant targets, where fewer than 50 models exceed the detection threshold, only the median expected luminosities for O-AGBs are presented.

\begin{figure*}[!t]
    \centering
    \includegraphics[width=0.9\textwidth]{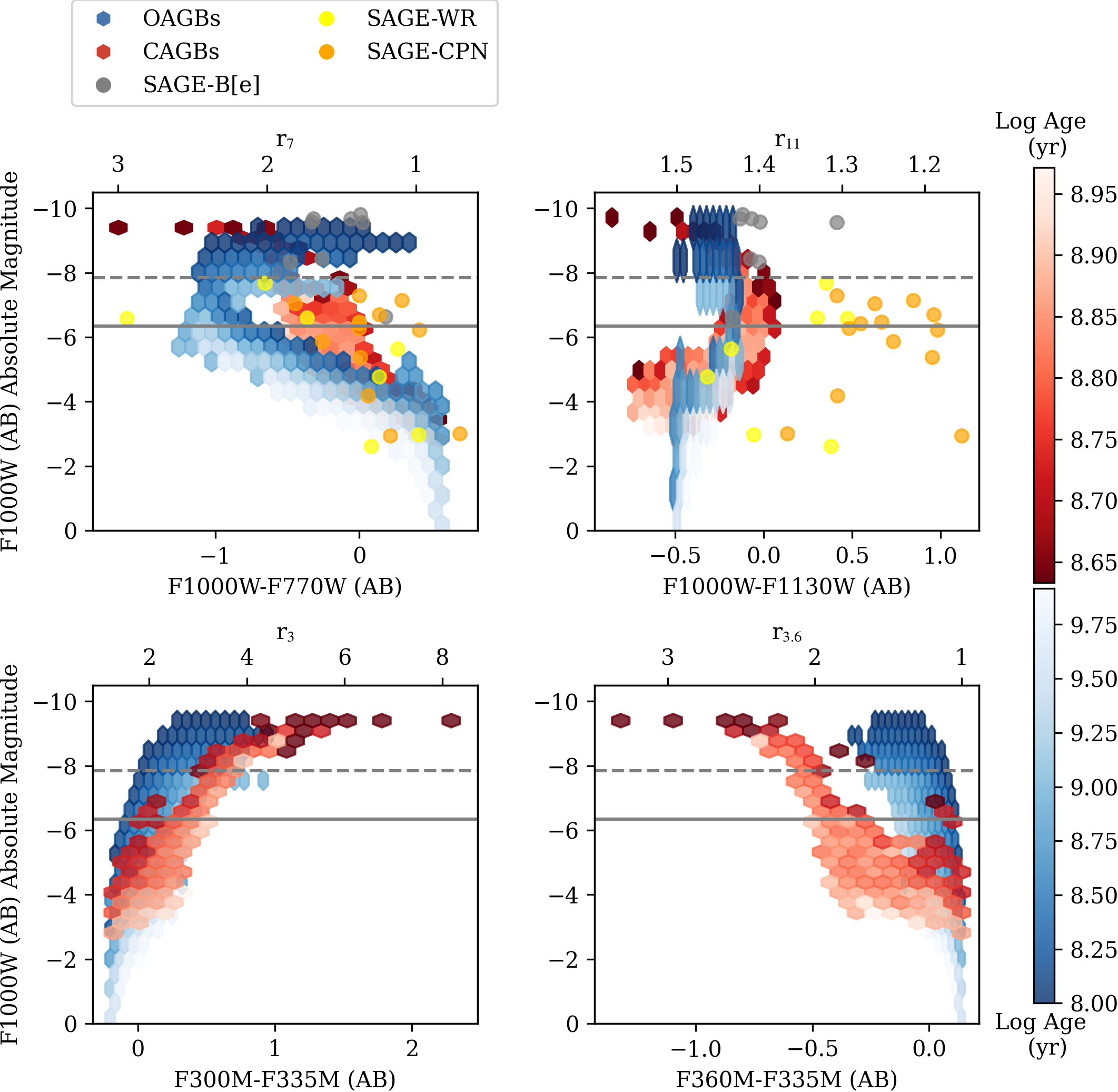}
\caption{Predicted color–magnitude diagrams for O-rich (blue) and C-rich (red) AGB stars based on PARSEC/Colibri models in hexagons. \added{We show only solar-metallicity models ($Z = 0.0152$) here, adopting dust compositions of 85\% amorphous carbon (AMC) and 15\% SiC for C-AGB stars, and 60\% silicate and 40\% AlOx for O-AGB stars, following \citet{Groenewegen2006} (see Section~\ref{sec:parsec}).} The top panel shows F1000W-F770W (left) and F1000W-F1130W (right), while the bottom panel presents F300M-F335M (left) and  F360M-F335M (right). The colorbar represents age in yr. Additionally, SAGE-LMC sources, including B[e] stars, CPNs, and WR stars, are highlighted in circles and distinct colors. The solid horizontal line indicates the detection limit of 5 $\mu$Jy at 5 Mpc, while the dashed line represents the limit at 10 Mpc, as adopted from Table \ref{tab:comp}.}  
\label{fig:models_miri}
\end{figure*}

\added{For O-AGBs, the PARSEC models predict $r_{2-3.6} < 2.5$ and $r_{2-7} < 6$. The models further predict $0.9 < r_{3} < 2$ and $r_{3.6} <1.1$. We also find that $r_{3-3.6}$ spans a range of approximately 0.4 to 1.3.}   Our tracks show $r_{7} < 1.3$ and $r_{11} < 1.2$ for O-AGBs, reaching values as low as 0.3, consistent with the SAGE sources from \citet{Jones2017}. \added{We also compare our MIRI flux ratio criteria with those from \citet{Jones2018}. Their synthetic photometry predicts $r_{7}$ values ranging from 0.2 to 1.6 and $r_{11} < 1$, which is in agreement with our classification criteria.}

\begin{figure*}[!t]
    \centering
    \includegraphics[width=0.82\textwidth]{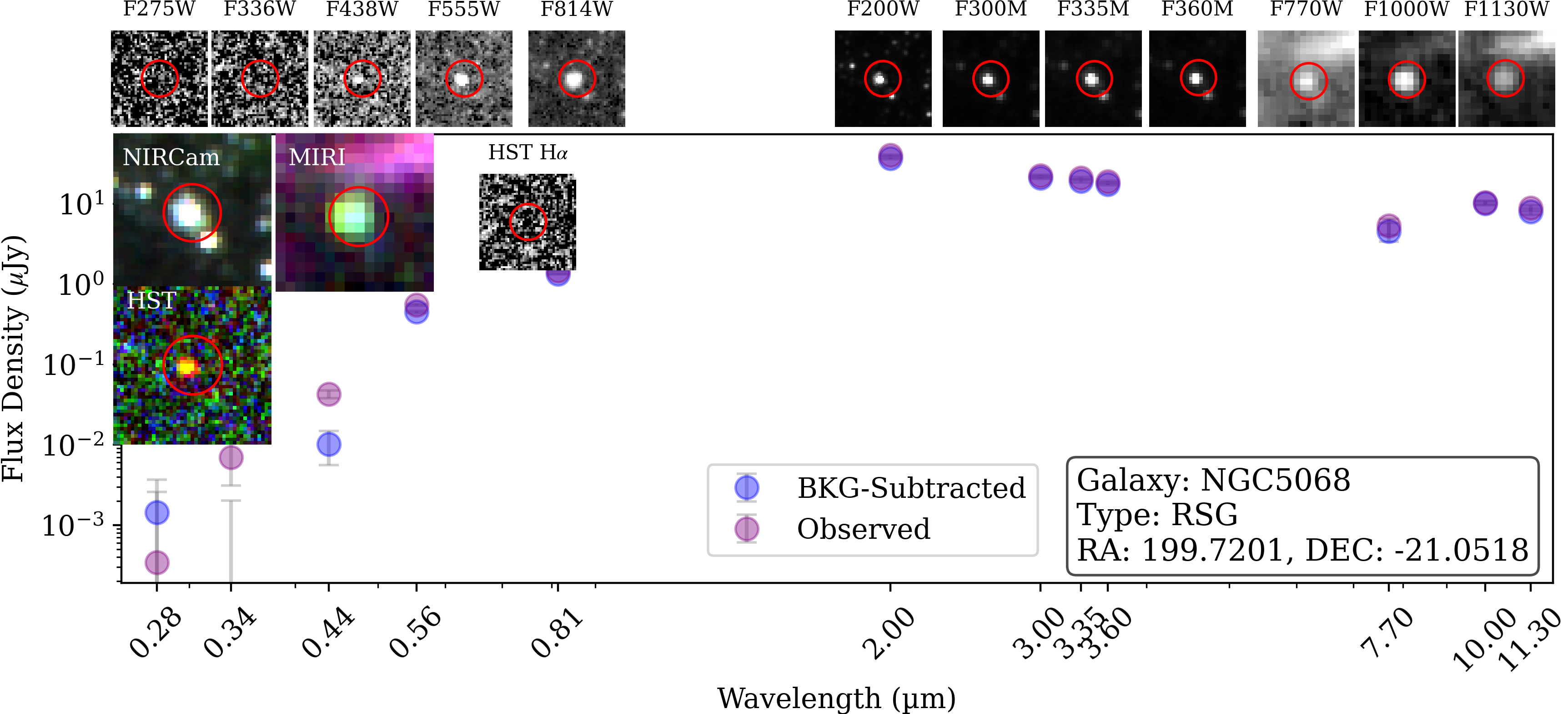}
    \caption{An example SED of a RSG from the 10~$\mu$m catalog using the same format as described in Figure \ref{fig:deep_emb_cluster}. The source is marked with a red circle of 0.33\arcsec\, radius. The physical resolution is approximately 10 pc.}
    \label{fig:rsg}
\end{figure*}

\begin{figure*}[!t]
    \centering
    \includegraphics[width=0.82\textwidth]{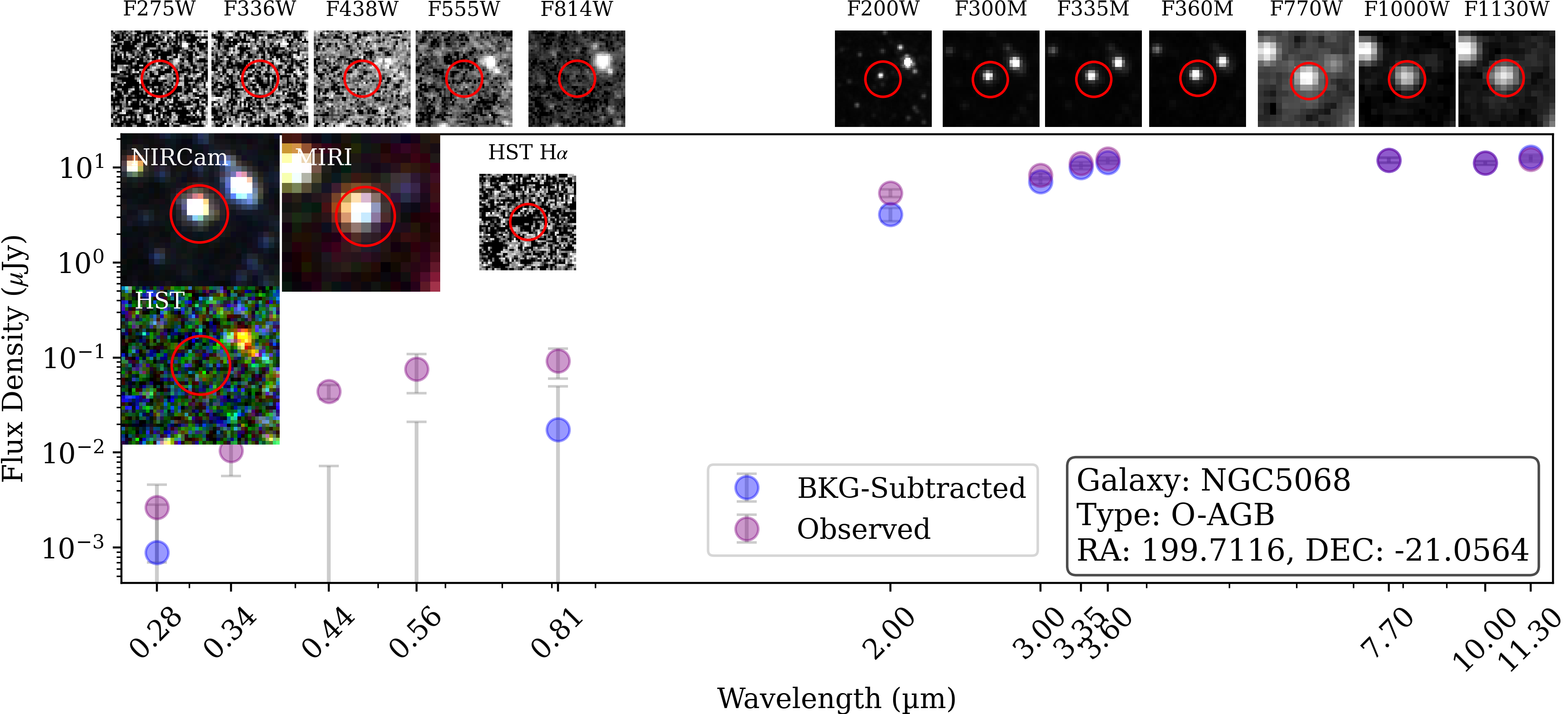}
    \caption{Similar to Figure \ref{fig:deep_emb_cluster}, but for a O-AGB stars.}
    \label{fig:oagb}
\end{figure*}

\paragraph{RSG and O-AGB detections} In total, we identified 1211 RSGs and 1267 O-AGBs in the 10~$\mu$m catalog, and 48 RSGs and 171 O-AGBs in the 21~$\mu$m catalog. \added{27} RSGs from the 21~$\mu$m catalog were also detected in the 10~$\mu$m catalog, and \added{76} O-AGBs from the 21~$\mu$m catalog were likewise found in the 10~$\mu$m catalog. \added{We also find that 365 sources are classified as both RSGs and O-AGBs in the 10~$\mu$m catalog, and 10 sources in the 21~$\mu$m catalog.} We illustrate the SEDs of an RSG and an O-AGB star in Figures~\ref{fig:rsg} and \ref{fig:oagb}, respectively. \added{We further cross-match our 10$\mu$m catalog with the strict RSG classification in PHANGS galaxies from \cite{sarbadhicary2026} using a matching radius of 0.36\arcsec, and find that 77 of the 136 RSGs in NGC~5068 are included in that catalog. We also find that, using a matching radius of 0.67\arcsec, 8 of the 10 RSGs in our 21,$\mu$m catalog for NGC~5068 have counterparts in \cite{sarbadhicary2026}.}

\begin{figure*}[!t]
    \centering
    \includegraphics[width=0.85\textwidth]{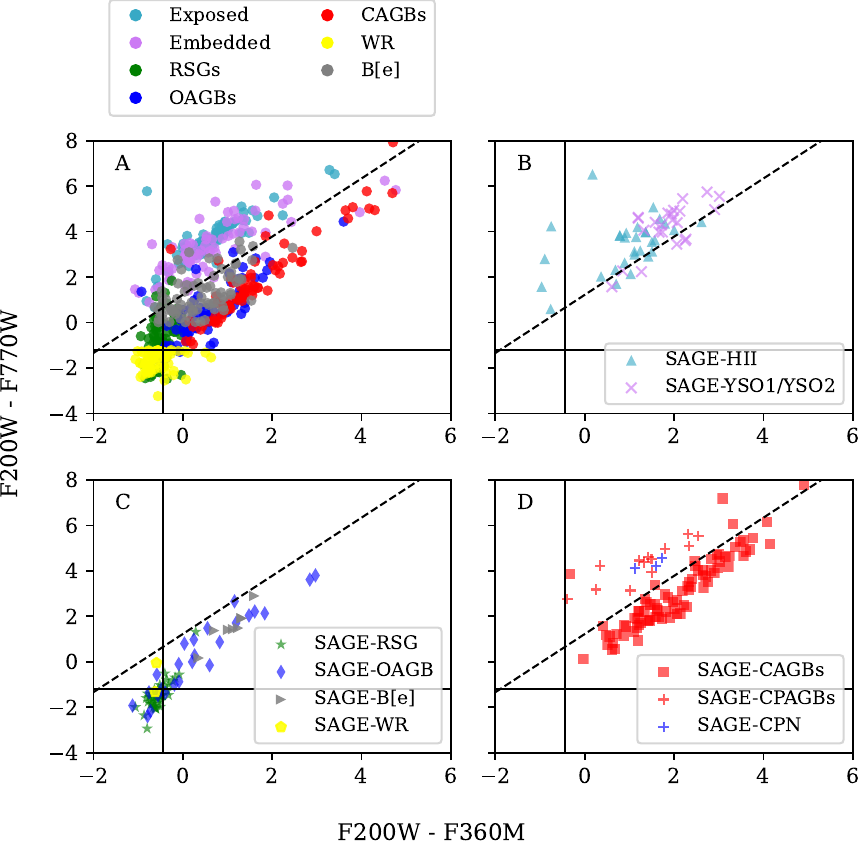}
\caption{Color–color diagram of classified sources using ratios from Table~\ref{tab:flux_ratios_sources}. Panel A: Sources in the PHANGS-JWST 10~$\mu$m catalog. To reduce crowding, we randomly sampled a maximum of 50 regions of each source type from all of the galaxies. Panels B, C, and D: Spectroscopically confirmed sources from the SAGE-LMC survey from \citep{Jones2017}\added{, where we use IRAC 1 band instead of F360M and only show sources that are detectable at 5.2 Mpc.}
\added{The dashed line serves to divide  ISM rich regions from other types of sources.}
The solid lines indicate the classification criteria used to separate RSGs and O-AGBs from C-AGBs. Overall, our color cuts would be able to distinguish the spectroscopically classified LMC sources.}
    \label{fig:class_figure}
\end{figure*}

We further compare the colors of our sources to SAGE-LMC sources in Figure~\ref{fig:class_figure}, finding good agreement for RSGs and O-AGBs using the color criteria F200W$-$F360M~$< -0.3$ and F200W$-$F770W~$< -1.8$ (see black solid lines in Figure~\ref{fig:class_figure}).  
We note the presence of a branch of SAGE-LMC O-AGBs sources beyond these criteria, which are likely either unclassified in our sample or blended with our C-AGBs.

\subsubsection{C-AGBs}
\label{sec:cagb_models}
C-AGBs are dusty stars characterized by $\mathrm{C/O} > 1$ in their atmospheres and can be divided into subgroups based on dust composition and mass-loss rates. Many relatively blue C-AGBs exhibit strong emission between 5–12~$\mu$m, with prominent $\mathrm{C_2H_2}$ absorption near the F770W PAH \added{\citep[See figure 6 in][]{Aringer}} band (7.7~$\mu$m) and SiC dust feature at 11.3~$\mu$m . In contrast, post-C-AGBs (C-PAGBs) show redder SEDs with more prominent molecular absorption features with a bright, broad 21~$\mu$m emission of uncertain origin \citep{Jones2017}. In this study, we do not distinguish between C-AGBs and C-PAGBs (especially given the C-PAGB color degeneracy with compact planetary nebulae discussed in Section \ref{sec:cpn}). This distinction requires spectroscopic data around key molecular absorption features. For blue C-AGBs, the mid-IR slope is negative \citep[see Figure 10 in][]{Jones2017}, while very red C-AGBs, also known as Very Red Objects (VROs), have redder SEDs with a positive mid-IR slope, indicating strong self-absorbed SiC features at 11.3~$\mu$m along \citep{Jones2017}.

In our modeling of C-AGB stars, we focus exclusively on the TP-AGB phase, selecting those with $C/O>1$. We note that the post-AGB phase is not included in our \texttt{PARSEC} models, limiting our ability to estimate the properties of these stars. These C-AGB stars span a mass range of 0.8–2 $M_{\odot}$ and have ages between 400~Myr and 1~Gyr. They also exhibit a similar luminosity to O-AGBs, but fainter in the range of $3.7< \log L/L_{\odot}<4.3$, with most detectable with $\log L_{\odot} > 4$. Figure \ref{fig:models} shows a C-AGB star with an age of approximately 700~Myr, illustrating the maximum C/O values (=2) achievable by the models.

%A notable feature occurs around 700~Myr, where stars can achieve the maximum C/O$=2$ through dredge up. These stars can exhibit mass-loss rates surpassing 250 $\times$ 10$^{-6}$ $M_{\odot}~\mathrm{yr}^{-1}$. Our models suggest that C-AGBs with a C/O ratio of 1 occur at an age of 400~Myr, increasing to a C/O ratio of 2 by 700~Myr. 

\begin{figure*}[!t]
    \centering
    \includegraphics[width=0.82\textwidth]{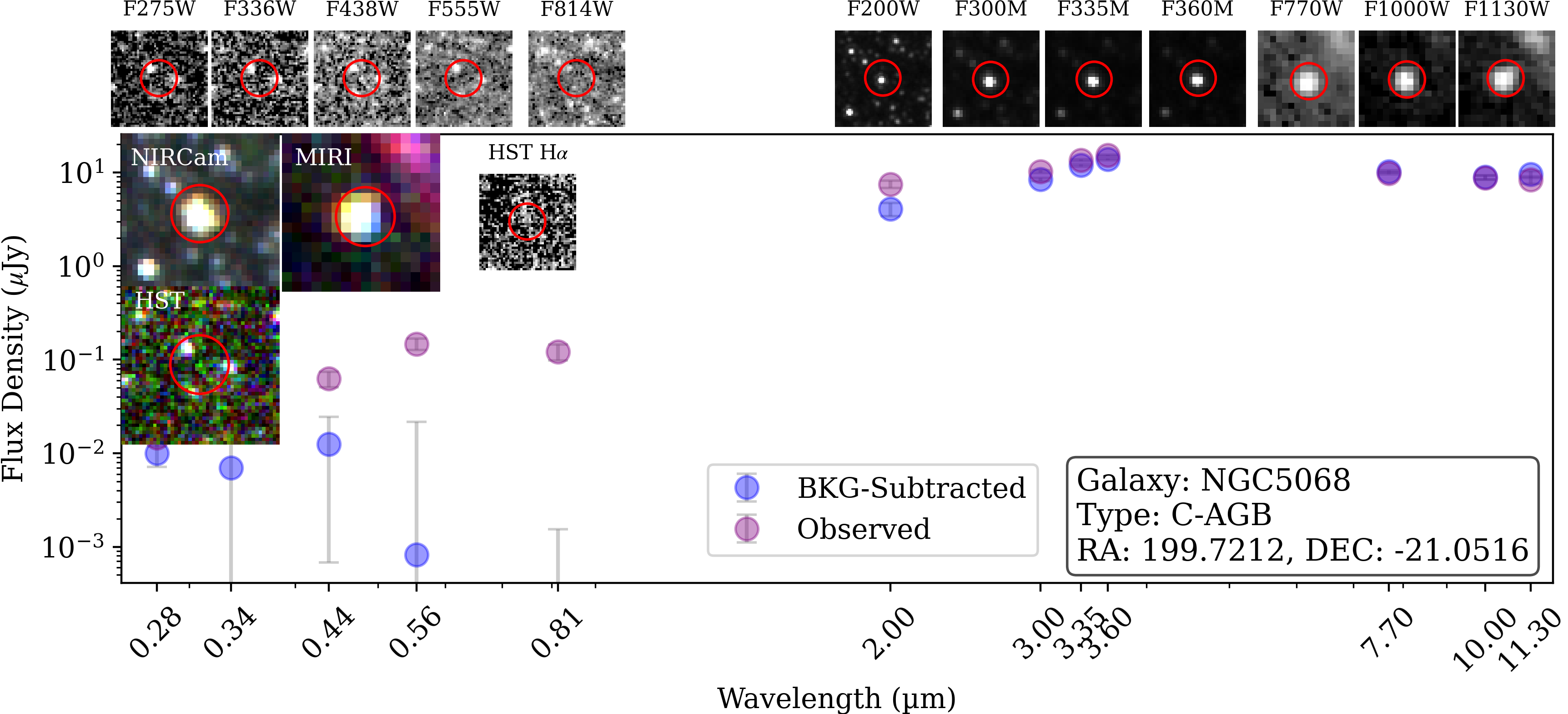}
    \caption{Similar to Figure \ref{fig:deep_emb_cluster}, but for a C-AGB stars.}
    \label{fig:cagb}
\end{figure*}

Figure \ref{fig:models_miri} shows color magnitude diagrams for both types of AGB stars with detection thresholds indicated with horizontal lines. \added{Our diverse models of C-AGBs, spanning different initial metallicities and dust chemistries, indicate that both stages are detectable in our nearest target, NGC~5068, with peak emission of}
\added{125~$\mu$Jy} and \added{75}~$\mu$Jy at 400~Myr and 1~Gyr, respectively, in the F1000W band, and they are also expected to appear bright in the F2100W band. At greater distances, such as in the NGC~1087, only lower-mass stars ($<$1 $M_{\odot}$) with ages $<$500~Myr \added{and sources with C/O $\approx 1$ are likely to be detected in the F1000W band, while being unlikely to be detected in the F2100W band.}

For C-AGBs, we adopt a lower stellar-to-PAH emission ratio threshold of $r_{2-7}< 3$ and $r_{2-3.6} < 1.5$. Our model tracks suggest that $r_{3}>0.9$ and $r_{3.6}<1.1$. \added{We also find that $r_{3-3.6}$ can reach values of up to $\sim 1$; however, the brightest sources in the $F1000W$ band exhibit ratios below 0.25, likely due to absorption by HCN and $\mathrm{C_2H_2}$, which suppresses the flux in the $F300M$ band.}

For the MIRI flux ratios, we adopt \added{$r_{7}<1.3$} to also account for the 7.5~$\mu$m molecular absorption feature. Motivated by the $r_{11}$ ratios classified as C-AGBs and C-PAGBs in the LMC from \cite{Jones2017}, which have a median ratio of 2, we restrict this ratio to vary between 0.5 and 2.5.

\added{We further use the $r_{3\text{–}7}$ versus $r_{3.6\text{–}11}$ diagram to better establish a separation between O-AGB and C-AGB stars, and show the detectable models at our nearest target distance of 5.2~Mpc in the F1000W band in Figure~\ref{fig:models_flux_ratio}. We additionally only include LMC sources from \citet{Groenewegen2018}, as they better match the lower end of our metallicity range. C-AGB stars are shown as red hexagonal bins. We find that the relation $r_{3\text{–}7} < 0.65\left(r_{3.6\text{–}11}\right)^{1.2}$ includes all C-AGB models, along with a small number of older ($\gtrsim$Gyr) O-AGB models. In contrast, the majority of O-AGB models lie above this boundary, as do most of the detectable LMC M-type stars from \citet{Groenewegen2018}.}

We identified \added{384} and \added{57} C-AGBs in the 10~$\mu$m and 21~$\mu$m catalogs, respectively. \added{We note that 12 out of the 57 C-AGBs detected in the 21~$\mu$m catalog ($\sim$20\%) have counterparts in the 10~$\mu$m catalog.} We show the SED of one such source in Figure \ref{fig:cagb} from NGC 5068. 

%C-AGBs detected at 21~$\mu$m are the brightest stellar population, rivaling embedded clusters.  removed. 

% In NGC~5068, we identify more than 400 C-AGBs, with a median absolute magnitude of $M_{{V}} = -0.5$~mag when detected, indicating that most of them are very faint and likely not detectable at in the optical F555W band.

\subsubsection{CPN/C-PAGB}
\label{sec:cpn}

\begin{figure*}[!t]
    \centering
    \includegraphics[width=0.8\textwidth]{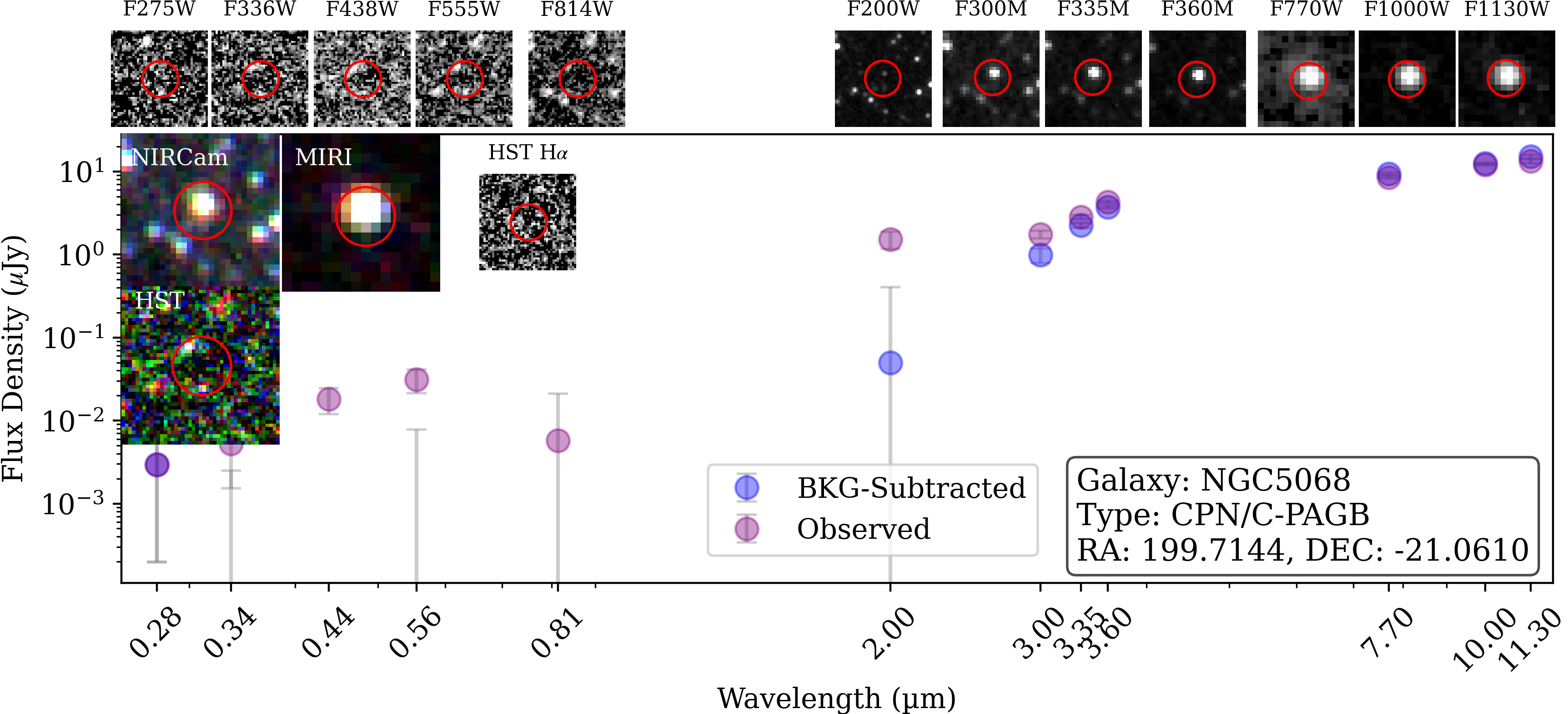}
    \caption{Similar to Figure \ref{fig:deep_emb_cluster}, but for a CPN/C-PAGB \added{candidate}. \added{This source is also flagged as a C-AGB star.}}
    \label{fig:cpn}
\end{figure*}

Carbon-rich Planetary Nebulae (CPNe) show prominent PAH features, and in some cases from the SAGE-LMC sample, they exhibit an additional SiC feature. These objects often display a positive mid-IR spectral slope beyond 5~$\mu$m \citep{Jones2017}. It is likely that we will detect planetary nebulae in our nearest galaxies, such as NGC5068 and IC5332. Figure \ref{fig:models_miri} shows the absolute magnitude of F1000W, along with the NIRCam and MIRI colors of CPNe, relative to our detection limits. The nearby targets (i.e. NGC 5068, IC5332) also have low metallicities ($12 + \log_{10}(\mathrm{O/H}) \sim 8.3 $) comparable to the LMC. The low-metallicity conditions in such environments promote the formation of CPNe \citep[and references therein]{Jones2017}.

To classify sources as planetary nebulae, we primarily rely on photometric colors of the SAGE-LMC compact CPNe sources from \citet{Jones2017}, which are more likely to be bright at infrared wavelengths. Based on SAGE-LMC sources, these objects show strong stellar emission, resulting in the criterion $r_{2.7} < 0.1$. Due to the presence of PAHs, they also exhibit \added{$r_{7} > 0.5$, $r_{11} > 1$, and $r_{7-11}>0.2$}.  We also find that $F_{2~\mu\mathrm{m}} / F_{\text{IRAC1}} < 0.5$ for most of these sources. Consequently, we apply the combined color criterion $\mathrm{F200W}/(\mathrm{F335M} + \mathrm{F360M}) < 0.5$. To avoid contamination from clusters exhibiting strong PAH emission in F335M, we adopt the criterion $r_{3.6}<1$, consistent with our selection for AGB stars. For the 21~$\mu$m catalog, we use the color ratio \added{$r_{21-11} > 1$}.  We do not impose any constraint on H$\alpha$ brightness, as some candidate CPNs may exhibit recombination lines.

Most of the selection criteria for CPNe sources align with those used to classify C-PAGB stars, with the exception of the $r_7$ ratio. Some bright C-PAGB sources can exhibit $r_7 > 1$, making it difficult to distinguish CPNs from C-PAGBs with the current NIRCam and MIRI filter sets. The F1000W/F2100W ratio is also similar between CPNe and C-PAGB sources, generally exceeding values of 1 or 2 \citep{Jones2018}. We note that the color of CPNe sources in the F200W$-$F770W versus F200W$-$\added{IRAC1} diagram largely overlap with those of CPAGBs (see Panel D). Given these overlaps, we treat CPNe and C-PAGB candidates as a blended category.

\begin{figure*}[!t]
    \centering
    \includegraphics[width=0.82\textwidth]{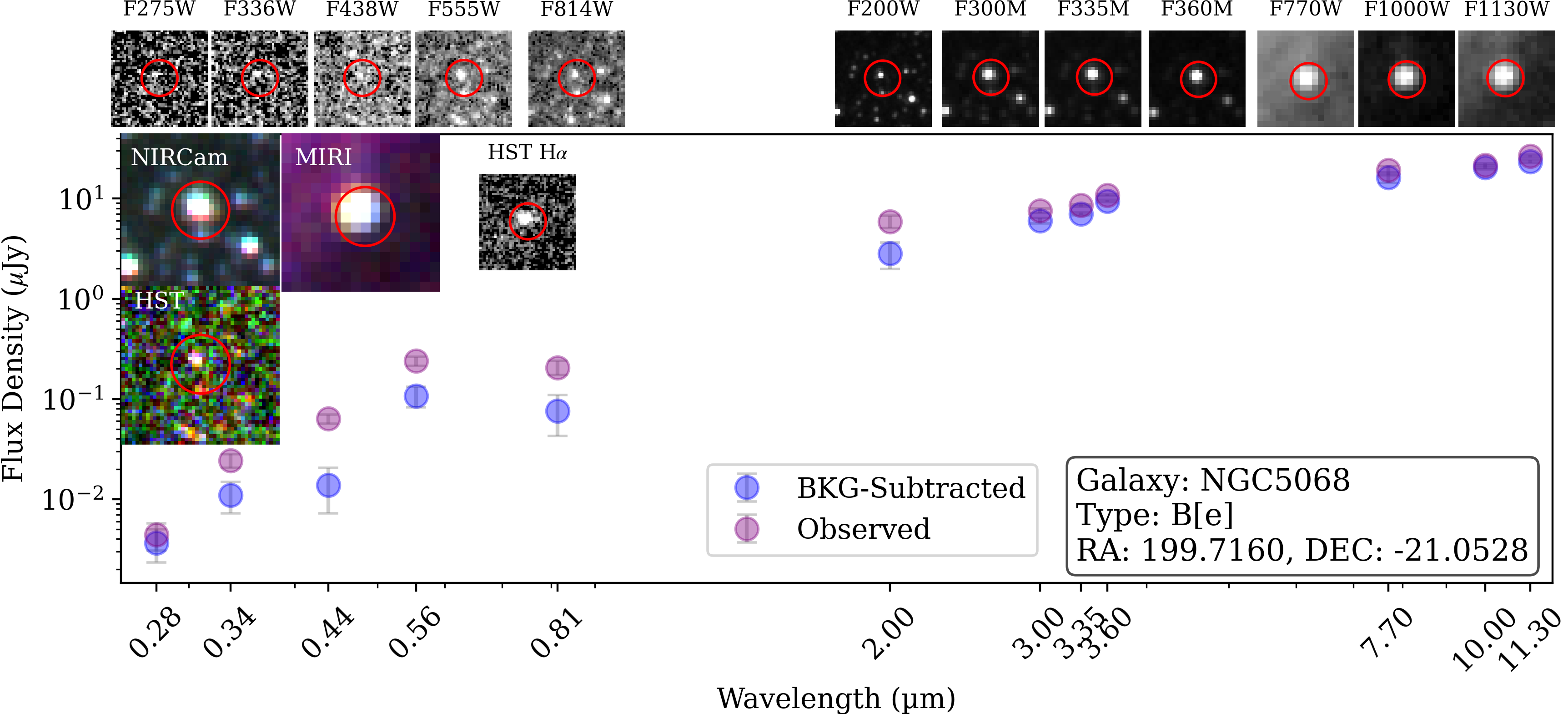}
    \caption{
    Similar to Figure \ref{fig:deep_emb_cluster}, but for a B[e] star candidate with prominent H$\alpha$ emission.}
    \label{fig:be}
\end{figure*}

\begin{figure*}[!t]
    \centering
    \includegraphics[width=0.82\textwidth]{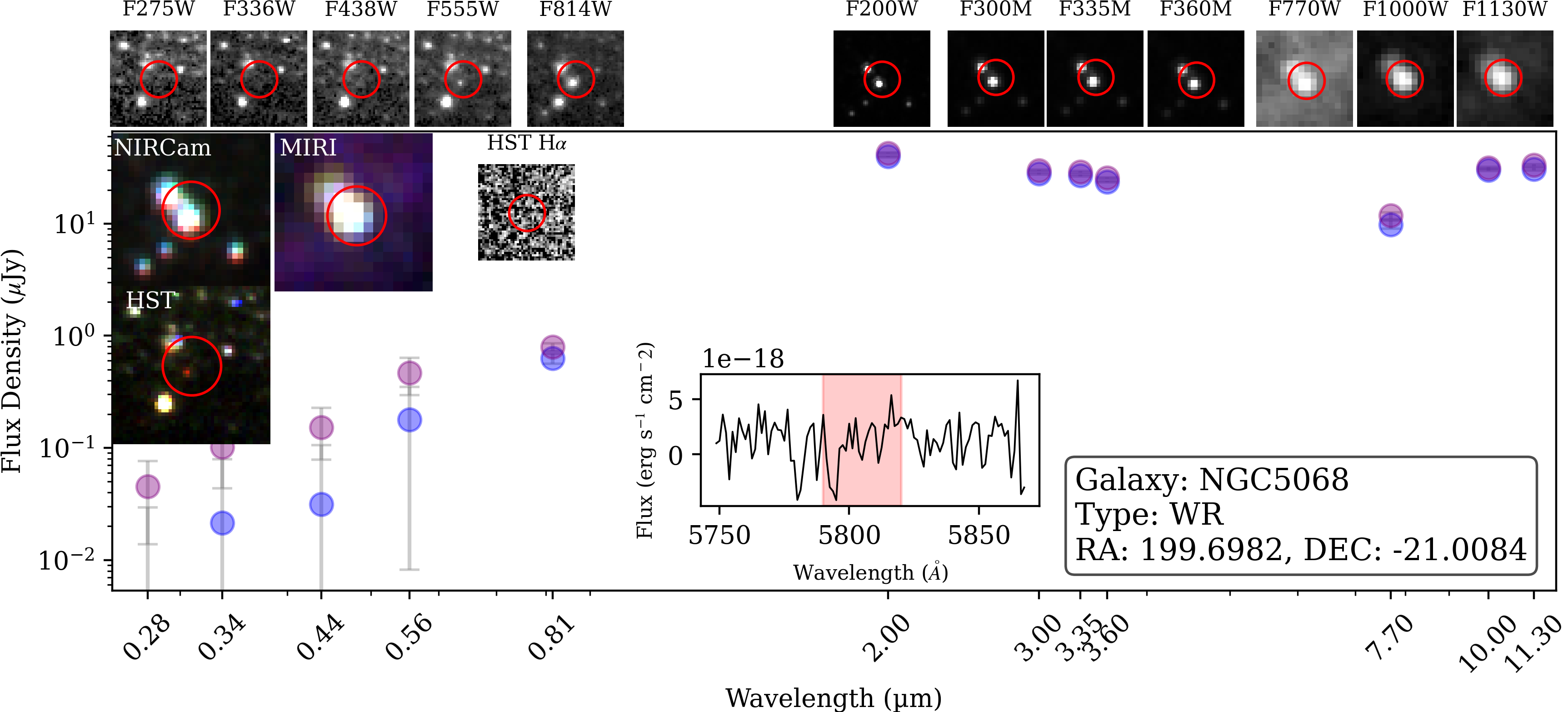}
    \caption{Similar to Figure \ref{fig:deep_emb_cluster}, but for a WR star. We also include the observed MUSE integrated spectrum from the central one-arcsecond region, zoomed in on the potential RB feature near $\sim$5808 {\AA}, which is highlighted in red.}
    \label{fig:wr}
\end{figure*}

We identified \added{1455} and \added{355} candidate CPNe/C-PAGB sources in the 10~$\mu$m and 21~$\mu$m catalogs, respectively, with 103 sources from the 21~$\mu$m catalog also present in the 10~$\mu$m catalog. Many of the common sources show PAH features at 7.7 and 11.3~$\mu$m, along with a rising slope in both the near- and mid-infrared, a characteristic also commonly seen in C-PAGBs. Figure~\ref{fig:cpn} presents the SED of a representative CPNe / C-PAGB candidate.

\subsubsection{B[e] Stars}
We identify B[e] stars using ratio cuts determined empirically. \citet{Rufflesmc} presented IRS spectroscopy of B[e] stars in the SMC, highlighting a broad 10~$\mu$m feature.  For B[e] stars, we apply a selection criterion requiring the H$\alpha$ flux to exceed five times the detection limit, as derived from Table~\ref{tab:comp}. Based on SAGE-LMC B[e] sources, we adopt the following color constraints: $r_{2.7} < 1$, $r_7 < 1.2$, $r_{11} < 1.5$, and $r_{7-11} < 1.5$. Additionally, the observed Ks/IRAC1 ratio is below 1, motivating the adoption of a color criterion of F200W/(F335M+F360M)~$<1$. For the 21~$\mu$m catalog, we further require $r_{21-10} < 3$. 

\begin{figure*}[!t]
    \centering
    \includegraphics[width=0.82\textwidth]{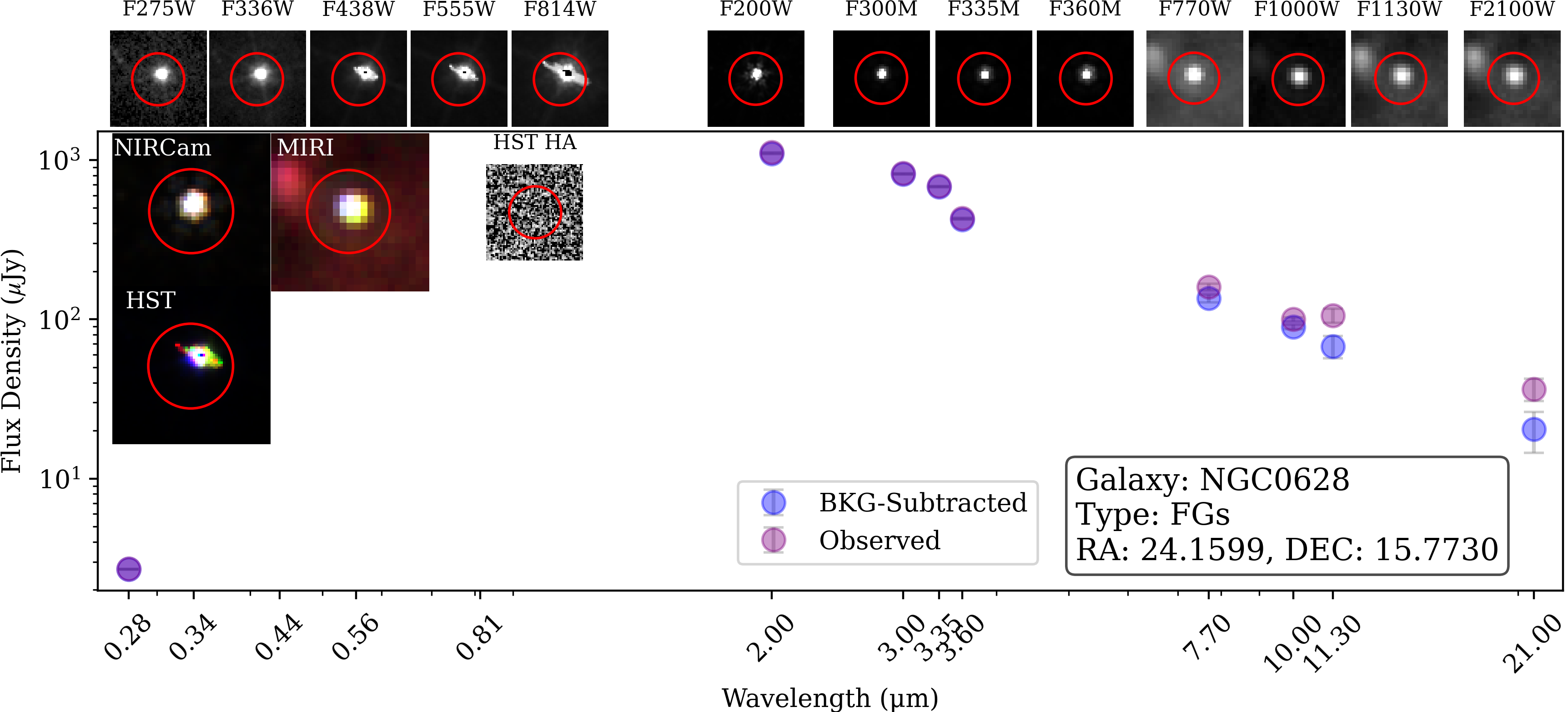}
    \caption{Similar to Figure \ref{fig:deep_emb_cluster}, but for a FG star from 21~$\mu$m catalog}
    \label{fig:fg}
\end{figure*}

We identified 98 and 11 B[e] star candidates in the 10~$\mu$m and 21~$\mu$m catalogs, respectively, with 5 sources from the 21~$\mu$m catalog also present in the 10~$\mu$m catalog. Figure~\ref{fig:be} shows an example of a B[e] star candidate. We note that only a few of these common sources appear point-like and show associated H$\alpha$ emission. We also note that for more distant targets, such as NGC~3627 ($D = 11.3$~Mpc) and beyond, HST H$\alpha$ measurements may be contaminated by emission from sources outside the aperture, potentially affecting the reliability of the flux measurements.

\subsubsection{WR Stars}

WR stars are expected to be bright in the mid-infrared continuum around 10~$\mu$m and, in some cases, show prominent PAH features \citep{Marchenko}, as well as strong stellar emission at 2~$\mu$m \citep{Cohen}, but are typically faint or undetected at 24~$\mu$m \citep{Bonanos}. We use the observed properties of SAGE-LMC WR stars as a reference for identifying WR candidates, adopting selection criteria of $r_{2.7} > 3$  $r_7 < 1.5$,\added{ r$_{11}>1$, and r$_{7-11}<0.6$}. We also use $r_{21-10} > 2$ for our 21 $\mu$m catalog. We did not put any constraint on H$\alpha$ flux as the emission we observe may originate from surrounding nebular regions rather than the WR stars themselves \citep{wr_muse}.

\begin{figure*}[!t]
    \centering
    \includegraphics[width=0.82\textwidth]{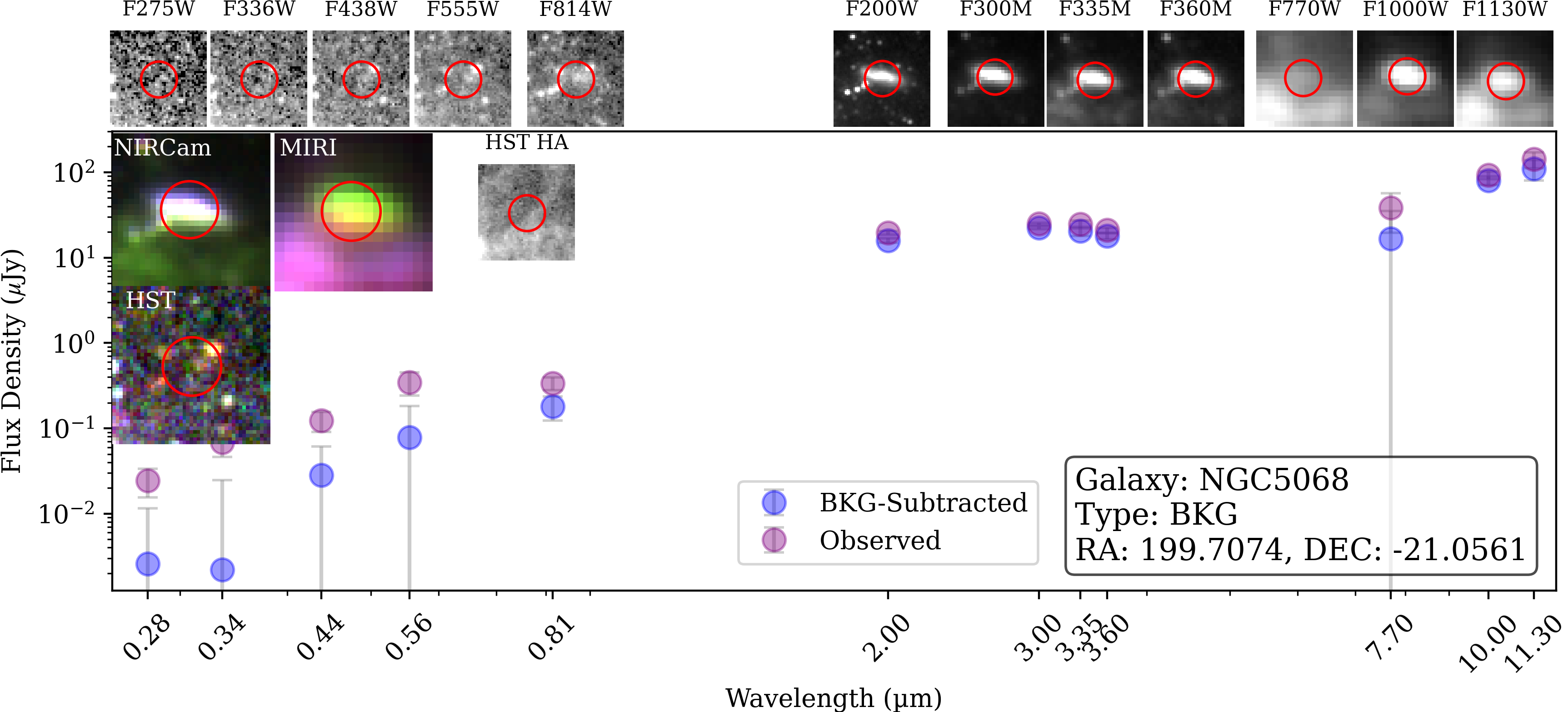}
    \caption{Similar to Figure \ref{fig:deep_emb_cluster}, but for a background galaxy.}
    \label{fig:bkg}
\end{figure*}

\begin{figure*}[!t]
    \centering
\includegraphics[width=\textwidth]{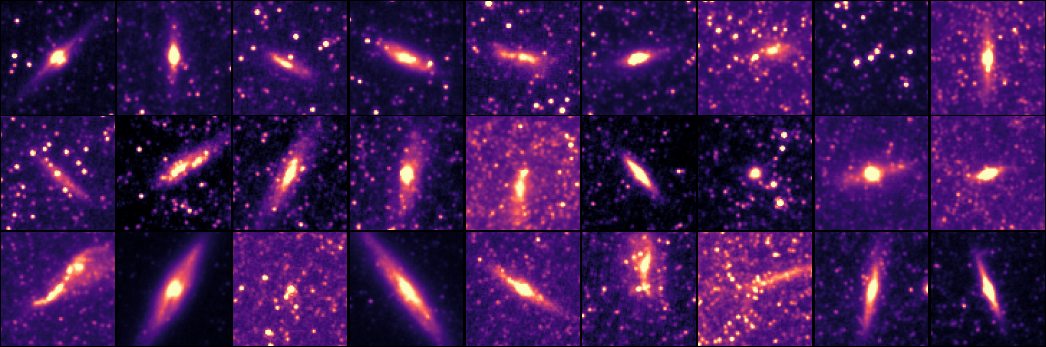}
\caption{Cutouts of BKGs observed in the F200W band, detected in F1000W peak catalog. Most of these objects have semi-major to semi-minor axis ratio higher than 2.5 (see Section \ref{sec:bkg}).}
\label{fig:bkgs}
\end{figure*}

Using the indicated infrared flux ratio cuts, we identify stars with strong stellar emission in the F200W band and a generally declining infrared spectral slope---features characteristic of optically confirmed WR stars in the LMC \citep{WR}, and consistent with model predictions from \citet[][see their Figure 15]{Bonanos}. 

In total, we identify \added{293} and \added{18} WR star candidates from our 10~$\mu$m and 21~$\mu$m catalogs, respectively. Among these, only three sources are matched between the two catalogs. We note that some of the 21\,$\mu$m-selected candidates exhibit extended structure in the NIRCam bands, suggestive of background galaxies. 

The most definitive diagnostic for WR classification comes from optical spectroscopy, which can detect blends of key ionic lines such as \ion{He}{2} and \ion{C}{4}. The former is among the lines that contribute prominently to the Blue Bump (BB) at $\sim$4686~\AA, while the latter contributes to the formation of the Red Bump (RB) at $\sim$5808~\AA\ \citep{wr_muse}. While we do have optical spectroscopy of our targets from MUSE, our spectral coverage is not well suited for WR identification. Ten of our targets (e.g.\ NGC~5068 and NGC~0628) were observed with the non-AO nominal (non-extended) spectral mode of MUSE, which does not cover the BB but only covers the RB. 
For the remaining nine targets observed under AO nominal mode, the BB window (4600–4700~{\AA}) is only partially covered (up to 50\%) at the spectral edge, while the RB window (5758–5858~{\AA}) is covered by at most 25\% due to the AO laser gap. Therefore, in this study, we primarily rely on infrared classification to identify WR candidates, and subsequently examine the MUSE spectroscopy for any evidence of the BB or RB features.

We highlight one 10~$\mu$m WR candidate for which we also include the observed MUSE spectrum (from \citealt{phangs-muse}), zoomed in on the $\sim$5808~\AA\ region where the RB, potentially containing C~IV emission, may be present (Figure \ref{fig:wr}). Some of our candidates exhibit a faint feature at this wavelength, although the RB in external galaxies is generally fainter than the BB and less reliable for WR classification \citep{wr_muse}. A more detailed analysis of WR features using MUSE spectroscopy and spectral modeling is underway and will be presented in Fu-Heng Liang et al. (in prep).

\subsection{Foreground Stars}
\label{sec:fg}

We cross-matched both catalogs with the \textit{Gaia} Data Release (DR3) survey to detect potential contamination from foreground stars (FGs) \citep{GaiaMission}. We applied the criterion $\texttt{parallax\_over\_error} > 3$ to identify foreground stars and performed a cross-match with our 10 and 21 $\mu$m catalogs using a 2$\arcsec$ separation to account for proper motion. 

The median flux of the foreground objects is about 22~$\mu$Jy in the 10~$\mu$m catalog and 56~$\mu$Jy in the 21~$\mu$m catalog.  Examining the different flux ratios of FGs, we find the following median values: $ r_{3} = 1.0 $, $ r_{3.6} = 0.9 $, $ r_{7} = 1.4 $, and $ r_{11} = 0.9 $. These results indicate that the SEDs of most FGs are characterized by the strongest emission at $2\,\mu\text{m}$ and an SED that generally follows a thermal spectrum consistent with a cool star in the Milky Way. Similar stars have also been identified among LMC stars and classified as both hot, massive stars and low-mass red giant stars \citep{Jones2017}. For the classification of these sources, we focus on cases where $ r_{2-3.6} > 1 $ and $ r_{2-7} > 1 $, suggesting a negative slope in the MIRI bands. This behavior is further defined by $ r_{7} > 1 $ and $ r_{11} < 1 $.

In the catalog, we label sources as foreground stars if they (1) cross match to a Gaia source with significant parallax or (2) match the typical colors of a foreground Milky Way star \added{, as mentioned above.} We thus identified 250 and 78 foreground objects in the 10~$\mu$m and 21~$\mu$m catalogs, respectively, accounting for $<1\%$ of the total sources in both catalogs. In the 10~$\mu$m catalog, most galaxies contain fewer than 10 foreground sources, with the highest number (32) found in NGC~2835.  We show one of the FGs from the 21$\mu$m catalog in Figure \ref{fig:fg}.

% \textbf{Based on defined criteria, we found 106 number of FGs in 10$\mu$m catalog and 15 in 21$\mu$m catalog. We show one of the FGs from the 21$\mu$m catalog in Figure \ref{fig:fg}.}

\subsection{Background Galaxies}
\label{sec:bkg}

Background galaxies are more challenging to classify as some of them resemble single stars in the MIRI bands and have red colors associated with dusty stars. We can usually discriminate these sources dusty, young clusters because their redshift moves the PAH emission features out of the bands where they are found at $z=0$. Many of these galaxies exhibit a nearly flat NIRCam slope, based on the visual inspection. We adopt the following ratios to classify background objects: $r_{2-3.6}<1.2$, $r_{3}<1.2$ and $r_{3.6}<1.2$. We present an example of the SED of a background galaxy in Figure \ref{fig:bkg}

After experimentation, we determined that the best route for confirming background galaxies after color selection was using the morphology of these objects in the F200W images. Figure \ref{fig:bkgs} shows example images associated with background galaxy in $2.5\arcsec$ cutouts at the location of the F1000W peaks. 
% Using the same source-finding algorithm described in Section \ref{sec:source_finder},
% The intensity-weighted centroid was used to construct the second moment matrix, with its eigenvalues defining the lengths of the axes. The orientation of the semi-major axis was determined as the angle of the principal axis relative to the image coordinate system. 
For this analysis, we apply our \texttt{AstroDendo}-based source finder to F200W images, which have a FWHM of 0.067\arcsec\, and considering a resolution element with 7 pixels, with a \texttt{minimum-beam=3}.  We select elongated F200W sources with a semi-major to semi-minor axis ratio greater than 1.5. The semi-major and semi-minor axes were calculated from the second moments of the intensity distributions within the dendrogram leaves. This approach identifies many extended, elongated background galaxies with structures clearly visible in F200W, as well as a few compact, bright sources with diffraction spikes, which could also be foreground stars. In total we find 321 and 980 BKGs in 10 and 21 $\mu$m catalogs, respectively.

% \textbf{We note that the SEDs of sources presented in Figures \ref{fig:rsg} through \ref{fig:bkg} may be affected by minor alignment issues, typically $\sim$0.1\arcsec, where the red circles in the center of cutouts are not centered on the sources for filters other than those used for detection (F1000W or F2100W).}

\subsubsection{Performance of Evolved Star Classification}
\label{sec:valiation}

\begin{figure}[!t]
    \centering
\includegraphics[width=1\linewidth]{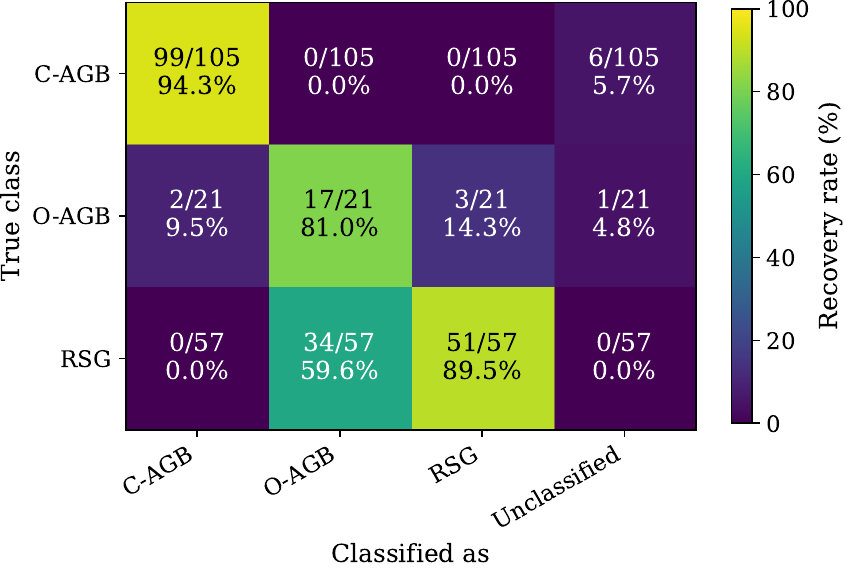}
\caption{\added{Confusion matrix for O-AGB, C-AGB, and RSG stars, showing the fraction of LMC sources detectable in the F1000W band from \citet{Groenewegen2018}, scaled to a distance of 5.2~Mpc, that satisfy each set of selection criteria. Rows correspond to the reference (true) stellar classes, while columns indicate whether sources meet the individual classification criteria listed in Table~\ref{tab:flux_ratios_sources}. The Unclassified column represents sources that do not satisfy any of the selection criteria. Because the criteria are evaluated independently, the predicted classes are not mutually exclusive, and row fractions do not necessarily sum to 100\%.}}
\label{fig:recovery}
\end{figure}

\added{We reclassify LMC sources from \citet{Groenewegen2018} to assess the performance of our classification criteria for the 10~$\mu$m catalog described in Table~\ref{tab:flux_ratios_sources}. We first select only those LMC sources that would be detectable at the distance of our nearest target galaxy (5.2~Mpc) in the F1000W band, corresponding to a scaled flux limit of $\sim$5~$\mu$Jy. We then reclassify these sources using our color cuts and evaluate the recovery fraction for each stellar class, summarizing the results in the confusion matrix shown in Figure \ref{fig:recovery}. Because the classification criteria are applied independently and are not mutually exclusive, individual sources may satisfy multiple criteria. We find that our method recovers more than 80\% of the detectable LMC sources from \citet{Groenewegen2018}. We additionally examine classification confusion by quantifying the fraction of sources from other stellar classes that contaminate each set of selection criteria. We find that our classification of C-AGB stars is robust with only 2 of O-AGB stars being incorrectly classified as C-AGB stars. A similar level of contamination is found for the RSG classification, with less than 15\% of O-AGB and C-AGB sources misclassified as RSGs. In contrast, the O-AGB classification shows a higher level of contamination, primarily from RSGs rather than C-AGBs. This reflects an intrinsic limitation of the currently available photometric bands in cleanly separating O-AGB stars from RSGs. 

Resolving this degeneracy will require additional spectroscopic or narrowband observations. We do not present confusion matrices for other stellar candidates, such as the compact CPN/C-PAGB, WR, or B[e] stars discussed later, as these classes are not included in our stellar model grids and are identified solely based on SAGE–LMC classifications from \citet{Jones2017}. In addition, these sources lack synthetic photometry in the F300M, F335M, and F360M bands and their classification are restricted to low-metallicity environments. While these results indicate a reasonably good distinction between C-AGB and O-AGB/RSG stars, our confidence is built on color cuts motivated, in part, from a catalog derived from the LMC and then evaluated using the same data. Verifying the utility of these color cuts will require further JWST observations of Local Group galaxies to establish the infrared colors in higher metallicity systems than the LMC.
}

\subsection{Summary of Source Classification}
\label{subsec:source_pro}
\added{Our final classification criteria are summarized in Table~\ref{tab:flux_ratios_sources}.} We have summarized our catalogs in Figure \ref{fig:class_figure}, which presents a color-color diagram ($ \mathrm{F200W - F770W} $ vs. $ \mathrm{F200W - F360M} $) of our 10~$\mu$m catalog. The figure illustrates our source classifications for a subset of objects in our catalogs and then compares these distributions to the colors for sources from the LMC.  RSGs, O-AGBs, and WRs are mainly found below $\mathrm{F200W–F770W = -1.8}$ and $\mathrm{F200W–F360M} = -0.3$, while other galaxy sources lie above these thresholds. 
\added{The dashed line, adapted from \citet{Sargent}, shows an empirical envelope for AGB stars and is included for reference.} Most C-AGBs, highlighted in red, fall below this line with $ \mathrm{F200W - F360M} > 0.3 $ and $ \mathrm{F200W - F770W} > -1.8 $. In contrast, both embedded and exposed are located above this threshold and are marked with distinct colors. Comparing our sources in PHANGS galaxies with SAGE sources shows good overall agreement.

\begin{figure*}[t!]
    \centering
    \includegraphics[width=0.98\textwidth]{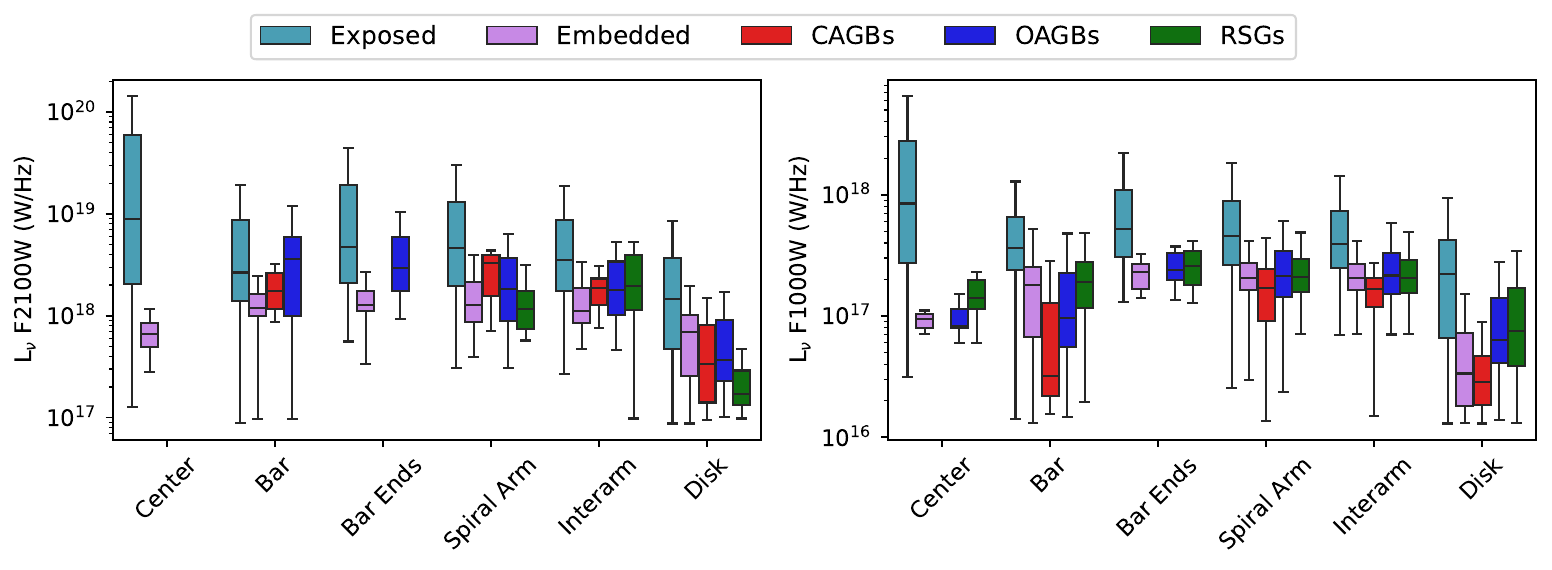}
    \caption{The luminosity distributions of different source types across galactic environments are shown for the 21~$\mu$m catalog at F2100W (left) and the 10~$\mu$m catalog at F1000W (right).}
    \label{fig:all_catalog}
\end{figure*}

\begin{figure*}[!t]
    \centering
\includegraphics[width=1\linewidth]{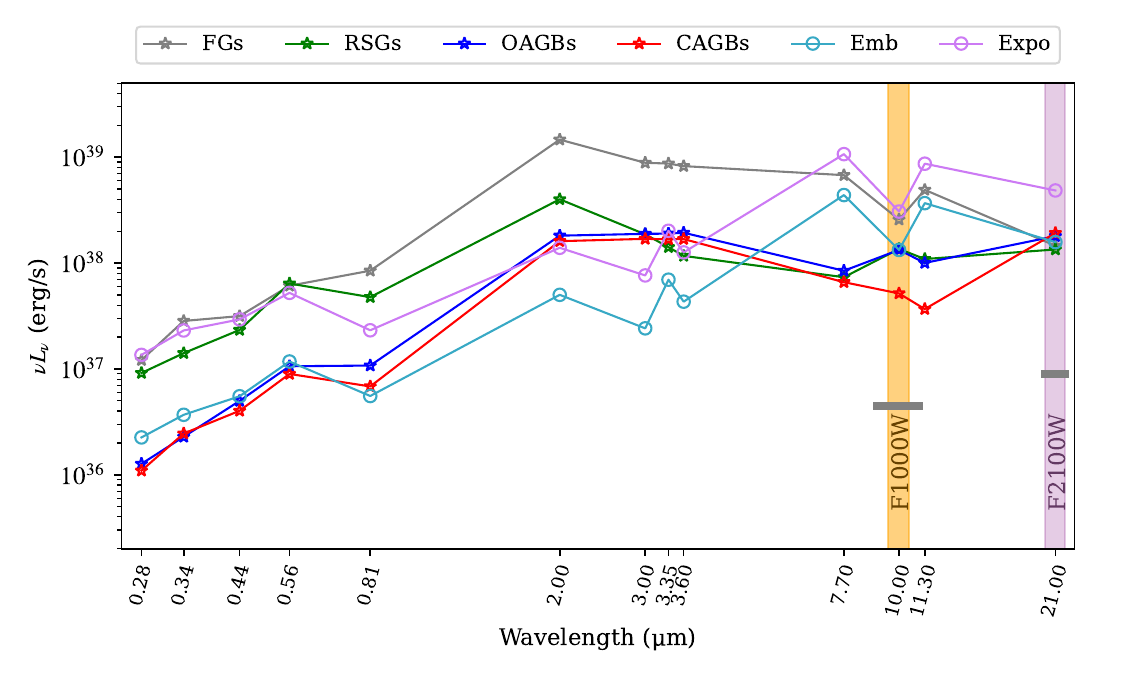}
   \caption{Average SEDs of 21~$\mu$m sources across all galaxies. The FGs presented here are solely selected through cross-matching with Gaia sources, as described in Section \ref{sec:fg}. The detection limits of NGC~5068 in F1000W and F2100W bands are shown as gray lines, as listed in Table \ref{tab:comp}.}
    \label{fig:mean_sed}
\end{figure*}

Figure \ref{fig:mean_sed} presents the average SED of 21~$\mu$m sources across all galaxies. In addition to FGs, RSGs and O-AGBs exhibit the brightest stellar emission (i.e. 2~$\mu$m to 3.6~$\mu$m), with a very similar SED from optical to near-infrared. As also noted in Figure~\ref{fig:mean_sed}, exposed clusters dominate the brightest peaks at 21~$\mu$m.

 % Following them, exposed clusters have the brightest SEDs, comparable to C-AGBs, and are followed by embedded clusters. removed

\begin{table}
\centering
\begin{tabular}{llll}
\toprule
Class & (Matched/Total) & F1000W& F2100W  \\
 & \%  &  ($\mu$Jy) & ($\mu$Jy) \\
\midrule
Expo & 85 (5371/6355) & 50–744 & 8–58 \\
Expo True$^{1}$ & 76 (3564/4714) & 53–848 & 9–66 \\
Emb & 17 (244/1411) & 29–90 & 6–15 \\
Emb True$^{1}$ & 21 (203/963) & 29–88 & 6–15 \\
RSG & 56 (27/48) & 29–99 & 13–67 \\
O-AGB & 44 (76/171) & 36–155 & 8–79 \\
C-AGB & 21 (12/57) & 40–133 & 5–24 \\
CPN & 29 (103/355) & 33–202 & 6–31 \\
WR & 17 (3/18) & 172–16276 & 32–2397 \\
B[e] & 45 (5/11) & 52–7382 & 48–3178 \\
bkg & 16 (160/980) & 38–156 & 5–45 \\
\bottomrule
\end{tabular}
\caption{Cross-matching results between F1000W and F2100W catalogs. Each row shows a stellar class with its percentage of matched sources (matched/total), flux range (16th–84th percentile) for matched F1000W sources, and flux range for matched F2100W sources.\\
$^{1}$HST H$\alpha$ SNR $>5$.\\
$^{2}$HST H$\alpha$ SNR $<5$.
\label{tab:cross_match_flux_ranges}}
\end{table}

In Table \ref{tab:cross_match_flux_ranges}, we present the population of sources that are cross-matched between the 10 and 21 $\mu$m catalogs. We report the number of sources in the 21$\mu$m catalog (i.e., total) and the number that have matched to the 10$\mu$m catalog, along with the corresponding percentage. In this table, we specifically separate ``True'' sources that have clear HST H$\alpha$ emission associated with each source (for exposed sources) or a non-detection (for embedded sources). We also report the typical flux densities associated with each category of sources.

\section{Results}
\label{sec:res}

Here, we present several different perspectives on the combined catalogs across these 19 galaxies. These figures demonstrate the utility of the catalogs and illustrate how this analysis compares to \citet{hassani23}.

\begin{figure*}[!t]
    \centering
\includegraphics[width=0.78\textwidth]{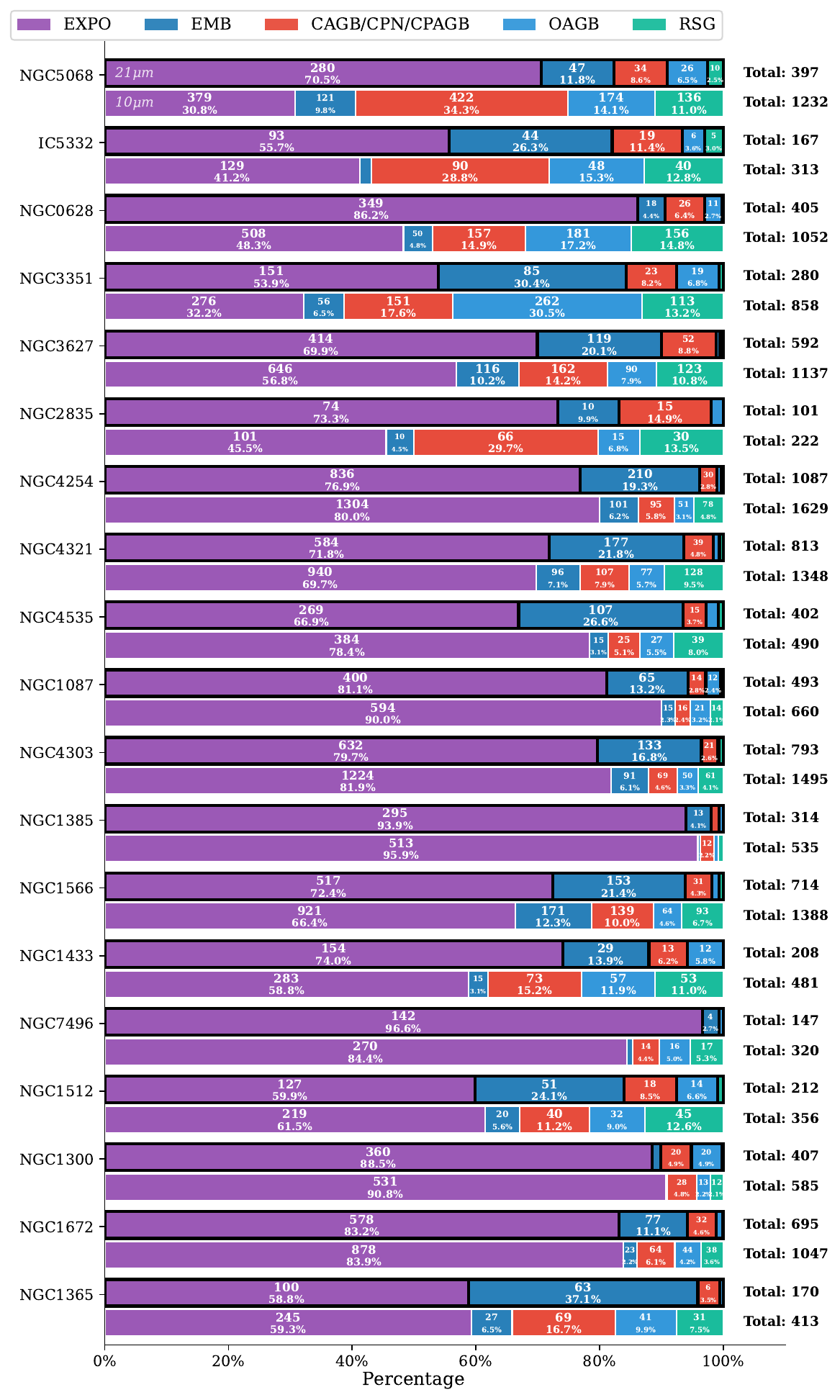}
\caption{Stellar population distributions across galaxies, ordered by distance (closest at top) for the 21~$\mu$m (top) and 10~$\mu$m catalogs (bottom). Data are presented as normalized percentages of the total stellar population for each galaxy (while neglecting small percentages of WRs, B[e], BKGs, and FGs), with absolute counts and percentages labeled within each segment.
}
\label{fig:stats}
\end{figure*}

Figure~\ref{fig:stats} shows the counts and percentages of exposed clusters, embedded O-AGBs, RSGs, and the combined C-AGB/CPNe class across 19 galaxies, sorted by distance. In most galaxies, exposed clusters are the most abundant population, comprising \added{$\sim$} 50–95\% of the total in different galaxies. Notably \added{in our 21$\mu$m catalog}, IC~5332, NGC~1365, and NGC~3351 show lower fractions of exposed clusters (below 60\%), whereas NGC~ 1385 and NGC~7496 exhibit about 90\% or more. Galaxies such as NGC~3351, NGC~3627, NGC~4321, NGC~4535, NGC~1566, and NGC~1365 have more than 20\% embedded clusters in the 21~$\mu$m catalog. 

% In the 10~$\mu$m catalog, NGC~3627 and NGC~4535 show more than 20\% of their total populations as embedded clusters.  REMOVED

This figure also illustrates how our detection limits for stellar populations in both catalogs change with distance to the target. Specifically, in the 10~$\mu$m catalog, we detect \added{about} 30\% of the population classified as C-AGBs/CPNs/CP-AGBs out to NGC~2835 ($D=12.22$~Mpc). \added{These stars remain detectable even in our most distant galaxy, NGC~1365. RSGs are detected in most of our targets, with total fractions below 10-15\%.}

% #\%. For more distant galaxies, the corresponding fractions are  $\lesssim 1\%$ for most targets. removed

\subsection{Source Environment}

\begin{table*}[t]
\centering
\label{tab:linear_fits}
\begin{tabular}{lccccccccccc}
\hline
Environment & $\langle\log(L_{\mathrm{H\alpha,corr}} / \nu L_{\nu,21\mu\mathrm{m}})\rangle$  & $\alpha$ & $\beta$ & $R^2$ & $\rho_s$ & $\sigma$ & $X_{\mathrm{norm}}$ & $Y_{\mathrm{norm}}$ & $N$ \\
&  &  &  &  &  &  & ($10^{38}~\mathrm{erg~s^{-1}}$) &($10^{38}~\mathrm{erg~s^{-1}}$)  &  \\
\hline
Center & 0.10 & $0.83 \pm 0.02$ & $-0.16 \pm 0.02$ & 0.88 & 0.95 & 0.31 & $12.21$ & $1.88$ & 231 \\
Bar & 0.16 & $0.91 \pm 0.02$ & $-0.13 \pm 0.01$ & 0.76 & 0.86 & 0.32 & $3.01$ & $0.61$ & 640 \\
Bar Ends & 0.13 & $0.70 \pm 0.04$ & $-0.10 \pm 0.03$ & 0.68 & 0.87 & 0.33 & $4.87$ & $0.87$ & 178 \\
Spiral Arm & 0.14 & $0.78 \pm 0.01$ & $-0.07 \pm 0.01$ & 0.76 & 0.88 & 0.26 & $5.24$ & $0.88$ & 2490 \\
Interarm & 0.15 & $0.80 \pm 0.01$ & $-0.08 \pm 0.01$ & 0.72 & 0.85 & 0.28 & $3.82$ & $0.66$ & 1755 \\
Disk & 0.15 & $0.91 \pm 0.02$ & $-0.11 \pm 0.01$ & 0.72 & 0.88 & 0.37 & $1.77$ & $0.30$ & 1030 \\
\hline
Overall & 0.14 & $0.84 \pm 0.01$ & $-0.10 \pm 0.00$ & 0.77 & 0.88 & 0.30 & $3.85$ & $0.67$ & 6324 \\
\hline
\hline
\end{tabular}
\caption{Linear fit results between $(L_{\mathrm{H\alpha,corr}})$ and $\log(\nu L_{21\mu\mathrm{m}})$ (see Equation \ref{eq:linregress}). The parameters $\alpha$ and $\beta$ represent the slope and intercept of the linear fit in log–log space, respectively. $R^2$ denotes the coefficient of determination, which quantifies the goodness of fit. $\rho_s$ is the Spearman rank correlation coefficient. $\sigma$ represents the standard deviation of the residuals in log space (dex). $X_{\mathrm{norm}}$ and $Y_{\mathrm{norm}}$ are the normalization factors, and $N$ is the number of data points.}
\label{tab:fit}
\end{table*}

In Figure \ref{fig:all_catalog}, we show the luminosity distribution of different classes of sources separated into different environments. In the 21~$\mu$m catalog across all environments, exposed clusters exhibit \added{luminosities} ranging from 10$^{18}$ to 10$^{20}$ W/Hz, but in the F1000W catalog, their \added{luminosities} are mostly below 10$^{18}$ W/Hz (except centers). Exposed \added{stellar clusters} regions are distributed across different galactic environments, from centers to disks, and are brighter than most other sources in galaxies.  Notably, exposed regions are especially bright in galaxy centers, bar ends, and spiral arms. Galactic bars host the fewest exposed clusters, accounting for less than 10\% of the total exposed cluster population in both catalogs. We also note that, within the full PHANGS sample, bar regions account for about 10\% of the total galactic area \citep{Querejeta2021}. Moreover, the exposed clusters found in bars are among the faintest in our sample. We also note that the number of clusters in bar ends is $<160$; however, only a small fraction of our targets are classified as having bar-end environments (e.g. NGC 3627). Unlike exposed regions, embedded sources do not show noticeable luminosity variations across different galactic environments, although sources are fainter in the disks.

We found that the most detected RSGs are located in the bar, arms, and disk, with the spiral arm and inter-arm regions hosting the most luminous RSGs. We further find that O-AGBs are common across all galactic environments in the 10~$\mu$m catalog. C-AGBs are also common in both catalogs. Furthermore, the median F1000W luminosity of C-AGBs in the 10~$\mu$m catalog is about 50\% lower than that of O-AGBs in the disk. 

\subsection{Correlation}
\label{sec:corr}

\begin{figure*}[!t]
    \centering
    \includegraphics[width=0.48\textwidth]{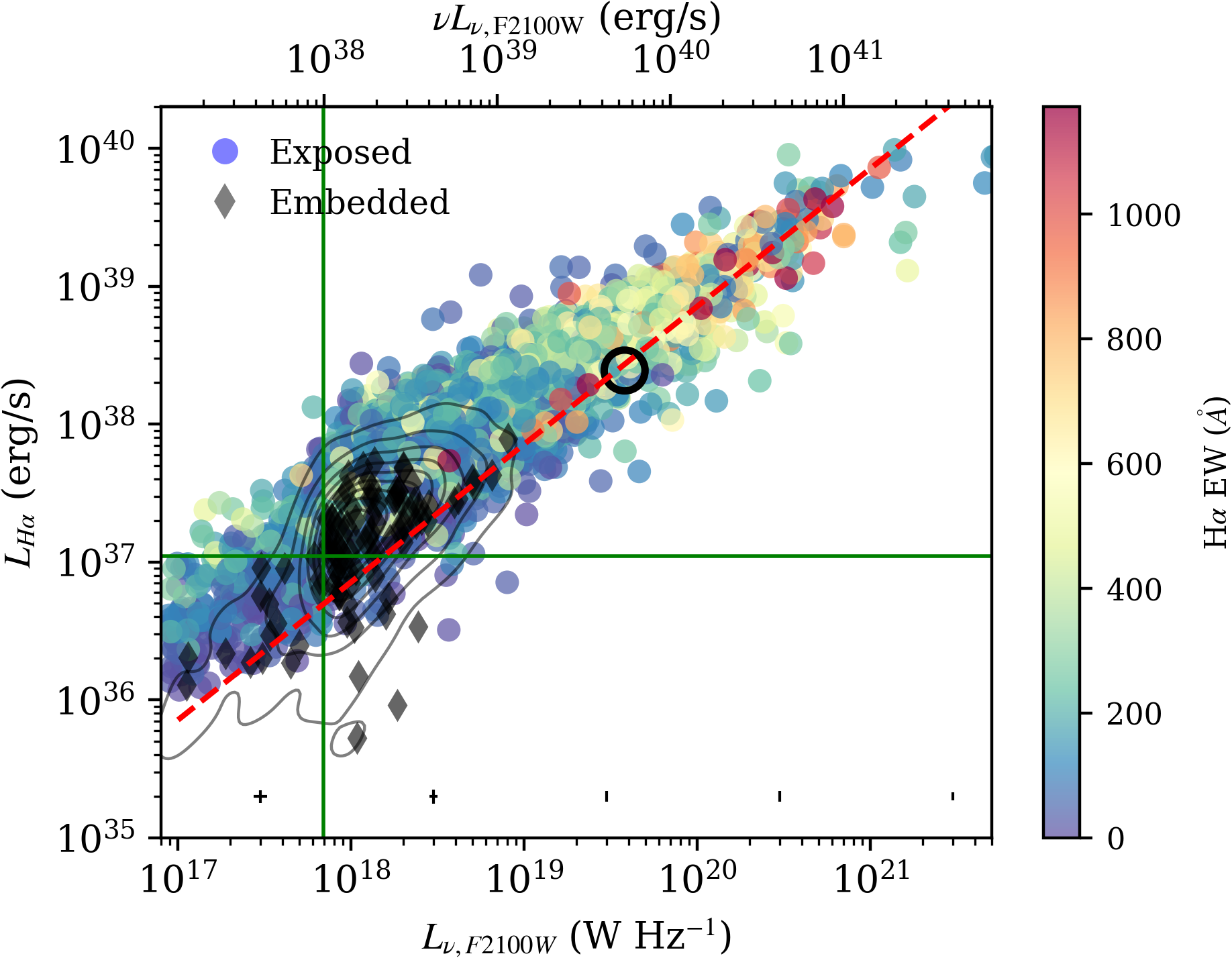}
    \includegraphics[width=0.48\textwidth]{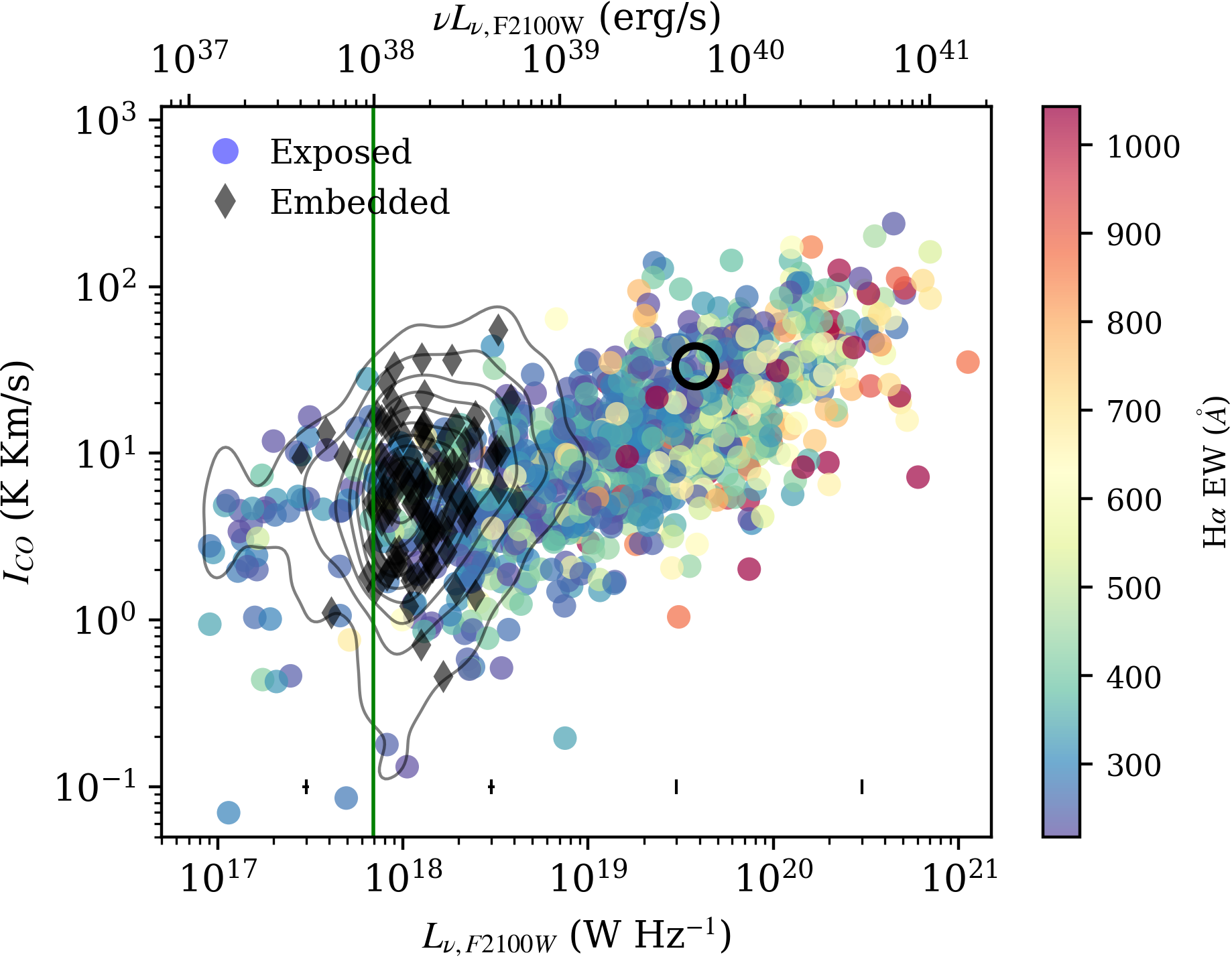}
\caption{Left: The extinction-corrected luminosity of the H$\alpha$ emission line versus the 21~$\mu$m luminosity, where the color bar represents the H$\alpha$ EW in unit of \AA. The red line indicates the mean value of $L_{\mathrm{H}\alpha}/\nu L_{\nu,\mathrm{F2100W}}=1/20$, adopted from \cite{hassani23}. The exposed and embedded regions are represented with different symbols, with only 100 embedded regions randomly displayed here. The black contours represent the kernel density estimation (KDE) with a threshold of 0.05. The green lines are the median of detection limit across sample, adopted from Table \ref{tab:comp}. We also show the representative fractional error bars in the bottom the plot. The black circle highlight the embedded source that nonetheless show H$\alpha$ emission in the aperture (see Section \ref{sec:ismsources} and Figure \ref{fig:deep_emb_cluster}). Right: The intensity of CO versus the 21~$\mu$m luminosity for exposed clusters with H$\alpha$ EW $>200$\AA .}
    \label{fig:sf_relation}
\end{figure*}

We also examine the correlation between F2100W luminosity with (1) attenuation-corrected HST H$\alpha$ emission using the Balmer decrement map from MUSE and (2) CO for exposed sources in Figure \ref{fig:sf_relation}. We also include embedded sources, which are marginally detected in H$\alpha$, so their estimated H$\alpha$ luminosity are likely closer to upper limits. We use the H$\alpha$ EW as the colorbar, as an indirect tracer of the stellar population age \citep{Leitherer99}, uncorrected for differential dust attenuation to the stars and ionized gas. Additionally, we compare with CO intensity, which shows a weaker correlation than H$\alpha$, becoming more scattered when considering only embedded sources. This is largely due to the sensitivity limits of CO observations, which primarily trace GMCs with masses around $ 10^5 $ M$_{\odot}$ \citep{phangsalma}.  Both plots illustrate that the more luminous 21~$\mu$m sources are associated with more H$\alpha$ and CO emission though the correlation with H$\alpha$ is stronger. Similar to \cite{hassani23}, we find a strong correlation ($\rho > 0.8$) spanning five orders of magnitude between the H$\alpha$ attenuation-corrected luminosity and the F2100W luminosity. 

We define $\log(L_{\mathrm{H\alpha,corr}} / \nu L_{21\mu\mathrm{m}})$ (hereafter the “log ratio”), which follows a log-normal distribution with a median value of $\sim$0.15 for exposed and embedded clusters. The median of the log ratio shows little variation across different galactic environments, ranging from 0.10 to 0.16. We also examine the correlation of the log ratio with CO intensity and equivalent width (EW) and find no strong correlation. However, in regions of intense central CO emission ($I_{\mathrm{CO}} > 100$ K km s$^{-1}$), the log ratio decreases to below 0.1, indicating a higher prevalence of clusters with relatively weak H$\alpha$ emission associated with bright CO intensity. This trend is also evident in the right panel of Figure~\ref{fig:sf_relation}. A weak trend is also observed between the log ratio and the F335M flux ($\rho = 0.4$), where the brightest F335M sources tend to have lower log ratios. No significant correlation is found between the log ratio and $r_3$, indicating that $r_3$ remains relatively constant whether a cluster is fully exposed with bright H$\alpha$ emission or more embedded with faint H$\alpha$. 

% This suggests that might F2100W traces a relatively narrow range of stellar population evolution. If this were not the case, clusters with low log ratios ($<0.1$, i.e., faint H$\alpha$ emission) and $r_3 > 1$ would be expected; however, such clusters are not common in our sample. 

We fit a linear relation in log space between the attenuation-corrected HST H$\alpha$ luminosity and the F2100W luminosity, expressed as
\begin{equation}
\log\left(\frac{L_{\mathrm{H\alpha,corr}}}{Y_{\mathrm{norm}}}\right) = 
\alpha \log\left(\frac{\nu L_{\nu,21\mu\mathrm{m}}}{X_{\mathrm{norm}}}\right) + \beta,
\label{eq:linregress}
\end{equation}
where $\alpha$ is the slope and $\beta$ is the intercept. Both luminosities are normalized by their respective medians (denoted as X$_{\mathrm{norm}}$ and Y$_{\mathrm{norm}}$). The slope varies between 0.70 and 0.91 across different galactic environments: it is flatter in spiral arms and bar ends (which host more dusty, young clusters), steeper in the bars and disks, and intermediate in the central regions. The correlation coefficient is high in all environments ($\rho > 0.85$) with a scatter around the linear fit of $\sim$0.3 dex. We find a slope of 0.84 and an intercept of -0.10 all environments. The results are summarized in Table~\ref{tab:fit}.  

For comparison, \citet{Belfiore2023} carry out the reverse analysis, examining the F2100W luminosity of \ion{H}{2} regions finding $I_\mathrm{F2100W} \propto I_{H\alpha}^{0.86}$. This would algebraically imply a steeper index of 1.16 in our analysis. This apparent discrepancy can be attributed in part to the objects of interest in the respective studies: we focus on 21 $\mu$m emitters where \citet{Belfiore2023} focus on the \ion{H}{2} regions detected by MUSE.  However, our finding that the brightest 21 $\mu$m emitters are relatively fainter in H$\alpha$ is also visible in Figure 3 of \citet{Belfiore2023}, which focuses on more typical regions.

 \begin{figure*}[!t]
    \centering
        \includegraphics[width=1\linewidth]{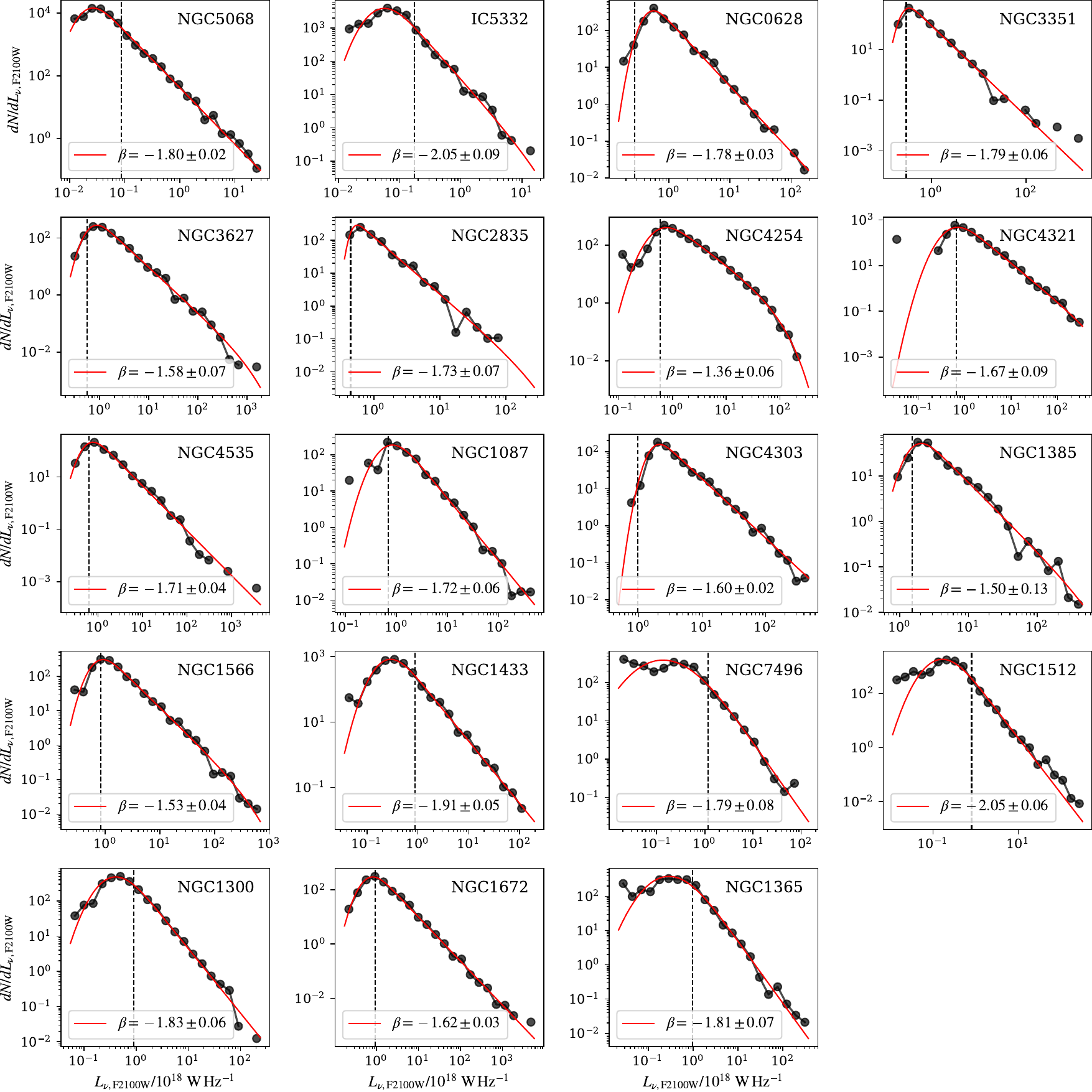}
\caption{The luminosity distribution function of the 21~$\mu$m exposed and embedded sources across different galaxies, sorted by distance. The black line represents the observed data distribution, while the line indicate the fitted model. The vertical dashed black lines mark the detection limits, as listed in Table~\ref{tab:comp}. We excluded data points with $L_{\nu,\mathrm{F2100W}} < 10^{16}$ W Hz$^{-1}$. }
    \label{fig:ldist_21}
\end{figure*}

 \begin{figure*}[!t]
    \centering
        \includegraphics[width=1\linewidth]{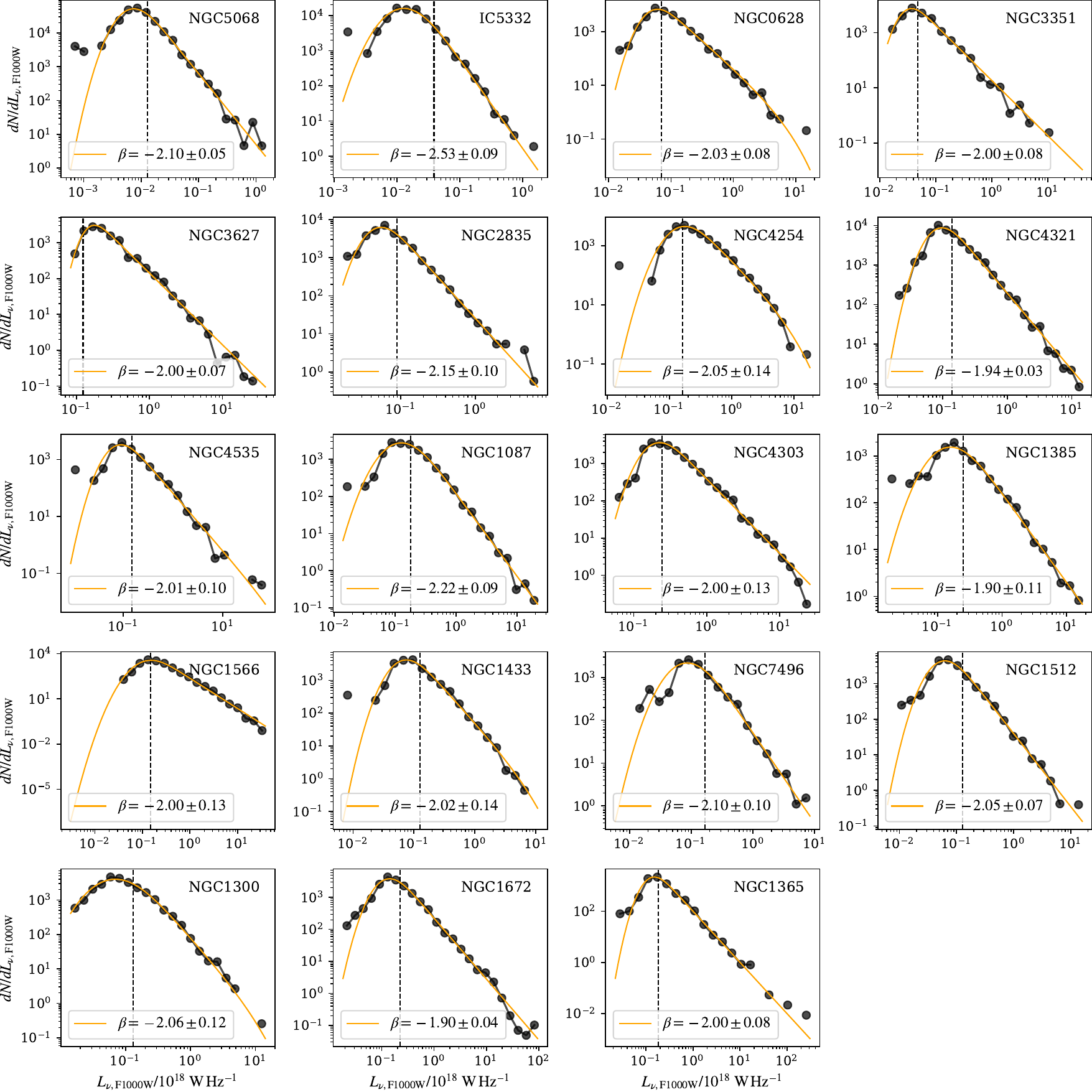}
\caption{Same as Figure \ref{fig:ldist_21}, but for the 10~$\mu$m catalog.}
    \label{fig:ldist_10}
\end{figure*}

\subsection{Luminosity Functions}
\label{sec:physics}

We present the Luminosity Functions (LF) of compact sources, defined as $dN/dL_{\nu}$, for all of the ISM sources in the 10~$\mu\mathrm{m}$ and 21~$\mu\mathrm{m}$ catalogs in Figures~\ref{fig:ldist_21} and \ref{fig:ldist_10}, respectively. The distributions exhibit a turnover at low luminosity. The turnover point generally aligns well with our completeness limit for most targets (see the vertical line in the figures). This turnover is more pronounced in the 10~$\mu$m catalog, which also contains a greater number of faint sources, as observed in galaxies such as NGC~1087, NGC~1365, and NGC~4254. To characterize the slope of the luminosity functions, we employ a Pareto-lognormal model to fit over distribution. This approach combines a Pareto (power-law) distribution to account for the heavy-tailed behavior of bright sources and a lognormal component to capture the turnover at the faint end \citep{basu2015}. Our model PDF also includes an exponential cutoff at high luminosities.  Our adoptied functional form is
%While a power-law tail could be replaced with a linear fit in logarithmic space, we find that some galaxies, particularly NGC~4254, NGC~4303, and NGC~1566, exhibit a more pronounced power-law behavior, especially in their 10~$\mu\mathrm{m}$ distribution. 
% A similar approach was also implemented in the modeling of the mass function of GMCs \citep{rosolowsky21}. 
\begin{equation}
p(L_{\nu}) \propto \alpha \cdot A \cdot \text{CDF}(B) \cdot \left(L_{\nu}\right)^{-\alpha - 1} \cdot \exp\left(-\frac{L_{\nu}}{\mu}\right),
\end{equation}
where $ \alpha $ is the slope of the Pareto tail describing the high-luminosity end, and $ \mu $ is the scale parameter controlling the exponential cutoff at the bright end. We also define $ \beta = -\alpha - 1 $, consistent with the definition reported in \cite{hassani23}. The normalization factor $ A $ and the transition factor $ B $ depend on parameters such as the distance to the galaxy, the faint part of distribution, and the width of the distribution. The term $\text{CDF}(B)$ represents the cumulative distribution function of the normal distribution, which ensures a smooth blending of the high-luminosity tail with the low-luminosity regime. The uncertainties in the slope ($\beta$) were estimated for each galaxy using bootstrap by resampling the luminosity, and the standard deviation of the slopes derived from the Pareto–lognormal fits was reported as the error in $\beta$.  Despite including the factor, we do not find strong evidence for a bright-end exponential cutoff and the parameter $\mu$ is typically not well constrained except in both 10 and 21 $\mu$m catalogs for NGC 4254.

The median slope ($\beta$) across our sample is $-1.73 \pm 0.10$ for 21~$\mu$m sources, while 10~$\mu$m sources exhibit slightly steeper slopes, averaging $-2.01 \pm 0.03$. When focusing exclusively on exposed sources and removing embedded sources from the distributions, the slopes are slightly steeper but remain consistent within the error ranges.  We did not attempt to show the luminosity function of embedded sources, as they exhibit less than one order of magnitude variation in luminosity.

Our results demonstrate that the power-law index of  $\sim -1.7$ is consistent with previous results from \cite{hassani23} for the 21~$\mu$m catalog. These slopes are in close agreement with the average \ion{H}{2} region luminosity function slope of $(-1.73 \pm 0.15)$ obtained from MUSE H$\alpha$ observations \citep{Santoro} of \ion{H}{2} regions, the FUV luminosity function slope of UV complexes $(-1.76 \pm 0.3)$ reported by \cite{Cook2016}, and the WISE 22~$\mu$m luminosity function slope for Milky Way \ion{H}{2} regions $(-1.71 \pm 0.02)$ \citep{Mascoop_2021}.  However, we note here that a direct comparison with the FUV results is not possible, as the emitting timescale of FUV is about 100~Myr, whereas for mid-infrared, it is much shorter, below 5~Myr. Our 10~$\mu$m slopes, which are steeper, around $-2.0$, are in agreement with the luminosity distribution of stellar clusters \citep{Cook2019,Messa2018}. We also compared our $\beta_{21\mu\mathrm{m}}$ slopes with those from \cite{Santoro} and found a Spearman rank correlation coefficient $\rho=0.75$. However, we did not find any significant correlation between our $\beta_{10,\mu\mathrm{m}}$ slopes and the luminosity function of \ion{H}{2} regions observed by MUSE.  We also compared our results with \cite{Pathak2024}, who studied the Probability Density Function (PDF) of galactic disks decomposed into diffuse and \ion{H}{2} region components, and found that the power-law indices estimated in that work are correlated with our estimates ($\rho = 0.7$ for the 21~$\mu$m sources, excluding NGC~4254, NGC 5068 and galaxies with slope errors higher than 0.07).  While both measurements are nominally parameterizing the luminosities for star-forming regions, \citet{Pathak2024} focus on the intensity distribution whereas this analysis is framed around discrete objects.  We also find a correlation ($\rho=0.71$) between the mean of the \citet{Pathak2024} log-normal distributions and the fraction of emission found at larger scales in our images $f_\mathrm{LS}$ for both of our catalogs.

\begin{table}[h!]
\centering
\begin{tabular}{lcc}
\hline
  Galaxy & $\beta_{21\mathrm{\mu m}}$ & $\beta_{10\mathrm{\mu m}}$ \\
\hline
  IC5332 & $-2.05 \pm 0.09$ & $-2.53 \pm 0.09$ \\
  NGC0628 & $-1.78 \pm 0.03$ & $-2.03 \pm 0.08$ \\
  NGC1087 & $-1.72 \pm 0.06$ & $-2.22 \pm 0.09$ \\
  NGC1300 & $-1.83 \pm 0.06$ & $-2.06 \pm 0.12$ \\
  NGC1365 & $-1.81 \pm 0.07$ & $-2.00 \pm 0.08$ \\
  NGC1385 & $-1.50 \pm 0.13$ & $-1.90 \pm 0.11$ \\
  NGC1433 & $-1.91 \pm 0.05$ & $-2.02 \pm 0.14$ \\
  NGC1512 & $-2.05 \pm 0.06$ & $-2.05 \pm 0.07$ \\
  NGC1566 & $-1.53 \pm 0.04$ & $-2.00 \pm 0.13$ \\
  NGC1672 & $-1.62 \pm 0.03$ & $-1.90 \pm 0.04$ \\
  NGC2835 & $-1.73 \pm 0.07$ & $-2.15 \pm 0.10$ \\
  NGC3351 & $-1.79 \pm 0.06$ & $-2.00 \pm 0.08$ \\
  NGC3627 & $-1.58 \pm 0.07$ & $-2.00 \pm 0.07$ \\
  NGC4254 & $-1.36 \pm 0.06$ & $-2.05 \pm 0.14$ \\
  NGC4303 & $-1.60 \pm 0.02$ & $-2.00 \pm 0.13$ \\
  NGC4321 & $-1.67 \pm 0.09$ & $-1.94 \pm 0.03$ \\
  NGC4535 & $-1.71 \pm 0.04$ & $-2.01 \pm 0.10$ \\
  NGC5068 & $-1.80 \pm 0.02$ & $-2.10 \pm 0.05$ \\
  NGC7496 & $-1.79 \pm 0.08$ & $-2.10 \pm 0.10$ \\
\hline
  Median & $-1.73 \pm 0.10$ & $-2.02 \pm 0.03$ \\
\hline
\end{tabular}
\caption{The table presents the results of the fits to the luminosity distribution function. }
\label{tab:l_dist}
\end{table}

\begin{figure}[!t]
    \centering
        \includegraphics[width=1.13\linewidth]{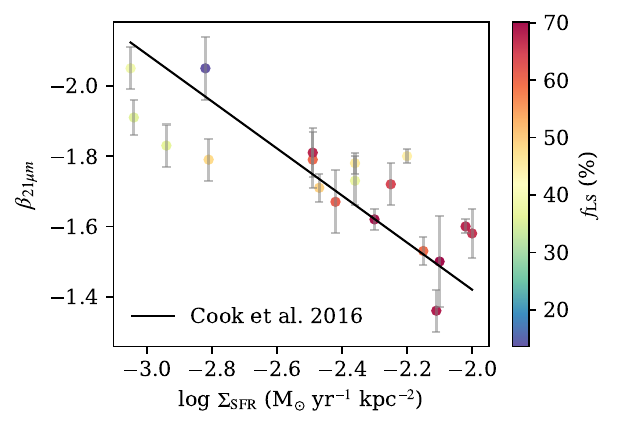 }
\caption{\added{The relation between $\beta_{21,\mu\text{m}}$, the slope of the luminosity function, and the star formation rate surface density ($\Sigma_{\text{SFR}}$) is shown. The black solid line represents the fit from \citet{Cook2016}, given by $\alpha = - 0.67 \times \Sigma_{\mathrm{SFR}} + 0.08$. }}
    \label{fig:betaplot}
\end{figure}

We find that the 21~$\mu$m LF slopes correlate well with the star formation rate surface density ($\Sigma_\mathrm{SFR}$), with $\rho=0.86$ ($p < 10^{-5}$).  Figure~\ref{fig:betaplot} shows the relationship between $\beta_{21\mu\text{m}}$ and $\Sigma_{\mathrm{SFR}}$, color-coded by $f_{\mathrm{LS}}$ from Table \ref{tab:comp}. This trend is primarily driven by the low-mass, low-SFR galaxies, which have fewer active star-forming regions and a smaller fraction of large-scale emission, also showing steeper slopes (e.g., IC~5332). In contrast, galaxies with shallower luminosity functions, which form a relatively larger number of massive young clusters, typically also exhibit a higher fraction of large-scale emission at F2100W, resulting in higher $\Sigma_\mathrm{SFR}$ \citep{Santoro}. We note that the correlation between the 10~$\mu$m LF slopes and $\Sigma_\mathrm{SFR}$ is weaker. In addition, we find a correlation coefficient of $\rho = 0.69$ with a $p$-value of $10^{-3}$ between $\beta_{10\mu\text{m}}$ and SFR of galaxies.  We also note that galaxies with high $\Sigma_\mathrm{SFR}$ also have larger values of $f_\mathrm{LS}$ as well as shallower indices.

% \cite{Johnson} found that a maximum GMC mass of $10^{4}~M_{\odot}$ is reached at $\log (\Sigma_\text{SFR} [M_{\odot}~\text{yr}^{-1}~ \text{kpc}^{-2}]) \approx -2.6$. At higher $\Sigma$SFR values, the slope of bright \ion{H}{2} regions, such as our 21~$\mu$m peaks, is expected to more closely resemble that of the GMC mass function, which is approximately $-1.6$ \citep{Santoro,Colombo}. We highlight this transition in the luminosity function slope of 21~$\mu$m in Figure \ref{fig:betaplot}. Again, we emphasize that while this transition is well pronounced in the luminosity distribution slopes observed by FUV observations from \cite{Cook2016}, it is less significant in our case, as mid-infrared observations capture a much narrower window of the stellar lifecycle.

\section{Conclusion}
\label{sec:conv}

In this study, we analyzed multiwavelength data from JWST NIRCam/MIRI, HST wide band, and narrow band H$\alpha$ to investigate the nature of 10 and 21~$\mu$m compact sources in 19 nearby galaxies. Building on the findings of \cite{hassani23}, we classified mid-infrared compact sources in our galaxies, from the earliest phases of star formation to the later stages of stellar evolution, such as AGB stars. Along with this study, we release our code for source finding, photometry, and the AST on GitHub\footnote{\url{https://github.com/hamidnpc/neloura}}.

To generate our catalogs, we first filter the maps using a constrained diffusion method \citep{Li2022}. We filtered out large-scale structures to isolate more bright mid-infrared compact sources while preserving total flux. This method allows us to recover faint, isolated sources that were otherwise blended with the background or large-scale filaments. We identify sources with a dendrogram-based source identification method and carry out aperture photometry at the peak locations in the filtered map. 

Our main findings are that:

\begin{itemize}
    \item We identified 24,945 sources at 21~$\mu$m and 55,581 sources at 10~$\mu$m. To ensure consistent resolution across different wavelengths, we convolved all maps from HST NUV to F1130W to the resolution of 11.3~$\mu$m (0.36\arcsec) for the 10~$\mu$m catalog. For the 21~$\mu$m catalog, all maps were convolved to the resolution of 21~$\mu$m (0.67\arcsec).
    
    \item We found that large-scale emission contributes significantly to the total flux in mid-infrared bands. Large-scale emission accounts for  $\sim 50\%$ of the total flux. This highlights the importance of isolating compact sources from large-scale (diffuse) emission to identify sources.

    \item The total flux of 21~$\mu$m compact sources accounts for about 20\% of the total galaxy flux, which is three times higher than the contribution from 10~$\mu$m sources.

    \item Using artificial star tests, we estimated the completeness limits for the 10 and 21~$\mu$m maps, finding a median of 5~$\mu$Jy in F1000W maps and about five times higher, at 24~$\mu$Jy, in F2100W maps. These limits show minimal variation with local background levels for 21~$\mu$m sources and remain consistent below 3~MJy/sr for the 10~$\mu$m catalog.

    \item We classified sources using flux ratios in the different bands based on models from CIGALE for young stellar clusters, \texttt{PARSEC} models for evolved stars, and empirical compilations of source classifications based on previous infrared mission of \textit{Spitzer}. We categorized dusty clusters into embedded and exposed populations, with the latter requiring a bright H$\alpha$ association. We also identify evolved stars and classify them as RSGs, O-AGBs, and C-AGBs. We also find candidate B[e] stars, WR stars, and CPNs / C-PAGBs where these latter two categories are indistinguishable in our observations.

    \item Because we are sensitive to lower source luminosities in nearer targets, the composition of our catalogs varies as a function of distance to the target. RSGs are expected to be observed only in galaxies within 10-15~Mpc, primarily in the F1000W band, if they are younger than 10~Myr. O-AGBs can be detected at greater distances but must be relatively young, with ages below 1~Gyr and C/O $<$ 0.5. C-AGBs, on the other hand, are detectable throughout the sample with F1000W band. We identified approximately 2500 RSGs and O-AGBs in the 10~$\mu$m catalog, with more than 350 C-AGBs. In the F2100W catalog, we classified about 220 RSGs and O-AGBs and approximately 60 C-AGBs. 

    \item The brightest mid-infrared sources are predominantly located in galaxy centers, followed by bar ends and spiral arms, where most exposed clusters are found in both catalogs. Other sources, such as embedded clusters, RSGs, O-AGBs, and C-AGBs, do not exhibit significant variations in their flux distributions across different galactic environments. \added{We find that the median luminosities of O-AGB stars are brighter than those of C-AGB stars in bars, spiral arms, interarm regions, and disks in our 10~$\mu$m catalog.}

    \item We found that ``embedded'' clusters make up $<5\%$ of our whole catalog and about 10 percent of sources with strong ISM emission. These likely-young stellar clusters are characterized by bright F2100W or F1000W emission, with little to no H$\alpha$ emission and a weak H$\alpha$ EW of 40 \AA, which is three times weaker than that of exposed clusters. The observed H$\alpha$ equivalent widths in exposed clusters likely correspond to very young ages---probably less than 5~Myr---in agreement with the 21\,$\mu$m peaks identified by \citet{hassani23,Whitmore2025}. We note that both embedded and exposed phases of star-forming regions exhibit similar PAH-to-continuum ratios (e.g., F335M ratios relative to F300M and F360M exceeds 1.5). This suggests minimal contamination from dusty stars in the classification of embedded sources, as they would not produce such strong PAH emission.

    \item We modeled the luminosity distribution of 10 and 21~$\mu$m ISM sources using a Pareto-lognormal function. We found that the power-law slope for 21~$\mu$m sources is $-1.7 \pm 0.1$, which is consistent with the slopes of the giant molecular cloud (GMC) mass function, bright H$\alpha$-emitting \ion{H}{2} regions, and FUV star-forming complexes. In contrast, the slope for 10~$\mu$m sources is steeper at $-2.0 \pm 0.1$, resembling more closely the distribution of stellar clusters. The slope of the 21~$\mu$m luminosity function correlates with both the total SFR and the SFR surface density, while this correlation is weaker for the 10~$\mu$m catalog. Given that the timescale of mid-infrared emission is less than 5~Myr, the shape of the luminosity function is directly related to global star formation properties in galaxies.

    \item We investigated the variation of $\langle\log(L_{\mathrm{H\alpha,corr}} / \nu L_{\nu,21\mu\mathrm{m}})\rangle$ across different galactic environments and found only a slight change, ranging from 0.10 to 0.17. We find a sub-linear relation between the attenuation-corrected H$\alpha$ luminosity and the F2100W luminosity, with a slope of $\sim$0.78 in spiral arms, steeper slopes in bars and disks ($\sim$0.91), and an intermediate slope in galaxy centers ($\sim$0.83). We find no strong correlation with CO intensity or with the MUSE H$\alpha$ equivalent width. Moreover, we detect no correlation between $r_{3}$ and the log ratio, suggesting that regardless of how optically exposed a cluster is (i.e., H$\alpha$-bright), the relative amount of F335M PAH emission compared to its continuum remains nearly constant.

\end{itemize}

Our study highlights the utility of mid-infrared emission in tracing both early and late stages of stellar evolution in galaxies. Warm dust emission serves as a key tracer of embedded star formation, where H$\alpha$ emission is not yet visible due to dust obscuration. Additionally, mid-infrared emission can be used to identify evolved stellar populations such as AGB stars. The connection between mid-infrared emission and different stellar populations provides a comprehensive view of star formation and stellar evolution in galaxies. In future work, we will estimate the physical properties of the dusty young stellar clusters.

\software{Python, Astropy \citep{astropy,astropy2,astropy2022}, Numpy \citep{numpy}, Spectral-cube \citep{spectral-cube}, Photutils \citep{photutils}, and Matplotlib \citep{matplotlib}, Neloura \citep{hamid_hassani_2026_18226636}, ChatGPT \citep{openai2024chatgpt}, Claude \citep{anthropic2024claude}}

\appendix

\section{Catalog Information}
\label{app:catalog}
We release our catalogs via the PHANGS data webpage\footnote{\url{https://www.phangs.org/data}} and at \url{https://dx.doi.org/10.11570/26.0003}. Table \ref{tab:fits_extensions} presents our the table format and associated units for the different quantities listed in the catalog.

\begin{table*}[t]
\centering
\begin{tabular}{l p{2cm} p{8cm}}
\hline
\textsc{Extension Name} & \textsc{Unit} & \textsc{Description} \\
\hline
\multicolumn{3}{l}{\textbf{Galaxy Information}} \\
galaxy & - & Galaxy name \\
ra, dec & deg & Right Ascension, Declination (J2000) \\
bmaj & arcsec & Resolution of the catalog (0.67\arcsec\ for 21$\mu$m catalog, 0.36\arcsec\ for 10$\mu$m catalog) \\
\hline
\multicolumn{3}{l}{\added{Observed Fluxes (broad-band)}} \\
F2100W, F1130W, F1000W, F770W & $\mu$Jy & MIRI fluxes \\
F360M, F335M, F300M, F200W & $\mu$Jy & NIRCam fluxes \\
F814W, F555W, F438W, F336W, F275W & $\mu$Jy & Optical/UV fluxes (HST) \\
HST\_Halpha & erg\,s$^{-1}$\,cm$^{-2}$ & H$\alpha$ flux (HST) \\
CO & K km/s & CO emission flux \\
EBV & mag & E(B-V) reddening \\
\hline
\multicolumn{3}{l}{\added{Observed Fluxes (emission lines, MUSE)}} \\
MUSE\_Halpha, HB4861, OIII4958, OIII5006, & erg\,s$^{-1}$\,cm$^{-2}$ & Integrated line fluxes \\
NI5197, NI5200, NII5754, HEI5875, OI6300, & \\
SIII6312, OI6363, NII6548, HA6562, NII6583, & & \\
SII6716, SII6730 & & \\
EW & \AA & Equivalent width of MUSE H$\alpha$ \\
\hline
\multicolumn{3}{l}{\added{Flux Uncertainties / SNR / Backgrounds}} \\
\_err & same as flux & Measurement uncertainties for each band/line \\
\_snr & - & Signal-to-noise ratio for each band/line \\
\_bkg & same as flux & Estimated background level for each band/line \\
\hline
\multicolumn{3}{l}{\added{Classification Flags}} \\
ISM\_source, ism\_source\_miri, ism\_source\_nircam & Boolean & ISM classification flags \\
stars\_bkg, stars\_bkg\_miri, stars\_bkg\_nircam & Boolean & Star/background flags \\
env & - & Environmental parameter (1--10, from \cite{Querejeta2021}) \\
emb, emb\_true & Boolean & Embedded cluster flag \\
expo, expo\_true & Boolean & Exposed cluster flag \\
deep\_emb$^{1}$ & Boolean & Deeply Embedded cluster flag \\
cpn, wr, be, bkg, FG & Boolean & Special classification flags (C-rich Planetary Nebula, \\
& & Wolf-Rayet, B[e] stars, background galaxies, foreground stars) \\
rsg, oagb, cagb & Boolean & Red supergiant, O-rich AGB, and C-rich AGB stars \\
\hline
\multicolumn{3}{l}{\added{Others}} \\
Ha\_compLim & erg\,s$^{-1}$ & H$\alpha$ completeness limit \\
21um\_compLim & $\mu$Jy & 21$\mu$m completeness limit \\
L\_Ha, L\_Ha\_corr & erg\,s$^{-1}$ & H$\alpha$ luminosity (observed and extinction-corrected) \\
L\_21um, nuL\_21um & W\,Hz$^{-1}$, erg\,s$^{-1}$ & 21$\mu$m luminosity \\
bkg\_major, bkg\_minor, bkg\_orr & - & Background galaxy major/minor axes and orientation \\
\hline
\end{tabular}
\caption{Description of catalog fields.\\
$^{1}$ Refer to clusters with H$\alpha$ luminosity above the H$\alpha$ detection limit, and F2100W luminosity ($L_{F2100W}$) at least three times higher than the detection limit at F2100W, with a F335W SNR greater than 3, but an HST H$\alpha$ SNR less than 3 (See section \ref{sec:emb_sources}).}
\label{tab:fits_extensions}
\end{table*}
% here are the plots

\added{
\section{Magnitude Offsets}
In this work, we present photometric measurements in terms of flux densities, usually measured in $\mu$Jy. These scale to the AB magnitude system using the usual relationship: $m_\mathrm{AB} = -2.5\log_{10} f_\nu + 8.90$.  Since many researchers still use the Vega magnitude system, we include the adopted offsets between these two systems in Table \ref{tab:abvega} based on the work of the JWST calibration team and embedded in the JWST pipeline \citep{jwst-pipeline}.
\begin{table}[]
    \centering
    \begin{tabular}{cc}
Filter & $m_\mathrm{AB}-m_\mathrm{Vega}$ \\
\hline
 F200W & 1.686 \\
 F300M & 2.481 \\
 F335M & 2.707 \\
 F360M & 2.857 \\
 F770W & 4.383 \\
F1000W & 4.955 \\
F1130W & 5.240 \\
F2100W & 6.532 \\      
    \end{tabular}
    \caption{Offsets between the AB and Vega Magnitude system.  These are extracted from the JWST pipeline calibration files \citep{jwst-pipeline} for context \texttt{jwst\_1464.pmap}.}
    \label{tab:abvega}
\end{table}
}

\section{Completeness Limits in Different Environments}
\label{app:comp}
\begin{figure*}[!t]
    \centering
        \includegraphics[width=1\linewidth]{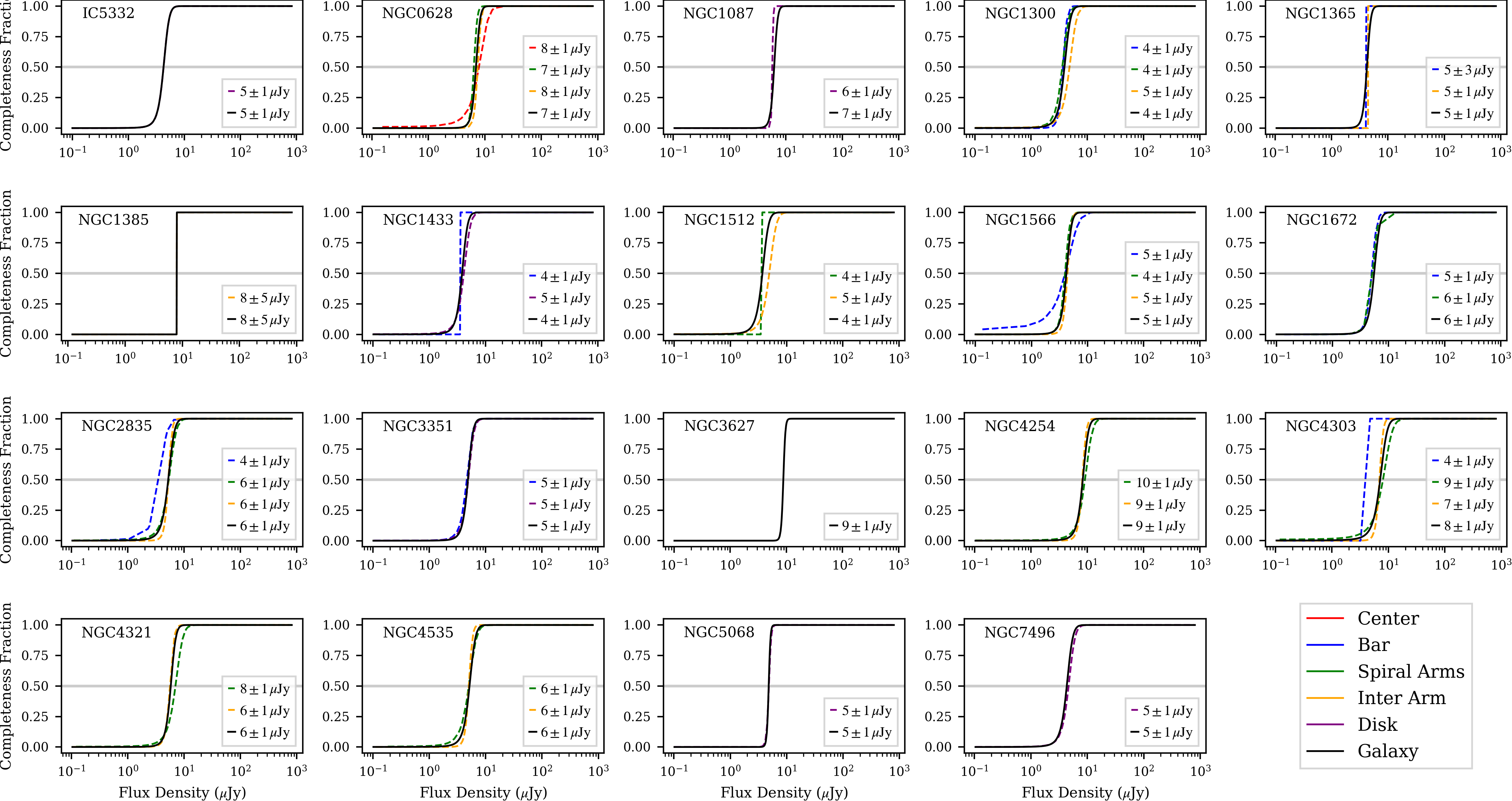}
\caption{Completeness limits for our sample are shown for the entire galaxy (black) and separately for each galactic environment (colored curves) in the F1000W band. Unlike Figure~\ref{fig:sources}, we display only the logistic fits to the data, rather than the individual data points, to improve visualization. Fits are shown only for environments with more than 10 data points.}
    \label{fig:10um}
\end{figure*}

\begin{figure*}[!t]
    \centering
        \includegraphics[width=1\linewidth]{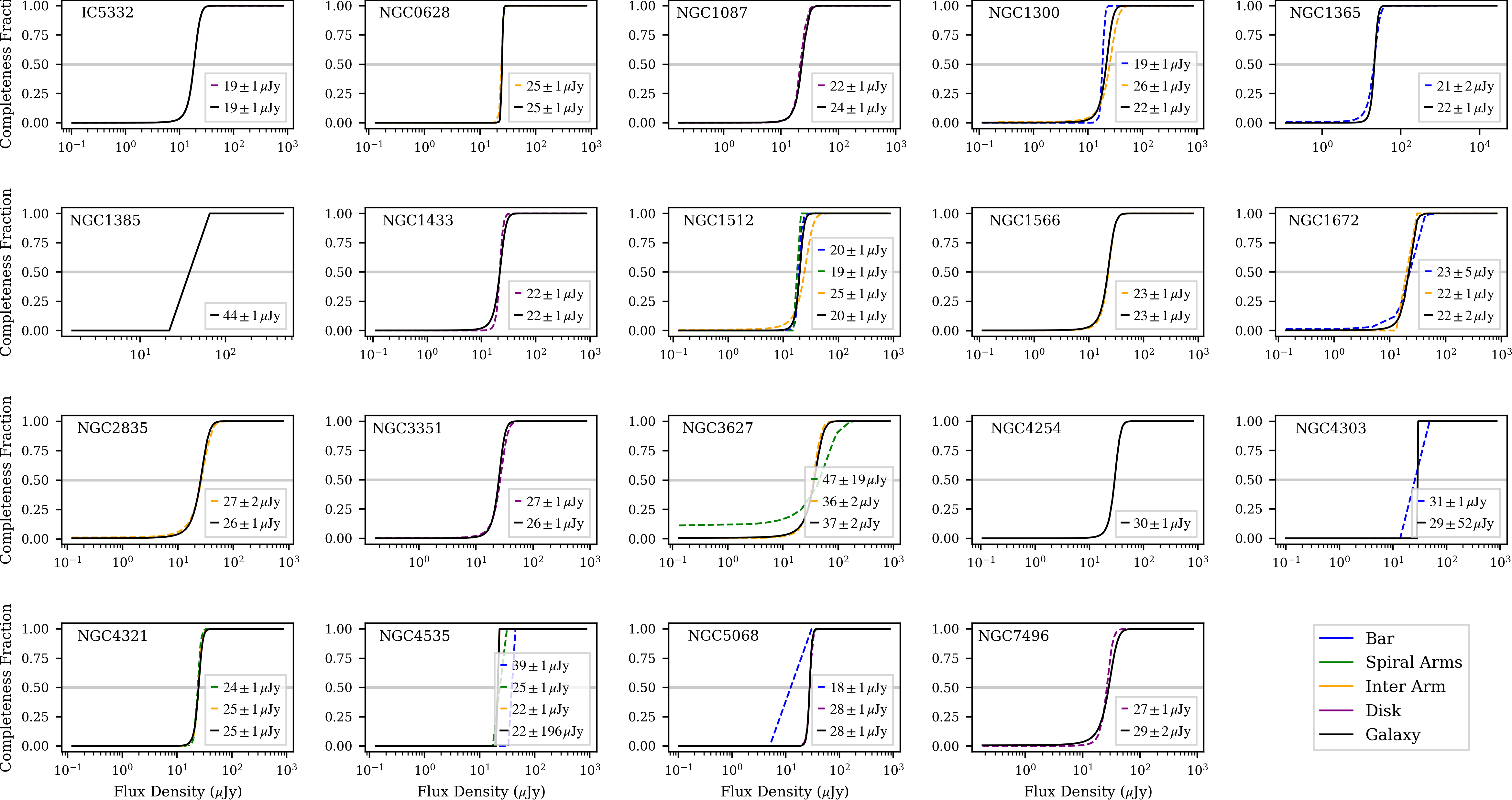}
\caption{Same as \ref{fig:10um}, but for F2100W band.}
    \label{fig:21um}
\end{figure*}

Here we present the completeness limits for all targets, both for the galaxies as a whole and subdivided by galactic environment, in the F1000W band (Figure~\ref{fig:10um}) and the F2100W band (Figure~\ref{fig:21um}).

\begin{acknowledgments}

This work has been carried out as part of the PHANGS collaboration.

HH acknowledges the use of the Canadian Advanced Network for Astronomy Research (CANFAR) Science Platform operated by the Canadian Astronomy Data Center (CADC) and the Digital Research Alliance of Canada (DRAC), with support from the National Research Council of Canada (NRC), the Canadian Space Agency(CSA), CANARIE, and the Canadian Foundation for Innovation (CFI).  HH also acknowledges the use of ChatGPT \citep{openai2024chatgpt} and Claude \citep{anthropic2024claude} to improve the structure and grammar of the text.

HH and ER acknowledge the support of the Natural Sciences and Engineering Research Council of Canada (NSERC), funding reference number RGPIN-2022-03499, and the Canadian Space Agency funding reference numbers JWSTAO2 and 23JWGO2A07. 

MB acknowledges support by the ANID BASAL project FB210003. This work was supported by the French government through the France 2030 investment plan managed by the National Research Agency (ANR), as part of the Initiative of Excellence of Université Côte d’Azur under reference number ANR-15-IDEX-01.

OE acknowledges funding from the Deutsche Forschungsgemeinschaft (DFG, German Research Foundation) -- project-ID 541068876.

LR gratefully acknowledges funding from the DFG through an Emmy Noether Research Group (grant number CH2137/1-1).

KG is supported by the Australian Research Council through the Discovery Early Career Researcher Award (DECRA) Fellowship (project number DE220100766) funded by the Australian Government. 

KG is supported by the Australian Research Council Centre of Excellence for All Sky Astrophysics in 3 Dimensions (ASTRO~3D), through project number CE170100013. 

A.K.L. gratefully acknowledges support from NSF AST AWD 2205628, JWST-GO-02107.009-A, and JWST-GO-03707.001-A and a Humboldt Research Award.

FHL acknowledges funding from the European Research Council’s starting grant ERC StG-101077573 (`ISM-METALS').

JPe acknowledges support by the French Agence Nationale de la Recherche through the DAOISM grant ANR-21-CE31-0010 and by the Thematic Action “Physique et Chimie du Milieu Interstellaire” (PCMI) of INSU Programme National “Astro”, with contributions from CNRS Physique \& CNRS Chimie, CEA,
and CNES.

IP acknowledges funding from the European Research Council (ERC) under the European Union's Horizon 2020 research and innovation programme (grant agreement 101020943, SPECMAP-CGM).

MQ acknowledges support from the Spanish grant PID2022-138560NB-I00, funded by MCIN/AEI/10.13039/501100011033/FEDER, EU.

This work is based on observations made with the NASA/ESA/CSA JWST. The data were obtained from the Mikulski Archive for Space Telescopes at the Space Telescope Science Institute, which is operated by the Association of Universities for Research in Astronomy, Inc., under NASA contract NAS 5-03127 for JWST. These observations are associated with programs 2107. 

This paper makes use of the following ALMA data, which have been processed as part of the PHANGS--ALMA CO~(2-1) survey: \\
\noindent ADS/JAO.ALMA\#2012.1.00650.S, \linebreak % (N628/M74)
ADS/JAO.ALMA\#2013.1.00803.S, \linebreak % (N5128/CenA)
ADS/JAO.ALMA\#2013.1.01161.S, \linebreak % (N1365 + N5236/M83)
ADS/JAO.ALMA\#2015.1.00121.S, \linebreak % (N5236/M83)
ADS/JAO.ALMA\#2015.1.00782.S, \linebreak % (N1313 + N7793)
ADS/JAO.ALMA\#2015.1.00925.S, \linebreak % (pilot low mass)
ADS/JAO.ALMA\#2015.1.00956.S, \linebreak % (pilot high mass)
ADS/JAO.ALMA\#2016.1.00386.S, \linebreak % (N5236/M83)
ADS/JAO.ALMA\#2017.1.00392.S, \linebreak % (low mass follow-up)
ADS/JAO.ALMA\#2017.1.00766.S, \linebreak % (early-type)
ADS/JAO.ALMA\#2017.1.00886.L, \linebreak % (large program)
ADS/JAO.ALMA\#2018.1.01321.S, \linebreak % (N253, N300, Circinus)
ADS/JAO.ALMA\#2018.1.01651.S, \linebreak % (main sample follow-up)
ADS/JAO.ALMA\#2018.A.00062.S, \linebreak % (ACA-only nearby)
ADS/JAO.ALMA\#2019.1.01235.S, \linebreak % (local sample follow up)
ADS/JAO.ALMA\#2019.2.00129.S, \linebreak % (N1068)
ALMA is a partnership of ESO (representing its member states), NSF (USA), and NINS (Japan), together with NRC (Canada), NSC and ASIAA (Taiwan), and KASI (Republic of Korea), in cooperation with the Republic of Chile. The Joint ALMA Observatory is operated by ESO, AUI/NRAO, and NAOJ. The National Radio Astronomy Observatory is a facility of the National Science Foundation operated under cooperative agreement by Associated Universities, Inc.

\end{acknowledgments}

\bibliographystyle{aasjournal}
\bibliography{main}



\end{document}